\documentclass[useAMS,usegraphicx,usenatbib]{mn2e}
\usepackage{times}

\newcommand{\Msun}{\ensuremath{\,{\rm M}_\odot}}                  
\newcommand{\Rsun}{\ensuremath{\,{\rm R}_\odot}}                  
\newcommand{\psun}{\ensuremath{\,\rho_\odot}}                     
\newcommand{\Mjup}{\ensuremath{\,{\rm M}_{\rm Jup}}}              
\newcommand{\Rjup}{\ensuremath{\,{\rm R}_{\rm Jup}}}              
\newcommand{\pjup}{\ensuremath{\,\rho_{\rm Jup}}}                 
\newcommand{\Mearth}{\ensuremath{\,{\rm M}_\oplus}}               
\newcommand{\Teff}{\ensuremath{T_{\rm eff}}}                      
\newcommand{\Teq}{\ensuremath{T_{\rm eq}^{\,\prime}}}             
\newcommand{\safronov}{\ensuremath{\Theta}}                       
\newcommand{\logg}{\ensuremath{\log g}}                           
\newcommand{\FeH}{\ensuremath{\left[\frac{\rm Fe}{\rm H}\right]}} 
\newcommand{\Porb}{\ensuremath{P_{\rm orb}}}                      
\newcommand{\ms}{\,m\,s$^{-1}$}                                   
\newcommand{\mss}{\,m\,s$^{-2}$}                                  
\newcommand{\as}{\ensuremath{^{\prime\prime}}}                    
\newcommand{\chir}{\ensuremath{\chi_\nu^{\,2}}}                   
\newcommand{\lten}{\ensuremath{\log_{10}}}                        

\newcommand{\mc}[1]{\multicolumn{2}{c}{#1}}
\newcommand{\mcc}[1]{\multicolumn{3}{c}{#1}}
\newcommand{\er}[3]{\ensuremath{#1^{+#2}_{-#3}}}

\newcommand{\erc}[3]{\mc{\ensuremath{#1^{+#2}_{-#3}}}}

\newcommand{\ercc}[3]{\mcc{\ensuremath{#1^{+#2}_{-#3}}}}
\newcommand{\ermcc}[5]{\mcc{\ensuremath{{#1\,^{+#2}_{-#3}}\,^{+#4}_{-#5}}}}
\newcommand{\kepler}{{\it Kepler}}
\newcommand{\spitzer}{{\it Spitzer}}
\newcommand{\corot}{CoRoT}
\newcommand{\reff}[1]{{#1}}
\newcommand{\wwo}[2]{{#1}}

\newcounter{appref}                                                     
\newcommand{\apptab}{\addtocounter{appref}{1}Table\,A\arabic{appref}}   
\newcommand{\apptabb}{Tables \addtocounter{appref}{1}A\arabic{appref} and \addtocounter{appref}{1}A\arabic{appref}}
\newcommand{\apptabbb}[1]{Tables \addtocounter{appref}{1}A\arabic{appref} to \addtocounter{appref}{#1}\addtocounter{appref}{-1}A\arabic{appref}}

\newcommand{\smfig}[2]{{#2}}                                            

\title[Studies of transiting extrasolar planets. V.]
      {Homogeneous studies of transiting extrasolar planets. V. New results for 38 planets}

\author[John Southworth]
       {John Southworth\thanks{E-mail: jkt@astro.keele.ac.uk} \\
        Astrophysics Group, Keele University, Staffordshire, ST5 5BG, UK}

\begin{document} \maketitle 

\begin{abstract}
I measure the physical properties of 38 transiting extrasolar planetary systems, bringing the total number studied within the {\it Homogeneous Studies} project to 82. Transit light curves are modelled using the {\sc jktebop} code, with careful attention paid to limb darkening, orbital eccentricity and contaminating light. The physical properties of each system are obtained from the photometric parameters, published spectroscopic measurements and five sets of theoretical stellar model predictions. Statistical errors are assessed using Monte Carlo and residual-permutation algorithms and propagated via a perturbation algorithm. Systematic errors are estimated from the interagreement between results calculated using five theoretical stellar models.

The headline result is a major upward revision of the radius of the planet in the OGLE-TR-56 system, from $1.23$--$1.38$\Rjup\ to $1.734 \pm 0.051 \pm 0.029$\Rjup\ (statistical and systematic errors, respectively). Its density is three times lower than previously thought. This change comes from the first complete analysis of published high-quality photometry. Significantly larger planetary radii are also found for Kepler-15, KOI-428, WASP-13, WASP-14 and WASP-21 compared to previous work.

I present the first results based on \kepler\ short-cadence data for Kepler-14, Kepler-15 and KOI-135. More extensive long-cadence data from the \kepler\ satellite is used to improve the measured properties of KOI-196, KOI-204, KOI-254, KOI-423 and KOI-428. The stellar component in the KOI-428 system is the largest known to host a transiting planet, at $2.48 \pm 0.17 \pm 0.20$\Rsun. Detailed analyses are given for HAT-P-3, HAT-P-6, HAT-P-9, HAT-P-14 and WASP-12, based on more extensive datasets than considered in previous studies.

\reff{Detailed analyses are also presented for the \corot\ systems 17, 18, 19, 20 and 23; Kepler- 7, 12 and 17; KOI-254; OGLE-TR- 111, 113, 132 and L9; and TrES-4.}

I revisit the correlations between orbital period and surface gravity, and orbital period and mass of the transiting planets, finding both to be significant at the $4\sigma$ level. I conclude by discussing the opportunities for follow-up observations, the sky positions and the discovery rate of the known transiting planets.
\end{abstract}



\section{Introduction}                                                                                                              \label{sec:intro}

Even though the first transiting extrasolar planet (TEP) was discovered only as recently as 1999 \citep{Henry+00apj,Charbonneau+00apj}, the number of recognised TEPs -- and the size of the literature on them -- has grown very quickly. The known planetary systems now display a bewildering diversity of qualities: from super-Earths with extremely short periods \citep[\corot-7\,b;][]{Leger+09aa} to massive planets on wide and highly eccentric orbits \citep[HD\,80606\,b;][]{Hebrard+10aa}; a system of six planets showing strong mutual gravitational perturbations \citep[Kepler-11;][]{Lissauer+11nat}; circumbinary planets \citep{Doyle+11sci,Welsh+12nat}; and ones with a size inexplicably approaching twice that of Jupiter \citep[WASP-17\,b;][]{Anderson+10apj,Me+12c}.

Part of the reason for the known TEP population being such a menagerie is the variety of methods deployed in catching them. Many of the TEPs orbiting brighter stars (including the first, HD\,209458\,b) were found by radial velocity surveys and later shown to transit. Others (such as the second known TEP OGLE-TR-56\,b; \citealt{Konacki+03nat}) were unearthed via deep photometric searches and subsequently confirmed spectroscopically. The dominant population of TEPs currently stems from large-scale photometric surveys of stars of intermediate brightness ($V=9$--$13$) coupled with spectroscopic follow-up programs, such as those of HAT \citep{Bakos+02pasp} and WASP \citep{Pollacco+06pasp}. This population is strongly biased towards larger planets with orbital periods $\Porb \la 10$\,d, as these {\it Hot Jupiters} show the greatest photometric variability. More recently, surveys such as \corot\ and \kepler\ have found success by obtaining uninterrupted photometry of stars using space satellites. The objects found in this way -- like the deep ground-based surveys -- mostly orbit fainter stars, which hinders attempts to obtain follow-up observations to refine their physical properties or even spectroscopically confirm the planetary nature of the transiting body.

An important problem with the study of TEPs is that their physical properties cannot be obtained simply by measuring some observable quantities and putting them through standard formulae: the number of quantities measurable directly from observations is one too few. An {\em additional constraint} is therefore needed, which is usually obtained by requiring the physical properties of the host star to match the predictions of stellar theory. In practise, there are several ways of implementing this constraint, and multiple sets of theoretical predictions from which to choose. As different researchers select different approaches, their results are inhomogeneous and therefore non-trivial to compare.

The diversity of the known TEPs, coupled with the varied methods used in their analysis, makes a consistent picture of their properties more difficult to attain. The current series of papers is an attempt to perform a uniform analysis for all suitable TEP systems, in order to produce homogeneous measurements of their physical properties. Such results are valuable in statistical investigations of the characteristics of this population \reff{(such as work similar to that by \citealt{Enoch+12aa})}, as well as providing an independent confirmation (or otherwise) of published numbers. In some cases it is possible to achieve improved results simply by considering all available data rather than concentrating on only one set of observations.

In Paper\,I \citep{Me08mn} I presented the methods used for analysing the transit photometry of TEP systems, and applied them to the fourteen object which had good light curves at that point. Paper\,II \citep{Me09mn} discussed the more thorny issue of how to turn the measured photometric and spectroscopic parameters into physical properties; the primary problem here is the need to incorporate the predictions of theoretical stellar evolutionary models (or some other additional constraint) into the process. This was achieved for the fourteen TEPs by using three sets of stellar models, plus an empirical stellar mass--radius relation as an alternative constraint. In Paper\,III \citep{Me10mn} the net was widened to include the full analysis of 30 TEPs with five sets of stellar models, plus a description of the methods used to incorporate contaminating light and orbital eccentricity into the analysis. Paper\,IV \citep{Me11mn} further extended the project to include 58 TEPs, concentrating on those which had space-based light curves from \corot, \kepler, or the NASA Epoxi satellites. An improved empirical constraint was also used (based on stellar density, temperature and metal abundance), calibrated on the measured properties of 180 stars in 90 detached eclipsing binaries.

In this work I present new photometric analyses of 30 more TEPs, and measure the physical properties of these plus a further eight objects previously considered in the current project. This brings the total number of TEPs within the {\it Homogeneous Studies} project to 82. Whilst this is by far the largest source of homogeneous physical properties, it still represents less than half of the known TEP systems. The physical properties, follow-up status, sky positions and discovery rate of the known TEPs are finally discussed in Sections \ref{sec:properties} to \ref{sec:discpos}. As with previous papers in this series, extensive tables of intermediate results are available for inspection in the online-only Supplementary Information.


\section{Light curve analysis}                                                                                                         \label{sec:lc}

I have studied the light curves of each TEP system using the methods discussed in detail in Paper\,I. The {\sc jktebop}\footnote{{\sc jktebop} is written in {\sc fortran77} and the source code is available at {\tt http://www.astro.keele.ac.uk/jkt/codes/jktebop.html}} code was used to find the best-fitting model for each light curve, from which the photometric parameters were extracted. {\sc jktebop} represents the star and planet as biaxial spheroids with shapes governed by the mass ratio. The results in this work are all very insensitive to the adopted mass ratio, in part because the projected surfaces of the stars suffer the least distortions from sphericity close to the times of transit and occultation. The correct terminology for transits (primary eclipses) and occultations (secondary eclipses) was discussed in sect.\,4 of Paper\,IV.

The main parameters for {\sc jktebop} are the orbital inclination ($i$), and fractional radii of the star ($r_{\rm A}$) and planet ($r_{\rm b}$), defined as:
\begin{equation}
r_{\rm A} = \frac{R_{\rm A}}{a} \qquad \qquad r_{\rm b} = \frac{R_{\rm b}}{a}
\end{equation}
where $a$ is the orbital semimajor axis, and $R_{\rm A}$ and $R_{\rm b}$ are the volume-equivalent stellar and planetary radii. In this work I fitted for the sum and ratio of the fractional radii:
\begin{equation}
r_{\rm A} + r_{\rm b} \qquad \qquad k = \frac{r_{\rm b}}{r_{\rm A}} = \frac{R_{\rm b}}{R_{\rm A}}
\end{equation}
as these are less strongly correlated. In \reff{all} cases the time of transit midpoint ($T_0$) was included as a fitted parameter. When modelling one transit, the orbital period ($\Porb$) was fixed to a value taken from the literature. When modelling multiple transits at the same time, \Porb\ was included as a fitted parameter.

Limb darkening (LD) was considered in detail. Every light curve was solved with each of five different LD laws (see Paper\,I) and with three different approaches to the LD coefficients: (1) both fixed (`LD-fixed'); (2) the linear coefficients ($u_{\rm A}$) fitted and the nonlinear coefficients ($v_{\rm A}$) fixed but perturbed\footnote{The size of the perturbation was incorrectly stated to be $\pm$0.05 in Paper\,IV, rather than the true value of $\pm$0.10. I thank the referee of a previous paper \citep{Me+12mn} for bringing this to my attention.} by $\pm$0.10 in the error analysis simulations (`LD-fit/fix'); (3) both coefficients fitted (`LD-fitted'). In a few cases where the photometric data are of limited quality the LD-fitted alternatives were not calculated. Initial or fixed values for the LD coefficients were obtained by bilinear interpolation to the correct \Teff\ and \logg\ within tabulated theoretical predictions\footnote{Interpolation within multiple tables of theoretical LD coefficients was performed using the {\sc jktld} code, written in {\sc fortran77} and available at {\tt http://www.astro.keele.ac.uk/jkt/codes/jktld.html}}. Once the best of the three options (LD-fixed, LD-fit/fix, LD-fitted) was determined, the final light curve parameters were taken to be the weighted mean of the relevant results for the four nonlinear laws.

Errorbars for the parameters were obtained from 1000 Monte Carlo (MC) simulations \citep{Me++04mn2}, for each of the solutions with the adopted approach to LD. The largest of these was retained, and an additional contribution added to reflect any variation in the parameter values from the four nonlinear LD laws. Finally, alternative errorbars were calculated using a residual permutation (RP) algorithm (as implemented in Paper\,I), which accounts for correlated noise. The RP errorbars were adopted if they were greater than the MC equivalents.

Any contaminating `third' light \citep[e.g.][]{Daemgen+09aa}, denoted by $L_3$, was included as a constraint following the procedure of \citet{Me+10mn}. A detailed investigation of the issue of third light was given in Paper\,III. Orbital eccentricity was treated similarly \citep{Me+09apj}, with constraints on the combinations $e\cos\omega$ and $e\sin\omega$ preferred over those for eccentricity ($e$) and periastron longitude ($\omega$) directly (see Paper\,III). Numerical integration over long exposure times is important for several TEPs, most notably those observed in long cadence by the \kepler\ satellite, and was dealt with in {\sc jktebop} following the scheme given in Paper\,IV.


\section{Calculation of physical properties}                                                                                       \label{sec:absdim}

The analysis of transit light curves gives the orbital ephemeris (\Porb, $T_0$) and the photometric parameters ($r_{\rm A}$, $r_{\rm b}$ and $i$). For each system we also have the orbital velocity amplitude of the host star ($K_{\rm A}$). These measured parameters alone do not lead to a unique solution for the physical properties of the system, so an additional constraint has to be sought from elsewhere. The possible constraints are outlined below\footnote{\reff{A small fraction of TEP systems have direct distance measurements from parallaxes, or stellar diameter measurements from interferometry \citep[e.g.][]{Vonbraun+12apj}. These observations can supply the crucial additional constraint, but are not possible for most TEP host stars due to their distance from the Earth.}}

In each case, the physical properties calculated were the mass, radius, surface gravity and density of the star ($M_{\rm A}$, $R_{\rm A}$, $\log g_{\rm A}$, $\rho_{\rm A}$) and of the planet ($M_{\rm b}$, $R_{\rm b}$, $g_{\rm b}$, $\rho_{\rm b}$), a surrogate for the planetary equilibrium temperature:
\begin{equation} \label{eq:teq}
\Teq = \Teff \left(\frac{R_{\rm A}}{2a}\right)^{1/2} = \Teff \left(\frac{r_{\rm A}}{2}\right)^{1/2}
\end{equation}
and the \citet{Safronov72} number:
\begin{equation}
\Theta = \frac{1}{2} \left(\frac{V_{\rm esc}}{V_{\rm orb}}\right)^2
       = \left(\frac{a}{R_{\rm b}}\right) \left(\frac{M_{\rm b}}{M_{\rm A}}\right)
       = \frac{1}{r_{\rm b}} \frac{M_{\rm b}}{M_{\rm A}}
\end{equation}

The set of physical constants used in the process above were tabulated in sect.\,3.3 of Paper\,IV\footnote{The radius of the Sun and of Jupiter were wrong by a factor of ten in sect.\,3.3 of Paper\,IV, due to a typographical error. The correct values are $6.95508 \times 10^{8}$\Rsun\ and $7.1492 \times 10^{7}$\Rjup, respectively. This does not affect any of the calculations in Paper\,IV or other analyses by the current author.}, and are very close to the values proposed to the astronomical community by \citet{HarmanecPrsa11pasp}.

\subsection{Additional constraint from theoretical stellar models}                                                            \label{sec:absdim:theo}

The standard approach in obtaining the additional constraint is to use the predictions of theoretical stellar evolutionary models. These provide the mass and radius of the host star, guided by determinations of its effective temperature (\Teff), metal abundance (\FeH) and light-curve-derived density \citep{SeagerMallen03apj}. The full physical properties of the system are then straightforward to calculate. I adopted the procedure introduced in Paper\,II, which begins with estimating the velocity amplitude of the {\it planet} ($K_{\rm b}$). The other measured quantities (\Porb, $K_{\rm A}$, $r_{\rm A}$, $r_{\rm b}$, $i$, $e$) \reff{were} used to determine the ensuing physical properties using standard formulae \citep[e.g.][]{Hilditch01book}. The value of $K_{\rm b}$ \reff{could then be} iteratively refined to minimise the figure of merit
\begin{equation}
{\rm fom} = \left[\frac{r_{\rm A}^{\rm (obs)}-(R_{\rm A}^{\rm (calc)}/a)}{\sigma{\rm (r_{\rm A}^{\rm (obs)})}}\right]^2 +
            \left[\frac{\Teff^{\rm (obs)}-\Teff^{\rm (calc)}}{\sigma(\Teff^{\rm (obs)})}\right]^2
\end{equation}
where $r_{\rm A}^{\rm (obs)}$ and $\Teff^{\rm (obs)}$ are quantities determined from observations. $R_{\rm A}^{\rm (calc)}$ and $\Teff^{\rm (calc)}$ were obtained for a given $K_{\rm b}$ value by interpolating in the tabulations of theoretical models for the calculated mass and observed \FeH\ of the star. The outcome of this procedure is a set of physical properties corresponding to the best agreement between observed and calculated quantities. The solution control parameter, $K_{\rm b}$, represents the entirety of the input from theoretical stellar models. Because the evolution of a star is highly non-linear with time, the above procedure was carried out for a range of ages starting from 0.01\,Gyr and incrementing in 0.01\,Gyr chunks until the star was significantly evolved (surface gravity $\logg < 3.5$). This led to a set of overall best physical properties plus a model-dependent estimate of the age of the system.

The statistical uncertainties on the input parameters were propagated using a perturbation analysis \citep{Me++05aa} implemented in the {\sc jktabsdim} code. This had the additional benefit of yielding a complete error budget for every output parameter, allowing identification of how best to improve our understanding of each TEP. The use of theoretical stellar models also incurs a systematic error due to our incomplete understanding of stellar physics. I therefore ran {\sc jktabsdim} solutions for each of five sets of theoretical models (see Paper\,III): {\it Claret} \citep{Claret04aa,Claret05aa,Claret06aa2,Claret07aa2}, {\it Y$^2$} \citep{Demarque+04apjs}, {\it Teramo} \citep{Pietrinferni+04apj}, {\it VRSS} \citep{Vandenberg++06apjs} and {\it DSEP} \citep{Dotter+08apjs}. This yielded five sets of values and statistical errorbars for each output parameter. For the final result for each parameter I have taken the unweighted mean of the five values. The largest of the individual statistical errors was adopted as the final statistical error, and the standard deviation of the five values was taken to represent the systematic error.

Three of the derived physical properties have no dependence on stellar theory so are quoted without systematic errorbars: $g_{\rm b}$ \citep{Me++07mn}; $\rho_{\rm A}$ to a good approximation \citep{SeagerMallen03apj}; and \Teq\ (Paper\,III).

\begin{table*} \caption{\label{tab:teps:lcpar} Final photometric parameters for the 38 TEPs
analysed in this work. Orbital periods are also given, and are either from this work or taken
from the literature. Systems for which a constraint on orbital eccentricity or third light was
required, are marked using a $\star$ in the column headed ``$e$?'' or ``$L_3$?''.}
\begin{tabular}{l l c c r@{\,$\pm$\,}l r@{\,$\pm$\,}l r@{\,$\pm$\,}l l}
\hline \hline
System & Orbital period & $e$? & $L_3$? & \mc{Orbital inclination,} & \mc{Fractional stellar}  & \mc{Fractional planetary} & Reference \\
       &     (days)     &      &        &    \mc{$i$ (degrees)}     & \mc{radius, $r_{\rm A}$} & \mc{radius, $r_{\rm b}$}  &           \\
\hline
\corot-2      & 1.7429935 (10)  & $\star$ & $\star$ & 87.45 & 0.34            & 0.1478 & 0.0023              & 0.02462 & 0.00035                & Paper\,IV        \\
\corot-17     & 3.7681 (3)      &         & $\star$ & \erc{89.3}{0.7}{4.9}    & \erc{0.157}{0.043}{0.012}    & \erc{0.0100}{0.0034}{0.0009}     & This work        \\
\corot-18     & 1.9000693 (28)  &         & $\star$ & 86.8 & 1.7              & 0.1502 & 0.0072              & 0.0209 & 0.0013                  & This work        \\
\corot-19     & 3.89713 (2)     &         &         & \erc{89.7}{0.3}{3.6}    & \erc{0.143}{0.028}{0.007}    & \erc{0.0111}{0.0024}{0.0006}     & This work        \\
\corot-20     & 9.24285 (30)    &         &         & 83.5 & 3.8              & 0.070 & 0.019                & 0.0062 & 0.0014                  & This work        \\
\corot-23     & 3.6313 (1)      & $\star$ & $\star$ & \erc{84.7}{3.4}{3.0}    & \erc{0.168}{0.022}{0.014}    & \erc{0.0117}{0.0018}{0.0011}     & This work        \\
HAT-P-3       & 2.8997360 (20)  &         &         & 86.15 & 0.19            & 0.1053 & 0.0020              & 0.01178 & 0.00030                & This work        \\
HAT-P-6       & 3.8530030 (14)  &         &         & 84.88 & 0.51            & 0.1346 & 0.0054              & 0.01271 & 0.00070                & This work        \\
HAT-P-9       & 3.922814 (2)    &         &         & 86.10 & 0.54            & 0.1177 & 0.0060              & 0.01250 & 0.00089                & This work        \\
HAT-P-14      & 4.6276691 (50)  & $\star$ &         & 83.01 & 0.27            & 0.1211 & 0.0036              & 0.00954 & 0.00045                & This work        \\
Kepler-7      & 4.8854922 (38)  &         & $\star$ & 85.32 & 0.16            & 0.1491 & 0.0013              & 0.01246 & 0.00015                & This work        \\
Kepler-12     & 4.4379637 (2)   &         & $\star$ & \erc{89.02}{0.25}{0.21}&\erc{0.12486}{0.00110}{0.00075}&\erc{0.014687}{0.000071}{0.000068}& This work        \\
Kepler-14     & 6.7901230 (43)  & $\star$ & $\star$ & 86.64 & 0.59            & 0.1260 & 0.0058              & 0.00698 & 0.00029                & This work        \\
Kepler-15     & 4.942782 (13)   &         & $\star$ & 86.144 & 0.051          & 0.10002 & 0.00048            & 0.010575 & 0.000086              & This work        \\
Kepler-17     & 1.4857108 (2)   &         & $\star$ & \erc{89.73}{0.27}{0.94} & \erc{0.1755}{0.0018}{0.0008} & \erc{0.02403}{0.00024}{0.00013}  & This work        \\
KOI-135       & 3.0240933 (27)  &         & $\star$ & 84.89 & 0.21            & 0.1409 & 0.0020              & 0.01212 & 0.00025                & This work        \\
KOI-196       & 1.8555588 (15)  &         & $\star$ & \erc{89.5}{0.5}{1.6}    & \erc{0.1508}{0.0091}{0.0020} & \erc{0.01406}{0.00107}{0.00029}  & This work        \\
KOI-204       & 3.2467220 (73)  &         & $\star$ & 84.30 & 0.83            & 0.1485 & 0.0090              & 0.01259 & 0.00098                & This work        \\
KOI-254       & 2.45524122 (65) & $\star$ & $\star$ & 87.48 & 0.38            & 0.0849 & 0.0040              & 0.01616 & 0.00091                & This work        \\
KOI-423       & 21.087168 (21)  & $\star$ & $\star$ & 89.06 & 0.21            & 0.0370 & 0.0015              & 0.00339 & 0.00017                & This work        \\
KOI-428       & 6.8731697 (96)  &         & $\star$ & 85.47 & 0.94            & 0.1437 & 0.0072              & 0.00859 & 0.00052                & This work        \\
OGLE-TR-56    & 1.21191096 (65) &         &         & 73.47 & 0.37            & 0.3292 & 0.0048              & 0.03378 & 0.00059                & This work        \\
OGLE-TR-111   & 4.0144477 (16)  &         &         & 88.50 & 0.35            & 0.0817 & 0.0018              & 0.01033 & 0.00030                & This work        \\
OGLE-TR-113   & 1.43247425 (34) &         &         & 87.80 & 0.86            & 0.1572 & 0.0020              & 0.02296 & 0.00039                & This work        \\
OGLE-TR-132   & 1.68986531 (64) &         &         & 83.83 & 0.96            & 0.2056 & 0.0078              & 0.01941 & 0.00096                & This work        \\
OGLE-TR-L9    & 2.48553417 (64) &         &         & 82.19 & 0.21            & 0.1722 & 0.0025              & 0.01928 & 0.00028                & This work        \\
TrES-4        & 3.5539268 (32)  &         & $\star$ & 82.33 & 0.47            & 0.1697 & 0.0064              & 0.01650 & 0.00059                & This work        \\
WASP-1        & 2.5199464 (8)   &         &         & 88.0 & 2.0              & \erc{0.1737}{0.0057}{0.0089} & \erc{0.0182}{0.0007}{0.0011}     & Paper\,I         \\
WASP-2        & 2.15222144 (39) &         & $\star$ & 84.81 & 0.17            & 0.1238 & 0.0018              & 0.01643 & 0.00030                & \citet{Me+10mn}  \\
WASP-4        & 1.33823144 (32) &         &         & 89.0 & 1.0              & \erc{0.1825}{0.0011}{0.0010} & \erc{0.02812}{0.00022}{0.00014}  & \citet{Me+09mn2} \\
WASP-5        & 1.6284246 (13)  &         &         & 85.8 & 1.1              & 0.1847 & 0.0061              & 0.02050 & 0.00091                & \citet{Me+09mn}  \\
WASP-7        & 4.9546416 (35)  &         &         & 87.03 & 0.93            & 0.1102 & 0.0061              & 0.01053 & 0.00070                & \citet{Me+11aa}  \\
WASP-12       & 1.0914222 (11)  &         &         & 83.3 & 1.1              & 0.3260 & 0.0052              & 0.03777 & 0.00088                & This work        \\
WASP-13       & 4.353011 (13)   &         &         & 84.88 & 0.45            & 0.1382 & 0.0049              & 0.01310 & 0.00059                & This work        \\
WASP-14       & 2.2437661 (11)  & $\star$ &         & 81.1 & 1.5              & 0.2084 & 0.0099              & 0.0210 & 0.0010                  & This work        \\
WASP-18       & 0.94145181 (44) & $\star$ &         & 85.0 & 2.1              & 0.2795 & 0.0084              & 0.0272 & 0.0012                  & \citet{Me+09apj} \\
WASP-21       & 4.3225060 (31)  &         &         & 86.77 & 0.45            & 0.1069 & 0.0037              & 0.01170 & 0.00054                & This work        \\
XO-2          & 2.6158640 (16)  &         &         & 88.8 & 1.2              & \erc{0.1237}{0.0024}{0.0047} & \erc{0.01300}{0.00033}{0.00070}  & Paper\,III       \\
\hline \hline \end{tabular} \end{table*}

\begin{table*} \caption{\label{tab:teps:spec} Spectroscopic quantities for the TEP host
stars, taken from the literature and adopted in the analysis in the present work.}
\begin{tabular}{l r@{\,$\pm$\,}l l r@{\,$\pm$\,}l l r@{\,$\pm$\,}l l}
\hline \hline
System & \multicolumn{3}{l}{Velocity amplitude (\ms)} & \mc{\Teff\ (K)} & Reference & \mc{\FeH} & Reference \\
\hline
\corot-2    &  603     &  18        & \citet{Gillon+10aa}         & 5598 &  50 & \citet{Schroter+11aa}      &    0.04 & 0.05 & \citet{Schroter+11aa}   \\
\corot-17   &  312.4   &  29.0      & \citet{Csizmadia+11aa}      & 5740 &  80 & \citet{Csizmadia+11aa}     &    0.0  & 0.1  & \citet{Csizmadia+11aa}  \\
\corot-18   &  590     &  14        & \citet{Hebrard+11aa}        & 5440 & 100 & \citet{Hebrard+11aa}       & $-$0.1  & 0.1  & \citet{Hebrard+11aa}    \\
\corot-19   &  126     &   6        & \citet{Guenther+12aa}       & 6090 &  70 & \citet{Guenther+12aa}      & $-$0.02 & 0.10 & \citet{Guenther+12aa}   \\
\corot-20   &  454     &   9        & \citet{Deleuil+12aa}        & 5880 &  90 & \citet{Deleuil+12aa}       &    0.14 & 0.12 & \citet{Deleuil+12aa}    \\
\corot-23   &  377     &  34        & \citet{Rouan+12aa}          & 5900 & 100 & \citet{Rouan+12aa}         &    0.05 & 0.1  & \citet{Rouan+12aa}      \\
HAT-P-3     &   89.1   &   2.0      & \citet{Torres+07apj}        & 5185 &  50 & \citet{Torres+07apj}       &    0.27 & 0.05 & \citet{Torres+07apj}    \\
HAT-P-6     &  115.5   &   4.2      & \citet{Noyes+08apj}         & 6570 &  80 & \citet{Noyes+08apj}        & $-$0.13 & 0.08 & \citet{Noyes+08apj}     \\
HAT-P-9     &   84.7   &   7.9      & \citet{Shporer+09apj}       & 6350 & 150 & \citet{Shporer+09apj}      &    0.12 & 0.20 & \citet{Shporer+09apj}   \\
HAT-P-14    &  218.9   &   5.8      & \citet{Torres+10apj}        & 6600 &  90 & \citet{Torres+10apj}       &    0.11 & 0.08 & \citet{Torres+10apj}    \\
Kepler-7    &   42.9   &   3.5      & \citet{Latham+10apj}        & 5933 &  50 & \citet{Latham+10apj}       &    0.11 & 0.05 & \citet{Latham+10apj}    \\
Kepler-12   & \erc{48.2}{4.3}{4.4}  & \citet{Fortney+11apjs}      & 5947 & 100 & \citet{Fortney+11apjs}     &    0.07 & 0.04 & \citet{Fortney+11apjs}  \\
Kepler-14   & \erc{683}{27}{25}     & \citet{Buchhave+11apjs}     & 6395 &  60 & \citet{Buchhave+11apjs}    &    0.12 & 0.06 & \citet{Buchhave+11apjs} \\
Kepler-15   & \erc{78.7}{8.5}{9.1}  & \citet{Endl+11apjs}         & 5595 & 120 & \citet{Endl+11apjs}        &    0.36 & 0.07 & \citet{Endl+11apjs}     \\
Kepler-17   &  399     &   9        & \citet{Bonomo+12aa}         & 5781 &  85 & \citet{Bonomo+12aa}        &    0.26 & 0.10 & \citet{Bonomo+12aa}     \\
KOI-135     &  375     &  13        & \citet{Bonomo+12aa}         & 6041 & 143 & \citet{Bonomo+12aa}        &    0.33 & 0.11 & \citet{Bonomo+12aa}     \\
KOI-196     &   85     &  11        & \citet{Santerne+11aa2}      & 5660 & 100 & \citet{Santerne+11aa2}     & $-$0.09 & 0.16 & \citet{Santerne+11aa2}  \\
KOI-204     &  124     &   5        & \citet{Bonomo+12aa}         & 5757 & 134 & \citet{Bonomo+12aa}        &    0.26 & 0.10 & \citet{Bonomo+12aa}     \\
KOI-254     &  110     &  10        & \citet{Johnson+12apj}       & 3820 &  90 & \citet{Johnson+12apj}      &    0.20 & 0.10 & \citet{Johnson+12apj}   \\
KOI-423     & \erc{1251}{30}{27}    & \citet{Bouchy+11aa2}        & 6260 & 140 & \citet{Bouchy+11aa2}       & $-$0.29 & 0.10 & \citet{Bouchy+11aa2}    \\
KOI-428     &  179     &  27        & \citet{Santerne+11aa}       & 6510 & 100 & \citet{Santerne+11aa}  &\erc{0.10}{0.15}{0.10}&\citet{Santerne+11aa}  \\
OGLE-TR-56  &  212     &  22        & \citet{Bouchy+05aa2}        & 6119 &  62 & \citet{Santos+06aa}        &    0.25 & 0.08 & \citet{Santos+06aa}     \\
OGLE-TR-111 &   78     &  14        & \citet{Pont+04aa}           & 5044 &  83 & \citet{Santos+06aa}        &    0.19 & 0.07 & \citet{Santos+06aa}     \\
OGLE-TR-113 &  267     &  34        & \citet{Torres++08apj}       & 4790 &  75 & \citet{Torres++08apj}      &    0.09 & 0.08 & \citet{Torres++08apj}   \\
OGLE-TR-132 &  167     &  18        & \citet{Moutou+04aa}         & 6210 &  59 & \citet{Gillon+07aa3}       &    0.37 & 0.07 & \citet{Gillon+07aa3}    \\
OGLE-TR-L9  &  510     & 170        & \citet{Snellen+09aa}        & 6933 &  58 & \citet{Snellen+09aa}       & $-$0.05 & 0.20 & \citet{Snellen+09aa}    \\
TrES-4      &   97.4   &   7.2      & \citet{Mandushev+07apj}     & 6200 &  75 & \citet{Sozzetti+09apj}     &    0.14 & 0.09 & \citet{Sozzetti+09apj}  \\
WASP-1      &   125    &   5        & \citet{Albrecht+11apj2}     & 6213 &  51 & \citet{Albrecht+11apj2}    &    0.17 & 0.05 & \citet{Albrecht+11apj2} \\
WASP-2      &   153.6  &   3.0      & \citet{Triaud+10aa}         & 5170 &  60 & Sect.\,\ref{sec:teps:other}&    0.04 & 0.05 & \citet{Albrecht+11apj2} \\
WASP-4      &   242.1  &   3.0      & \citet{Triaud+10aa}         & 5540 &  55 & \citet{Maxted++11mn}       & $-$0.03 & 0.09 & \citet{Gillon+09aa}     \\
WASP-5      &  268.7   &   1.8      & \citet{Triaud+10aa}         & 5770 &  65 & \citet{Maxted++11mn}       &    0.09 & 0.09 & \citet{Gillon+09aa}     \\
WASP-7      &   97     &  13        & \citet{Hellier+09apj}       & 6520 &  70 & \citet{Maxted++11mn}       &    0.00 & 0.10 & \citet{Hellier+09apj}   \\
WASP-12     &  226     &   4        & \citet{Hebb+09apj}          & 6250 & 100 & \citet{Fossati+10apj2}     &    0.32 & 0.12 & \citet{Fossati+10apj2}  \\
WASP-13     &   55.7   &   5.5      & \citet{Skillen+09aa}        & 5826 & 100 & \citet{Skillen+09aa}       &    0.0  & 0.2  & \citet{Skillen+09aa}    \\
WASP-14     &  990     &   3        & \citet{Blecic+11xxx}        & 6475 & 100 & \citet{Joshi+09mn}         &    0.0  & 0.2  & \citet{Joshi+09mn}      \\
WASP-18     & 1816.7   &   1.9      & \citet{Triaud+10aa}         & 6455 &  70 & \citet{Maxted++11mn}       &    0.00 & 0.09 & \citet{Hellier+09nat}   \\
WASP-21     &   37.2   &   1.1      & \citet{Bouchy+10aa}         & 5800 & 100 & \citet{Bouchy+10aa}        & $-$0.46 & 0.11 & \citet{Bouchy+10aa}     \\
XO-2        &    92.2  &   1.7      & \citet{Narita+12pasj}       & 5340 &  50 & \citet{Burke+07apj}        &    0.45 & 0.05 & \citet{Burke+07apj}     \\
\hline \hline \end{tabular} \end{table*}

\subsection{Additional constraint from eclipsing binary relations}                                                              \label{sec:absdim:eb}

In Paper\,II I found a way to avoid the use of stellar models entirely, by defining an empirical mass--radius relation based on well-studied eclipsing binaries. This approach was not very successful because it did not allow for either the chemical composition or evolution of the host star. It was also subject to the known phenomenon that the components of low-mass eclipsing binaries tend to have larger radii and lower \Teff s than those predicted by stellar theory \citep[e.g.][]{Lopez07apj,Ribas+08conf}.

An alternative approach is to explicitly include a dependence on \Teff\ and \FeH. \citet{Enoch+10aa} proposed a set of calibration equations giving $\log M$ or $\log R$ as functions of $\rho$, \Teff\ and \FeH, where $\rho$ was chosen because a very good approximation to this quantity is calculable directly from transit light curves \citep{SeagerMallen03apj}. In Paper\,IV I adopted this approach, but with modified calibration coefficients obtained from a larger and better-conditioned sample of eclipsing binary systems\footnote{See the catalogue of well-studied detached eclipsing binary star systems: {\tt http://www.astro.keele.ac.uk/$\sim$jkt/debcat/}}. A similar re-calibration has been performed by \citet{Gillon+11aa}.

In the current work I have calculated the physical properties for each TEP system using equation\,7 in Paper\,IV with the coefficients of the calibration for $\lten R$ covering stellar masses of 0.2--3.0\Msun. These results are presented as an alternative to the default solutions incorporating the predictions of theoretical stellar models.


\section{Data acquisition}                                                                                                           \label{sec:data}

The \corot\ data used here are the N2 public version obtained in FITS format from the public archive\footnote{\tt http://idoc-corot.ias.u-psud.fr/}. The data for each object exist in short and/or long cadence flavours, with effective integration times of 32\,s and 512\,s respectively. I used the total flux as given by the ``WHITEFLUX'' column. The chromatic light curves were not considered as they add little to the results and have passbands which vary from star to star. I rejected any datapoints which were flagged as being unreliable (i.e.\ I required STATUS $=$ 0). Numerical integration was used for the long-cadence data, adopting $N_{\rm int} = 3$ as the number of integration points.

The \kepler\ data used were the public data obtained from the Multimission Archive at STScI (MAST\footnote{\tt http://archive.stsci.edu/kepler/data\_search/ search.php}). These also come in two forms: short cadence (effective integration time 58.84876\,s; \citealt{Gilliland+10apj}) and long cadence (29.4244\,min; \citealt{Jenkins+10apj}). Numerical integration is crucial for the interpretation of the latter type of data (see Paper\,IV for a demonstration) and I adopted $N_{\rm int} = 10$ in all cases. The \kepler\ data cover Quarter 0 (Q0) to Quarter 2 (Q2) with additional data up to Quarter 6 (Q6) available for some objects.

In order to account for slow variations in these data, both astrophysical and instrumental, all data more than one transit duration away from a transit were discarded. A straight line was then fitted to the data around each transit (not including the datapoints in transit) and divided out to normalise the transit to unit flux. A straight line was in almost all cases sufficient to remove the slow variations, and is also unable to distort the shape of the transit. A few instances where a polynomial was used instead are noted below. In those cases where a large number of datapoints remained, they were converted into orbital phase, sorted, and then phase-binned into a much smaller number of normal points. The amount of binning was carefully chosen to avoid smearing out the transit shape. Details of this process are given below when relevant.

Other data were obtained from public archives, from published papers, and by personal communication when necessary.


\section{Results for individual systems}                                                                                             \label{sec:teps}

In this section I present {\sc jktebop} and {\sc jktabsdim} analyses of 38 TEPs based on published high-quality data. The results were obtained using the same methods as those in previous papers, leading to homogeneous measurements for a sample of 82 TEPs. The final photometric parameters of the 38 TEPs considered in the current work are aggregated in Table\,\ref{tab:teps:lcpar}, and their adopted spectroscopic parameters are given in Table\,\ref{tab:teps:spec}. I have enforced minimum errorbars of $\pm$50\,K for \Teff\ and $\pm$0.05\,dex for \FeH, as the stellar effective temperature and metallicity scales set a lower limit on how precisely we can measure these quantities. Such minimum errorbars may still be optimistic \citep[see e.g.][]{Bruntt+10mn,Bruntt+12mn}, and this will be revisited in future.


\subsection{\corot-17}                                                                                                       \label{sec:teps:corot17}

\begin{figure} \includegraphics[width=\columnwidth,angle=0]{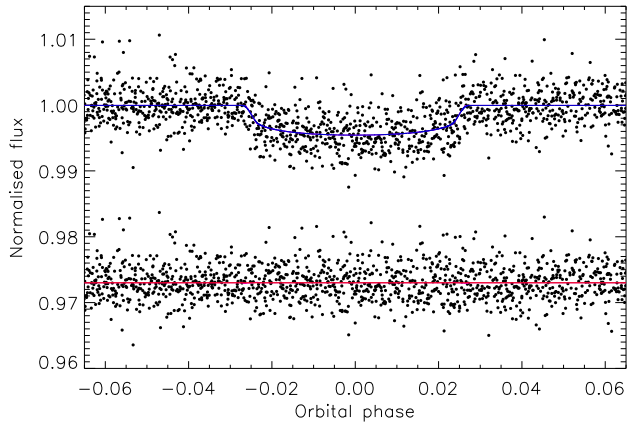}
\caption{\label{fig:corot17:lc} Phased \corot\ 512\,s cadence light curve of \corot-17
compared to the best fit found using {\sc jktebop} and the quadratic LD law.
The residuals are plotted at the base of the figure, offset from unity. The
purple line through the data shows the best-fitting model without numerical
integration, and through the residuals shows the difference between this
model with and without numerical integration. In this case the purple
lines are barely visible behind the other lines.} \end{figure}

Discovered by \citet{Csizmadia+11aa}, the \corot-17 system harbours a fairly massive planet ($2.46 \pm 0.25$\Mjup) around an old star (\er{8.0}{1.0}{3.9}\,Gyr). The only available light curve is that from the \corot\ satellite, which is comparatively poor due to the faintness of the star ($V = 15.46$). \citet{Csizmadia+11aa} found that their observations were consistent with a circular orbit, and that the \corot\ light curve is contaminated with a third light of $L_3 = 8 \pm 4$\,\%. 

The \corot\ data comprise 10\,248 512\,s cadence datapoints covering 23 transit events. 1571 datapoints in the vicinity of transits were retained (see Sect.\,\ref{sec:data}), of which ten were subsequently rejected by a $4\sigma$ clip. I modelled the light curve using numerical integration ($N_{\rm int} = 3$) and with \Porb\ and $T_0$ included as fitted parameters. LD-fixed and LD-fit/fix solutions were attempted, but LD-fitted solutions were avoided due to the modest quality of the light curve. The Monte Carlo algorithm implemented in {\sc jktebop} includes the application of random perturbations to the starting parameters for each simulation, in order to sample the parameter space well. I had to turn this option off as the perturbations yielded many poorly conditioned results, a situation also encountered for Kepler-4 in Paper\,IV (sect.\,6.2). The RP errorbars were therefore adopted, as they are significantly larger than the MC errorbars. The full light curve solutions are given in \apptab.

A summary of the light curve solutions is given in \apptab\ and the best fit to the \corot\ light curve is plotted in Fig.\,\ref{fig:corot17:lc}. The parameters found by \citet{Csizmadia+11aa} agree with my own to within the errorbars, although I find a solution with slightly larger $i$, smaller $r_{\rm A}$ and $r_{\rm b}$, and asymmetric errorbars. The physical properties of \corot-17 are given in \apptab\ and again show good agreement with those of \citet{Csizmadia+11aa}, except that the stellar surface gravity found in that work ($\logg = 4.40 \pm 0.10$) is incompatible with the star's mass and radius ($M_{\rm A} = 1.04 \pm 0.10$\Msun\ and $R_{\rm A} = 1.59 \pm 0.07$\Rsun). I provide the first measurement of the surface gravity of the planet.

\corot-17 would benefit from new photometric and spectroscopic observations, in order to refine its physical properties and check the spectroscopic \Teff\ value. The orbital ephemeris is imprecise, and will be unable to accurately predict transit midpoints by 2012 August (see Sect.\,\ref{sec:followup}).


\subsection{\corot-18}                                                                                                       \label{sec:teps:corot18}

\begin{figure} \includegraphics[width=\columnwidth,angle=0]{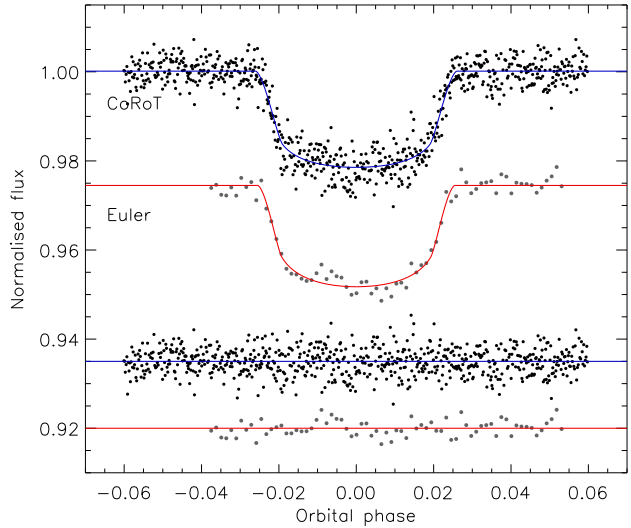}
\caption{\label{fig:corot18:lc} The \corot\ and Euler telescope light curves of
\corot-18 compared to the {\sc jktebop} best fits. Other comments are
the same as Fig.\,\ref{fig:corot17:lc}.} \end{figure}

The discovery of \corot-18\,\reff{b} was presented by \citet{Hebrard+11aa}, and contains a relatively massive planet ($M_{\rm b} = 3.27 \pm 0.17$\Mjup) orbiting an active star with a rotation period of $5.4 \pm 0.4$\,d. Its relative faintness ($V=15.0$) made RV measurements difficult. This was not a problem for the $K_{\rm A}$ value, which is large, but observations of the Rossiter-McLaughlin effect \citep{Rossiter24apj,McLaughlin24apj} only constrain the orbit to be prograde and probably aligned.

The \corot\ data are all of 32\,s cadence, and cover 13 consecutive transits. Of the 56\,819 original datapoints, 6932 in the vicinity of a transit were retained and then phase-binned by a factor of ten to yield 694 normal points. \citet{Hebrard+11aa} modelled these data after taking the unusual approach of binning them in time by a factor of 16 to be equivalent to 512\,s cadence, to improve computational tractibility. They therefore substantially lower the information content of the data they analysed. By comparison, the phase-binned data I worked on have a cadence equivalent to 32.2\,s. These were modelled with the assumptions of $L_3 = 2.0 \pm 0.1$\,\% and $e=0$ \citep{Hebrard+11aa}. The MC errors were found to be larger than the RP ones, and the full solutions are in \apptab.

\citet{Hebrard+11aa} also presented ground-based transit observations of \corot-18, from the Swiss Euler Telescope and EulerCam. This shows intriguing evidence around phase $-0.008$ for the passage of the planet over a starspot, but as noted by \citet{Hebrard+11aa} the deviation cannot conclusively be assigned to this phenomenon. It was conservatively treated as correlated noise in my analysis. G.\ H\'ebrard (private communication) confirmed that these data were obtained through an $R$ filter. They were modelled as with the \corot\ light curve, except that third light was not included (\apptab).

The light curve solutions are summarised in \apptab\ and show that the \corot\ and Euler data agree well. The values from \citet{Hebrard+11aa} differ by about $1\sigma$, probably due to their approach to binning the data. The best fits are shown in Fig.\,\ref{fig:corot18:lc}.

The physical properties of \corot-18 are shown in \apptab\ and show that I find a smaller system scale than \citet{Hebrard+11aa}, meaning that the masses and radii of both components are slightly smaller. \citet{Hebrard+11aa} were unable to constrain the age of the star, as different age indicators were inconsistent with each other. The more precise light curve parameters derived in the current work point to a large age of roughly 10\,Gyr, which disagrees with the comparatively fast rotation and strong lithium abundance indicative of an age of $<$1\,Gyr. The evolutionary status of the star is therefore uncertain. I provide the first measurement of $g_{\rm b}$. The system would benefit from a better light curve, which would allow refined physical properties to be obtained and also shed light on the star's age via a more precise measurement of its density.


\subsection{\corot-19}                                                                                                       \label{sec:teps:corot19}

\begin{figure} \includegraphics[width=\columnwidth,angle=0]{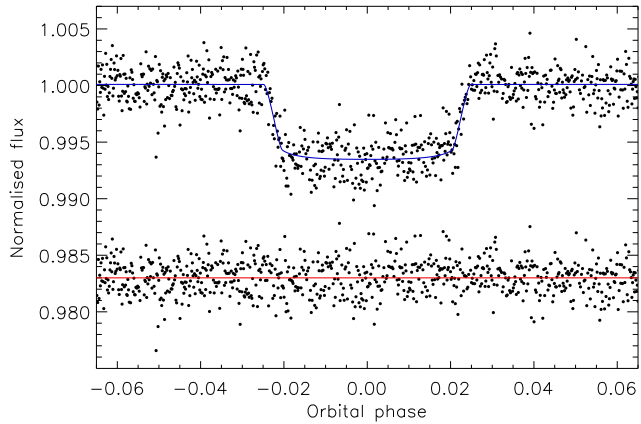}
\caption{\label{fig:corot19:lc} The \corot\ light curve of \corot-19
\citep{Guenther+12aa} compared to the {\sc jktebop} best fit. Other
comments are the same as Fig.\,\ref{fig:corot17:lc}.} \end{figure}

\reff{The planetary nature of this} system was discovered by \citet{Guenther+12aa}. The \corot\ data comprise 56\,837 datapoints at 32\,s cadence covering seven consecutive transits, of which 9229 points near transit were binned by a factor of ten to give 924 normal points (Fig.\,\ref{fig:corot19:lc}). \citet{Guenther+12aa} constrained $L_3 < 0.32$\,\% and $e < 0.15$ ($3\sigma$) so third light and orbital eccentricity were neglected in my analysis (\apptab). The RP errors were used, as they are modestly larger than the MC ones. From \apptab\ it is clear that I find a slightly different solution to the light curve compared to \citet{Guenther+12aa}, with much larger and asymmetric errorbars. The two solutions are compatible within my errors, but not within those of \citet{Guenther+12aa}.

The physical properties I find for the \corot-19 system (\apptab) are imprecise and agree with those of \citet{Guenther+12aa}. A discrepant solution with a young stellar age (zero-age main sequence versus $\sim$5\,Gyr) was found when using the $Y^2$ models, but restricting the possible solutions to older ages yielded results which were consistent with those from the other four model sets. I provide the first measurement of $g_{\rm b}$. Further photometry and spectroscopy of \corot-19 are needed to refine measurements of its physical properties, including the orbital ephemeris.


\subsection{\corot-20}                                                                                                       \label{sec:teps:corot20}

\begin{figure} \includegraphics[width=\columnwidth,angle=0]{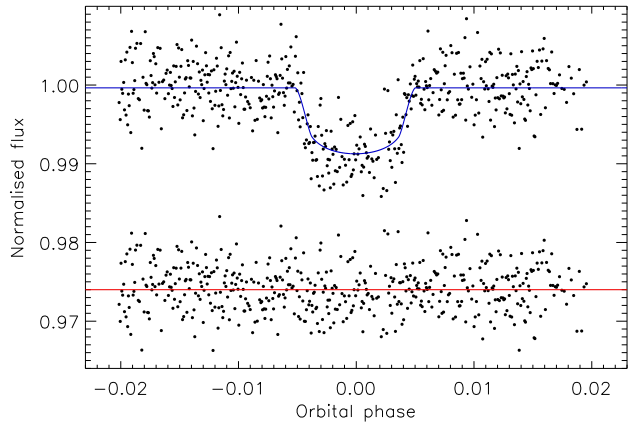}
\caption{\label{fig:corot20:lc} The \corot\ light curve of \corot-20 compared to the
{\sc jktebop} best fit. Other comments are the same as Fig.\,\ref{fig:corot17:lc}.}
\end{figure}

\reff{As found} by \citet{Deleuil+12aa}, \corot-20 contains a rather massive planet ($5.06 \pm 0.34$\Mjup) in an eccentric ($e = 0.562 \pm 0.013$) and long-period (9.24\,d) orbit around a G2\,V star. Third light is constrained to be less than 0.6\% so was neglected in my analysis. I adopted $e\cos\omega = 0.312 \pm 0.022$ and $e\sin\omega = 0.468 \pm 0.017$ \citep{Deleuil+12aa}. The \corot\ data comprise 56\,855 datapoints at 32\,s cadence but cover only three transits. The 2479 datapoints near these transits were binned by a factor of five into 496 normal points and modelled using {\sc jktebop} (\apptab). The light curve is of limited quality (Fig.\,\ref{fig:corot20:lc}) so I obtained imprecise results. Compared to \citet{Deleuil+12aa} I find a solution with $1\sigma$ lower $i$ and correspondingly larger $r_{\rm A}$ and $r_{\rm b}$, with much larger errorbars (\apptab).

A similar problem was found, when calculating the physical properties of the system, as for \corot-19 above. The $Y^2$ models gave a discrepant solution with a very young age (\apptab). This was solved in the same way, by artifically restricting the possible ages for the $Y^2$ models to $>$1\,Gyr. \citet{Deleuil+12aa} opted for the low-age solution, which is more consistent with their different photometric parameters and with their detection of the Li 6708\,\AA\ absorption line. The discrepancy supports the much larger errorbars I find, and suggests further observations of \corot-20 would be valuable. I find significantly different physical properties for the planet: \wwo{$M_{\rm b} = 5.06 \pm 0.36 \pm 0.04$}{$M_{\rm b} = 5.06 \pm 0.36$}\Mjup\ (\citealt{Deleuil+12aa} obtained $4.24 \pm 0.23$\Mjup), $R_{\rm b} = 1.16 \pm 0.26$\Rjup\ ($0.84 \pm 0.04$\Rjup) and $\Teq = 1100 \pm 150$\,K ($1002 \pm 24$\,K). I provide the first published measurement of $g_{\rm b}$.


\subsection{\corot-23}                                                                                                       \label{sec:teps:corot23}

\begin{figure} \includegraphics[width=\columnwidth,angle=0]{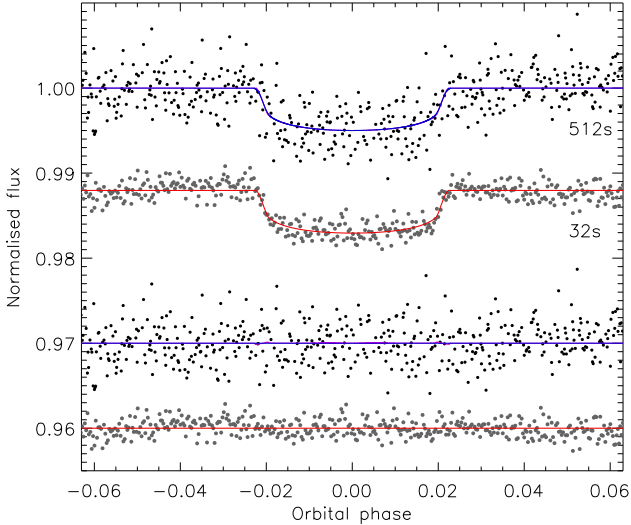}
\caption{\label{fig:corot23:lc} The \corot\ light curve of \corot-23
compared to the {\sc jktebop} best fit. Other comments are the same
as Fig.\,\ref{fig:corot17:lc}.} \end{figure}

\corot-23 \citep{Rouan+12aa} contains another massive and eccentric planet ($M_{\rm b} = 3.06 \pm 0.31$\Mjup\ and $e = 0.16 \pm 0.02$), orbiting a somewhat evolved star ($\log g_{\rm A} = \er{4.01}{0.08}{0.11}$). \corot\ observed it at 512\,s cadence initially (3598 datapoints covering nine transits) and then switched to 32\,s cadence (130\,544 datapoints covering 15 transits) after it was identified as a planet candidate. The 584 datapoints at 512\,s cadence in the region of transits were modelled using $N_{\rm int} = 3$. The 20\,614 datapoints at 32\,s cadence near transit were phased and binned by a factor of 40 to yield 516 normal points. Both sets of data were modelled adopting $e = 0.16 \pm 0.02$, $\omega = 52^\circ \pm 9^\circ$, and with $L_3 = 7.2$\%. \citet{Rouan+12aa} did not give an errorbar for $L_3$, so I used an estimate of 3\%. For the 512\,s data (\apptab) the RP errorbar was larger than the MC one for $i$, whereas for the 32\,s data (\apptab) the RP errorbars were all smaller than the MC equivalents.

The light curve solutions are summarised in \apptab\ and the best fits plotted in Fig.\,\ref{fig:corot23:lc}. Compared to \citet{Rouan+12aa}, I find larger errorbars, and $1.5\sigma$ larger $r_{\rm A}$ and $r_{\rm b}$. The solutions to both datasets are ill-defined and have asymmetric errorbars. Despite these issues, my physical properties of \corot-23 are in good agreement with those of \citet{Rouan+12aa} both for values and errorbars (\apptab). A lower \Teff\ would match the photometric parameters better, and an improved light curve is important for refining these parameters plus improving the orbital ephemeris. \corot-23 would be an interesting system to study further, due to its evolved host star, but its faintness may be an issue ($V = 15.63 \pm 0.07$). I provide the first published measurement of $g_{\rm b}$.


\subsection{HAT-P-3}                                                                                                           \label{sec:teps:hatp3}

\begin{figure} \smfig{\includegraphics[width=\columnwidth,angle=0]{plot-hatp03.eps}}{\includegraphics[width=\columnwidth,angle=0]{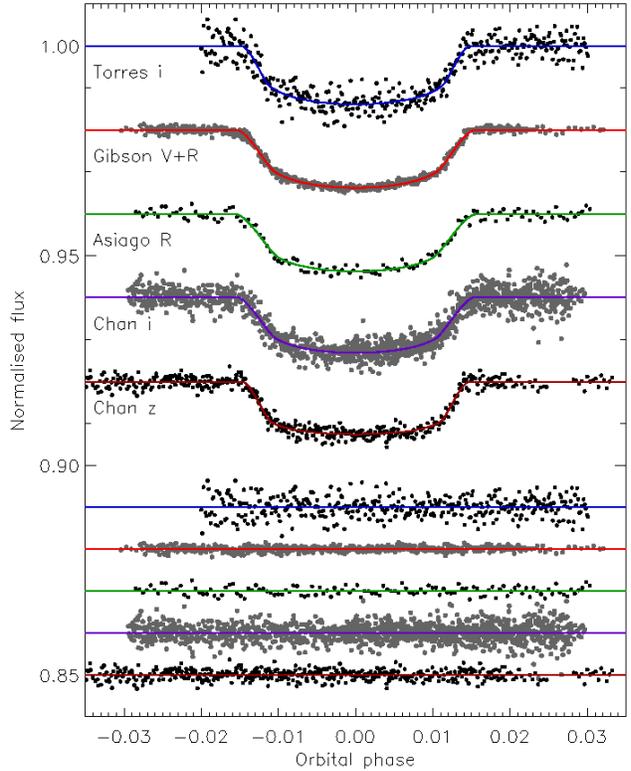}}
\caption{\label{fig:hatp3:lc} Light curves of HAT-P-3 compared to the best fit
found using {\sc jktebop} and the quadratic LD law. The residuals are plotted
at the base of the figure, offset from unity. The light curve sources and
passbands are labelled.} \end{figure}

\reff{Presented} by \citet{Torres+07apj}, HAT-P-3 is a fairly typical TEP system except that it is metal-rich ($\FeH = 0.27 \pm 0.05$) and the planet is unexpectedly small, two qualities which may be connected. \citet{Torres+07apj} found that a heavy-element core of mass $\sim$75\Mearth\ was needed for theoretical models to match the planet's radius; \citet{Chan+11aj} suggested $\sim$100\Mearth.

The discovery paper \citep{Torres+07apj} included follow-up photometry of one transit obtained with the KeplerCam imager on the FLWO 1.2\,m telescope. The dataset comprises 386 datapoints obtained in the $i$-band. I multiplied the errorbars of the data by $\sqrt{8.503}$ to obtain a reduced $\chi^2$ of $\chir=1.0$. The data were modelled with {\sc jktebop}, with the findings that the LD-fit/fix results are best, RP errors are similar to MC errors, and the errorbars are significantly asymmetric (\apptab).

\citet{Gibson+08aa} published an extensive analysis of HAT-P-3 based on photometry of seven transits obtained with LT/RISE, with the intention of probing for transit-timing variations (TTVs). To solve these data I phase-binned them by a factor of 40 to reduce the 19\,681 original points to 493 normal points, with the rejection of 24 points by a $4\sigma$ clip, and multiplied the errorbars by $\sqrt{2.358}$ to obtain $\chir=1.0$. The LD-fit/fix solutions are best, MC and RP errors are similar, and the errorbars are symmetric (\apptab).

\citet{Nascimbeni+11aa} obtained high-speed photometry of one transit of HAT-P-3 (exposure time 2\,s, cadence 5\,s) using AFOSC on the Asiago 1.8\,m telescope. The 2247 original datapoints were already condensed to 122 binned points in the datafile lodged with the CDS; I modelled these after multiplying the errors by $\sqrt{1.985}$. The results (\apptab) show insignificant correlated noise and that the LD-fit/fix solutions are best.

Finally, \citet{Chan+11aj} observed HAT-P-3, again using KeplerCam, over three transits in the $i$-band and three more in the $z$-band. The two sets of data (1258 and 506 points respectively) were solved separately with fixed \Porb. I found that the RP errors were similar to the MC ones and that the LD-fit/fix solutions are best (\apptabb). The scatter of the $z$-band data is around 1.3\,mmag, usefully lower than the 2.2\,mmag for the $i$-band data. The best fits to all datasets are shown in Fig.\,\ref{fig:hatp3:lc}.

The light curve solutions are summarised in \apptab, and show a comparatively poor agreement between datasets. Ths is primarily due to the $z$-band data from \citet{Chan+11aj} but also to the data from \citet{Torres+07apj}. The \chir\ of the model ``the values from different light curves agree'' is 1.7--2.1 for $r_{\rm A}+r_{\rm b}$, $k$, $r_{\rm A}$ and $r_{\rm b}$, and 1.0 for $i$. If the $z$-band data are discounted then I instead find a much better $\chir = 0.3$--$1.0$. The source of this discrepancy is not clear, so I have adopted the results ignoring the $z$-band data. The agreement between my final photometric parameters and those of previous studies is not great; my parameter values mostly lie in the middle of the distribution of those from previous studies. This situation is understandable in that I have included multiple datasets which are in modest disagreement, whereas previous researchers have tended to concentrate on only their own data rather than the full set of available observations.

The results from my {\sc jktabsdim} analysis (\apptab) show good agreement with previous work, except that I find modestly larger radii for both star and planet. The larger stellar radius shifts the age of the system from 0.4--1.6\,Gyr to \er{7.5}{4.2}{3.8}\,Gyr. My new solution should be the most reliable as I have based it on all available photometry (albeit with one dataset rejected). Further photometry would be useful in helping to understand the two discrepant datasets, but HAT-P-3 is already well-characterised so such effort is better spent on other TEPs.


\subsection{HAT-P-6}                                                                                                           \label{sec:teps:hatp6}

\begin{figure} \smfig{\includegraphics[width=\columnwidth,angle=0]{plot-hatp06.eps}}{\includegraphics[width=\columnwidth,angle=0]{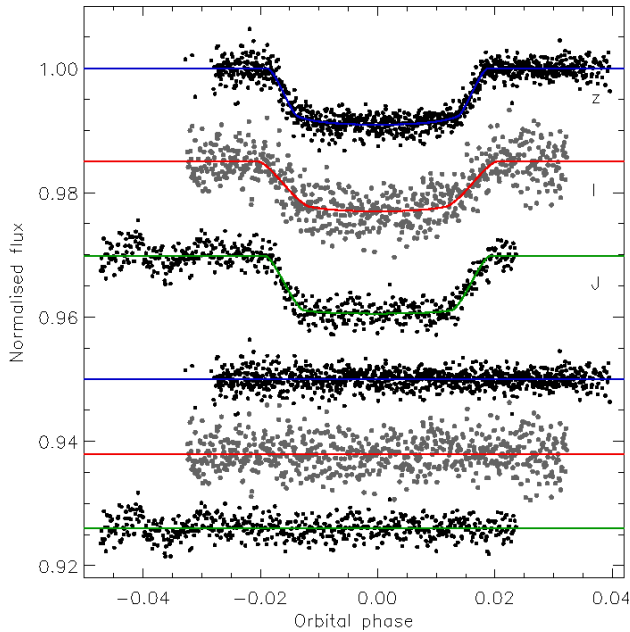}}
\caption{\label{fig:hatp6:lc} Light curves of HAT-P-6 compared to the {\sc jktebop}
best fits. From top to bottom the datasets are $z$-band \citep{Noyes+08apj},
$I$-band \citep{Todorov+12apj} and $J$-band \citep{Todorov+12apj,Sada+12pasp}.
Other comments are the same as Fig.\,\ref{fig:hatp3:lc}.} \end{figure}

HAT-P-6\,\reff{b} was discovered by \citet{Noyes+08apj} and is known to have a retrograde orbit from Rossiter-McLaughlin observations \citep{Hebrard+11aa2}, in agreement with the high \Teff\ of the host star. \citet{Todorov+12apj} have presented \spitzer\ observations of occultations at 3.6 and 4.5\,$\mu$m which provide a useful constraint on the orbital shape: $e\cos\omega < 0.004$ (3$\sigma$).

Two $z$-band follow-up transits of HAT-P-6 were presented by \citet{Noyes+08apj}, observed with the FLWO 1.2\,m and KeplerCam on two nights separated by 312\,d. I used the decorrelated version of the data, as is plotted in fig.\,1b of that paper. The number of datapoints is 499 and 515, respectively, and I multiplied the errorbars by $\sqrt{3.76}$ to obtain $\chir = 1.0$. The results (\apptab) show that the LD-fit/fix solutions are good and that the MC errors are larger than the RP alternatives.

Aside from their \spitzer\ photometry, \citet{Todorov+12apj} presented two transits, one observed in the $I$-band using a 0.36\,m telescope and one observed in the $J$-band using the KPNO 2.1\,m telescope with FLAMINGOES infrared imager and spectrograph. The latter dataset was also presented by \citet{Sada+12pasp}, who required a fifth-order polynomial to normalise the light curve to unit flux due to the increased complications involved in infrared versus optical observations.

The $i$-band light curve comprises 649 datapoints which suffer from a large scatter, requiring the use of fixed LD coefficients as well as a fixed out-of-transit brightness in order to arrive at a `reasonable' model for the data (\apptab). For this reason I did not use the results from these data when calculating the final photometric parameters.

For the 560 $J$-band datapoints, I multiplied the errorbars by $\sqrt{2.65}$ and calculated {\sc jktebop} solutions \reff{using a fifth-order polynomial flux normalisation} (\apptab). The RP errors are twice as big as the MC errors, indicating that these data contain significant correlated noise (see in particular around phase $-0.04$ in Fig.\,\ref{fig:hatp6:lc}. I adopted the LD-fit/fix results and RP errors. The best-fitting LD is weaker than theoretically predicted, but only by $1\sigma$.

The $z$-band and $J$-band results show a reasonable agreement (\apptab), except for $k$ where $\chir=2.1$. This is likely due to the lower reliability of the $J$-band transit, and the errorbars of the final photometric parameters have been inflated accordingly. Further photometry of HAT-P-6 would be useful. \citet{Noyes+08apj} and \citet{Torres++08apj} find almost identical photometric parameters to each other, which are also in agreement with my own.

I find physical properties for the HAT-P-6 system which are in good ageement with previous work (\apptab) but indicate a sightly larger star and smaller planet. My new results are based on more extensive data than other studies of HAT-P-6 so should be the most reliable. Further improvement could be achieved by obtaining new photometry, as the light curves are the current bottleneck in our understanding of this TEP system.


\subsection{HAT-P-9}                                                                                                           \label{sec:teps:hatp9}

\begin{figure} \smfig{\includegraphics[width=\columnwidth,angle=0]{plot-hatp09.eps}}{\includegraphics[width=\columnwidth,angle=0]{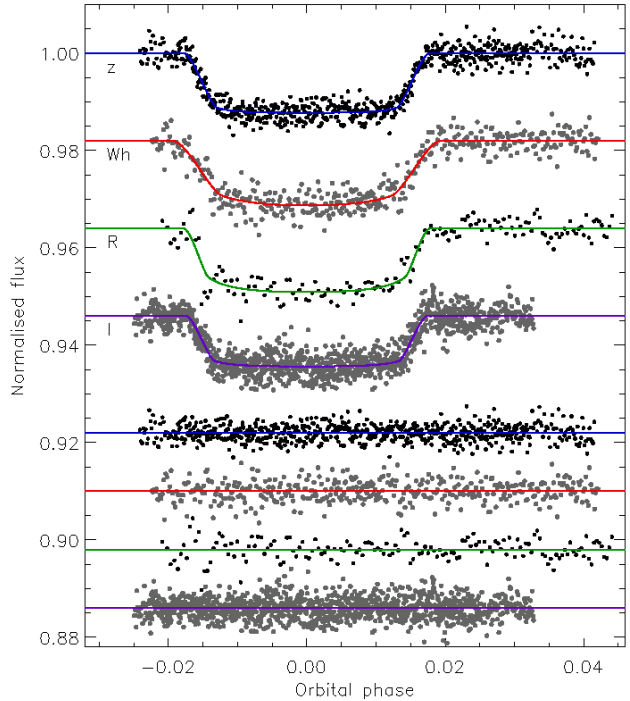}}
\caption{\label{fig:hatp9:lc} The light curves of HAT-P-9 compared to the
{\sc jktebop} best fits. The top three datasets are from \citet{Shporer+09apj2}
and the fourth is from \citet{Dittmann+12newa}. Other comments are the same as
Fig.\,\ref{fig:hatp3:lc}.} \end{figure}

The discovery of HAT-P-9\,\reff{b} was announced by \citet{Shporer+09apj2}, who also obtained two follow-up transits with the FLWO 1.2\,m ($z$-band), one with the WISE 0.46\,m (unfiltered) and one with the WISE 1.0\,m ($R$-band) telescopes. \citet{Moutou+11aa} measured the Rossiter-McLaughlin effect to find $\lambda = -16 \pm 8$, which is consistent with alignment between the orbital axis of the planet and the spin axis of the star. \citet{Dittmann+12newa} obtained two more transit observations in the $I$-band using the 1.6\,m Kuiper telescope.

For the $z$-band data, 675 points over two transits, I fitted for \Porb. The RP errors were smaller than the MC errors and full solutions are in \apptab. For the unfiltered (399 datapoints) and $R$-band (154) data I did not fit for \Porb.

For the unfiltered data I adopted LD coefficients appropriate for the \corot\ white light curves as these are close enough for the current purposes, and found that the RP errors were slightly bigger than the MC ones (\apptab). For the $R$-band data I multiplied the errorbars by $\sqrt{5.45}$ to get $\chir=1.0$, and found that the data prefer a central transit (\apptab) in disagreement to the other datasets for this star. The different light curve shape can clearly be seen in Fig.\,\ref{fig:hatp9:lc}. 

The two $I$-band transits appear to have slightly different depths (see fig.\,2 in \citealt{Dittmann+12newa}) but individual solutions showed that the difference is not significant. I therefore modelled all 1144 datapoints together, with \Porb\ included as a fitted quantity (\apptab).

The light curve solutions are summarised in \apptab. Agreement between the three datasets ($z$, $R$, $I$) is poor ($\chir=2.7$ for $r_{\rm A}+r_{\rm b}$, $\chir=8.7$ for $k$) and not easily explicable. The $R$-band results appear to be somewhat different to the $z$ and $I$ ones, but rejection of these does not improve matters ($\chir=13.7$ for $k$). I therefore calculated the final photometric parameters from the $z$, $R$ and $I$ individual results, with the errors inflated to account for the disagreement. Further photometry of HAT-P-9 is recommended to help understand this discrepancy and improve the measured system parameters. The results from \citet{Shporer+09apj2} agree with my values to within my errorbars, but not to within their errorbars.

The {\sc jktabsdim} results (\apptab) show a very good agreement between the values obtained via different stellar models, and with the values found by \citet{Shporer+09apj2}. HAT-P-9 would benefit from further photometric and spectroscopic (\Teff\ and \FeH) study.


\subsection{HAT-P-14}                                                                                                         \label{sec:teps:hatp14}

\begin{figure} \smfig{\includegraphics[width=\columnwidth,angle=0]{plot-hatp14.eps}}{\includegraphics[width=\columnwidth,angle=0]{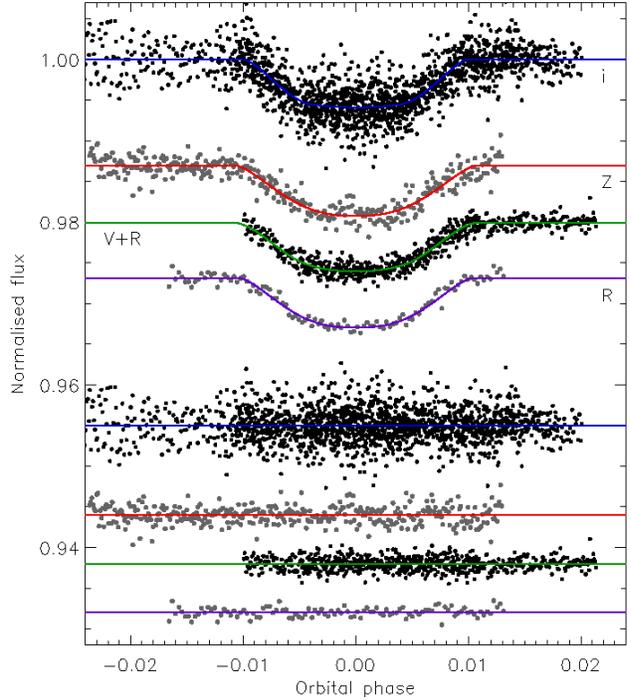}}
\caption{\label{fig:hatp14:lc} Light curves of HAT-P-14 compared to the {\sc jktebop}
best fit. From top to bottom the datasets are $i$-band from \citet{Torres+10apj}, FTN
$Z$ and RISE $V$$+$$R$ from \citep{Simpson+11aj2} and $R$ from \citet{Nascimbeni+11aa}.
Other comments are the same as Fig.\,\ref{fig:hatp3:lc}.} \end{figure}

Discovered by \citet{Torres+10apj}, the HAT-P-14 system contains a relatively massive planet ($2.27 \pm 0.08$\Mjup). Its low orbital inclination (i.e.\ high impact parameter) is potentially helpful in measuring the light curve parameters to high precision, and in probing for variations in these quantities. It was independently discovered by the WASP survey \citep{Simpson+11aj2}, from whom additional photometry and spectroscopy is available. The spectroscopic properties from the two consortia (\Teff\ and \FeH) agree. One more transit light curve has been presented by \citet{Nascimbeni+11aa}, and \citet{Winn+11aj} have found it to be in a retrograde orbit from observing its Rossiter-McLaughlin effect.

Both \citet{Torres+10apj} and \citet{Simpson+11aj2} find a significant orbital eccentricity. This is expected given the correlation between the presence of a relatively massive planet and an eccentric orbit \citep{Me+09apj}. I adopted $e = 0.107 \pm 0.013$ and $\omega = 94^\circ \pm 4^\circ$ \citep{Torres+10apj,Winn+11aj,Pont+11mn} in the analysis below. Due to the limited quality of individual light curves I did not attempt LD-fitted solutions.

\citet{Torres+10apj} presented $i$-band observations of five transits using the FLWO 1.2\,m, for which I multiplied the errorbars by $\sqrt{9.82}$ to get $\chir = 1.0$. The full solutions are in \apptab\ and the MC errors are larger than the RP ones.

The work by \citet{Simpson+11aj2} includes two $V$$+$$R$ band transits obtained using LT/RISE, of which one is only patially covered, and one $Z$-band transit using FTN/Spectral. I put the former onto a common flux zeropoint and multiplied the errorbars by $\sqrt{5.15}$ before the {\sc jktebop} analysis. For both datasets (\apptab\ and \apptab), the LD-fit/fix solutions returned unphysical LD coefficients so I adopted the LD-fixed results. The RP errors are significantly larger than the MC errors, showing that correlated noise is important for both datasets.

The data of \citet{Nascimbeni+11aa} were originally 2247 datapoints but the file deposited at the CDS is binned to 109 normal points. Solutions of these data (\apptab) once more yielded unphysical LD coefficients so the LD-fixed solutions had to be adopted.

A summary of the light curve solutions can be found in \apptab\ and a plot of the best fits in Fig.\,\ref{fig:hatp14:lc}. Agreement between the four light curves is within the uncertainties for $r_{\rm A}+r_{\rm b}$, $i$ and $r_{\rm A}$, and slightly greater than the uncertainties for $k$ and $r_{\rm b}$. Literature value agree with my own, but the errorbars are in some cases too small. This situation persists for the physical properties of the system (\apptab). More photometry and RV measurements would be useful, especially in reducing the reliance on theoretical LD coefficients, but HAT-P-14 is a well-understood system.


\subsection{Kepler-7}                                                                                                        \label{sec:teps:kepler7}

\begin{figure} \smfig{\includegraphics[width=\columnwidth,angle=0]{plot-kepler07.eps}}{\includegraphics[width=\columnwidth,angle=0]{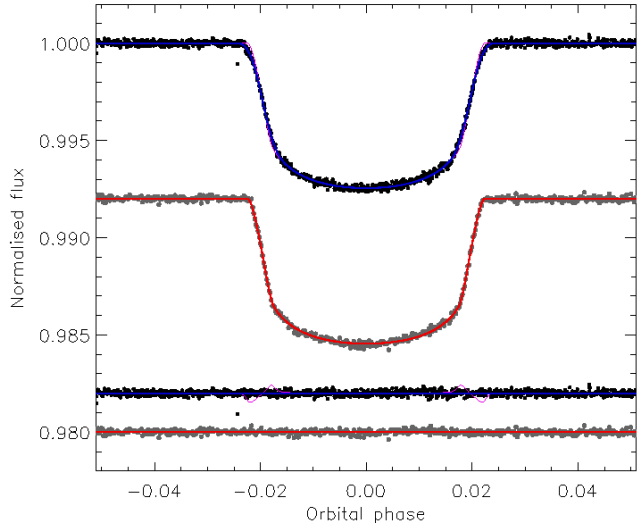}}
\caption{\label{fig:kepler7:lc} The \kepler\ light curves of Kepler-7 compared
to the {\sc jktebop} best fits. The top light curve is the Q0-Q3 long-cadence
data and the bottom one is the Q3 short-cadence data. Other comments are the
same as Fig.\,\ref{fig:corot17:lc}.} \end{figure}

Kepler-7 contains a low-density TEP ($0.094 \pm 0.009$\pjup) and a slightly evolved star ($\logg = 3.97 \pm 0.02$). The discovery paper \citep{Latham+10apj} presented an analysis of the \kepler\ Q0 and Q1 data. The studies of \citet{KippingBakos11apj} and Paper\,IV had access also to the Q2 data, although none of the three datasets were short-cadence. \citet{Demory+11apj} presented a study of the Q0-Q4 data, of which Q3 and Q4 are short-cadence, and found the planet to have a high albedo from consideration of the occultation visible in the \kepler\ data. Their analysis included asteroseismological constraints, which are not used here in order to avoid inhomogeneity.

In the current work I have analysed the Q3 data, which are the only short-cadence data publicly available for Kepler-7. The improved time sampling of these data means that my results are better than those from Paper\,IV, which were based on long-cadence data. A third light value of $L_3 = 1.4$\% is listed in the Kepler Input Catalogue \citep[KIC;][]{Brown+11aj}, but without uncertainties. I therefore adopted an uncertainty of 0.5\%.

A new orbital ephemeris was derived from the Q0-Q3 long-cadence data, fitting only those data near one of the 42 transits and using $N_{\rm int} = 10$. This yielded:
$$ T_0 = {\rm BJD(TDB)} \,\, 2\,455\,074.756840 (53) \, + \, 4.8854922 (38) \times E $$
where the reference time of minimum has been carefully chosen to be near the midpoint of the data interval, thus minimising its covariance with the orbital period. $E$ represents an integer cycle count, and the bracketed quantities give the uncertainty in the last digit of the preceding number. Errorbars were calculated using MC and RP simulations and the larger of the two options is quoted. The two sets of errorbars were very similar, and also close to the formal errors calculated by {\sc jktebop} from the covariance matrix of the best-fitting solution.

The Q3 data comprise 125\,082 datapoints covering 17 full transits. One transit had a brightening feature affecting six datapoints around time BJD(TDB) $55133.35$ which could indicate a stellar flare or instrumental artefact, so the data for this transit were ignored. 16\,897 datapoints around the remaining transits were phase-binned by a factor of 25 to give 676 normal points. These data were modelled using {\sc jktebop} (\apptab). The LD-fitted solutions are good enough to be adopted, and the RP errors are smaller than the MC ones (as usual for phase-binned data).

The light curve solutions are summarised in \apptab\ and the best fit to the data is plotted in Fig.\,\ref{fig:kepler7:lc}. The new results from the Q3 short-cadence data are in excellent agreement with those from Paper\,IV but are more precise. I combined the two to get the final photometric parameters. The values found by \citet{KippingBakos11apj} are in agreement with my new results, whereas the ones from \citet{Latham+10apj} are in marginal disagreement.

When determining the physical properties of Kepler-7 I found that the results using the different stellar models are in comparatively poor agreement (\apptab), as often happens when the host star is evolved. My final results give a more massive star and planet than previously found, but within the errorbars. \kepler\ continues to observe this system, so even better results will available in the future.


\subsection{Kepler-12}                                                                                                      \label{sec:teps:kepler12}

\begin{figure} \smfig{\includegraphics[width=\columnwidth,angle=0]{plot-kepler12.eps}}{\includegraphics[width=\columnwidth,angle=0]{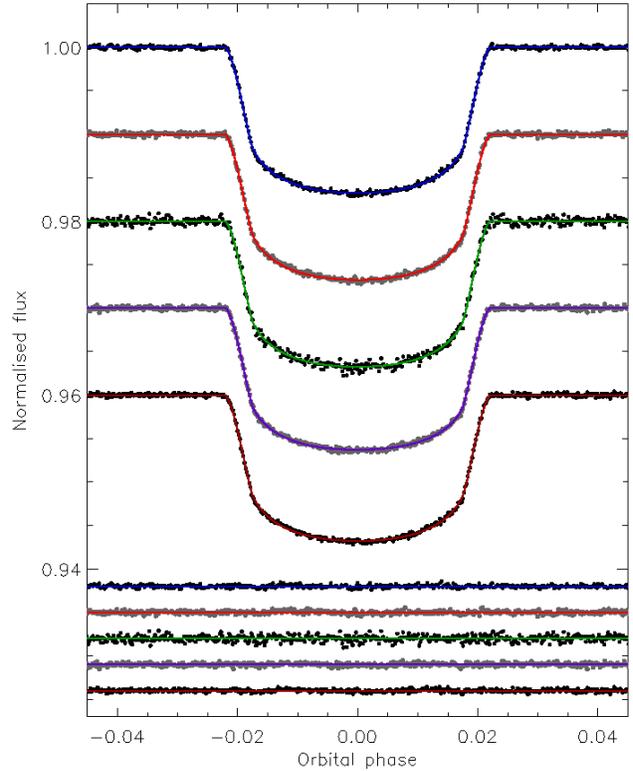}}
\caption{\label{fig:kepler12:lc} The \kepler\ light curves of Kepler-12 compared to the
{\sc jktebop} best fits. From top to bottom are the phase-binned short-cadence light
curves from Q2 to Q6. Other comments are the same as Fig.\,\ref{fig:hatp3:lc}.} \end{figure}

Discovered by \citet{Fortney+11apjs}, the Kepler-12 system contains the second-least-dense planet after WASP-17 \citep{Anderson+10apj,Anderson+11mn,Me+12c} and ahead of Kepler-7 (see above). \citet{Fortney+11apjs} obtained the constraint $e < 0.09$ \reff{at} the $3\sigma$ level. Data from Q0 to Q6 are publicly available, but the Q0 and Q1 data are only long-cadence so I did not consider them further. Q4 is split into three periods, of which the Q4.1 data are available but the Q4.2 and Q4.3 data are not.

I analysed the Q2 to Q6 data, all of which are short-cadence, using the season-dependent third light values given in the KIC and assuming an uncertainty of 0.5\%. The data from different quarters had to be analysed separately in order to be able to assign the appropriate third light values. In each case the transits were extracted from the full data, normalised to unit flux, and then phase-binned by a factor of 25 (a factor of seven for Q4 as it contains three times fewer datapoints). The solutions are given in \apptabbb{5}. In each case the MC and RP errors are similar. The light curve solutions are summarised in \apptab\ and the best fits are plotted in Fig.\,\ref{fig:kepler12:lc}. The five sets of solutions agree well with each other, and the combined results also agree well with those of \citet{Fortney+11apjs}.

The resulting physical properties (\apptab) are also in very good agreement with those of \citet{Fortney+11apjs}. Whilst \kepler\ continues to observe this system, the precision of its physical properties is limited by the spectroscopic properties of the host star and by imperfect interpolation within tabulated predictions from theoretical stellar models.


\subsection{Kepler-14}                                                                                                      \label{sec:teps:kepler14}

\begin{figure} \includegraphics[width=\columnwidth,angle=0]{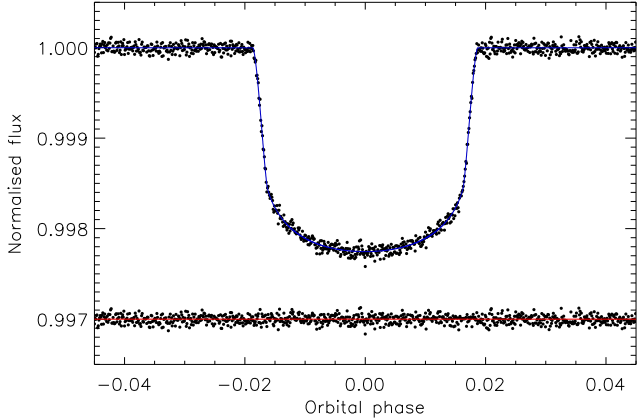}
\caption{\label{fig:kepler14:lc} The \kepler\ Q3-Q6 short-cadence light curve
of Kepler-14 compared to the {\sc jktebop} best fit. Other comments are the
same as Fig.\,\ref{fig:hatp3:lc}.} \end{figure}

Discovered by \citet{Buchhave+11apjs}, the system holds a massive ($7.7 \pm 0.4$\Mjup) hot Jupiter orbiting a comparatively hot star ($\Teff = 6395 \pm 60$\,K). The \kepler\ light curve is strongly contaminated by the light from a similarly bright star within 0.29\as\ of system. \citet{Buchhave+11apjs} measured a magnitude difference in the \kepler\ passband of $\Delta m_{\rm Kp} = 0.44 \pm 0.10$, and corrected both the light curves and RVs for this contamination. Given the very low contamination levels from KIC (0.007--0.010 for the four \kepler\ seasons) it seems that the only relevant third light comes from the very close star. Its closeness means that the data from different seasons will be affected in the same way, so it is reasonable to adopt $L_3 = 0.333 \pm 0.062$ when fitting all the \kepler\ data.

\kepler\ observed Kepler-14 in long cadence in Q0-Q6, and in short cadence in Q3-Q6. \citet{Buchhave+11apjs} opted to use the full set of long-cadence data only, for homogeneity and simplicity and because the transit is relatively long (L.\ Buchhave, private communication). Results based on short-cadence data stand a good chance providing improved results due to their much better time resolution. I have therefore modelled the Q3 to Q6 short-cadence data.

These data comprise a total of 507\,006 datapoints covering 50 complete transits. 58\,161 of these datapoints are close to transits, and I phase-binned them by a factor of 50 to get 1164 normal points. I adopted $e\cos\omega = 0.0006 \pm 0.0099$ and $e\sin\omega = 0.0350 \pm 0.0170$ \citep{Buchhave+11apjs}, whch suggests that eccentricity is significant at the $2\sigma$ level. The RP errorbars were smaller than the MC ones, and I was able to adopt LD-fitted solutions (\apptab).

Given the extremely high quality of the light curve I tried fitting for $L_3$ directly. Investigations of this possibility in Paper\,III showed that third light is almost completely degenerate with other photometric parameters ($r_{\rm A}$, $r_{\rm b}$ and $i$), so may not be obtainable even from exceptionally good data. I obtained $L_3 = 0.223 \pm 0.085$ which is within $1\sigma$ of the expected value but still poorer than using outside constraints. I also checked whether it was possible to fit for $e\cos\omega$ and/or $e\sin\omega$, and confirmed the result of \citet{Kipping08mn} that such parameters cannot be extracted. The best-fitting values were $e\cos\omega = 0.350 \pm 0.065$ and $e\sin\omega = 0.149 \pm 0.019$, which are inconsistent with the known results determined by \citet{Buchhave+11apjs} from RVs and a {\it Spitzer} occultation.

A summary of the light curve solutions is given in \apptab\ and the best fit is plotted in Fig.\,\ref{fig:kepler14:lc}. I find somewhat different photometric parameters from those of \citet{Buchhave+11apjs}, which is likely down to the use of short-cadence data in the current work. The third light has a strong effect on the results, as the photometric parameters are all strongly correlated with it. It thus sets a limit on the quality of the solution, so improved measurements of $L_3$ would be the best way to improve our understanding of the system.

I find a rather different set of physical properties for Kepler-14 (\apptab), with a less massive but more evolved star. Compared to \citet{Buchhave+11apjs} I find a stellar mass of \wwo{$M_{\rm A} = 1.318 \pm 0.052 \pm 0.029$}{$M_{\rm A} = 1.32 \pm 0.06$}\Msun\ rather than $1.512 \pm 0.043$\Msun\ (a change of $2.6\sigma$), a very similar stellar radius and a $1.4\sigma$ different surface gravity (\wwo{$\log g_{\rm A} = 3.918 \pm 0.040 \pm 0.003$}{$\log g_{\rm A} = 3.92 \pm 0.04$} versus \er{3.994}{0.028}{0.036}). The $\log g_{\rm A}$ derived from spectral analysis is $4.11 \pm 0.10$ \citep{Buchhave+11apjs}. These changes propagate to the planet, which is found to be less massive but the same size as before. \kepler\ will provide further photometry to help investigate this system, but obtaining revised \Teff, \FeH\ and $L_3$ values would be the best way to improving our understanding of the Kepler-14 system.


\subsection{Kepler-15}                                                                                                      \label{sec:teps:kepler15}

\begin{figure} \smfig{\includegraphics[width=\columnwidth,angle=0]{plot-kepler15.eps}}{\includegraphics[width=\columnwidth,angle=0]{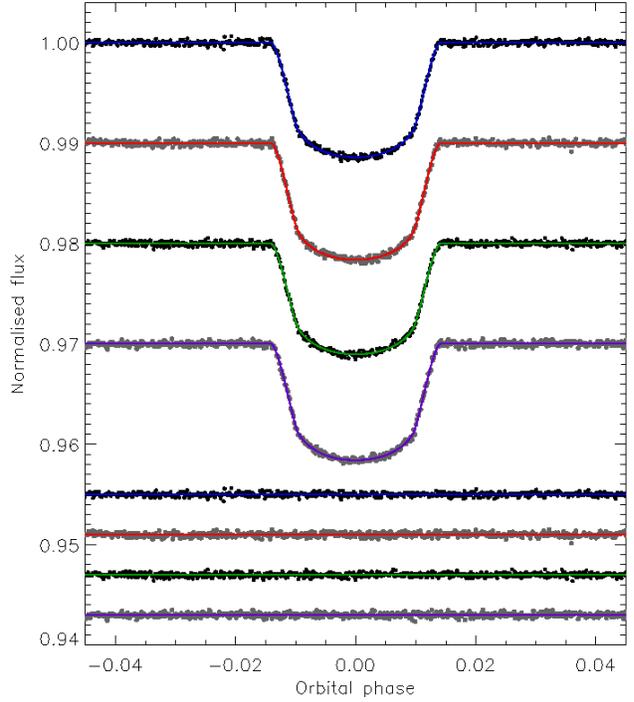}}
\caption{\label{fig:kepler15:lc} The \kepler\ Q3 to Q6 short-cadence light curves
of Kepler-15 compared to the {\sc jktebop} best fits. Other comments are the
same as Fig.\,\ref{fig:hatp3:lc}.} \end{figure}

\citet{Endl+11apjs} presented the discovery of this planetary system containing a comparatively low-mass ($0.70 \pm 0.10$\Mjup) planet around a cool ($5595 \pm 120$\,K) star. \citet{Endl+11apjs} used the \kepler\ long-cadence photometry from quarters Q1 to Q6 plus RVs from two telescopes to characterise the system. However, \kepler\ short-cadence data are also available for quarters Q3 to Q6, and their much greater time resolution should yield better results than for the long-cadence data.

The \kepler\ Q3 to Q6 short-cadence data were modelled individually quarter-by-quarter, using the third light values from the KIC and adopting an uncertainty of 0.5\% in these values. Eccentricity was not found to be significant by \citet{Endl+11apjs} so I assumed $e=0$. Each dataset was phase-binned by a factor of 20, giving 613, 563, 685 and 644 normal points respectively. The full solutions are in \apptabbb{4} and the best fits are shown in Fig.\,\ref{fig:kepler15:lc}.

\apptab\ summarises the light curve solutions, and shows that their interagreement is good (except for $k$ where $\chir = 1.8$). The final photometric parameters are extremely precise. However, their ageement with the findings of \citet{Endl+11apjs} is very poor: differences of $24\sigma$ occur for $r_{\rm A}$ and $i$, $7\sigma$ for $k$ and $33\sigma$ for $r_{\rm b}$. It is unclear how such a large discrepancy could arise, although one possibility to investigate would be the use of long-cadence versus short-cadence data. I was unable to reproduce the results of \citet{Endl+11apjs} by fitting the long-cadence data either with or without numerical integration. My new results should be preferred because the short-cadence data are much better for measuring planet parameters than the long-cadence data (see Paper\,IV).

As would be expected from the rather different photometric parameters, my calculated physical properties (\apptab) are at odds with those of \citet{Endl+11apjs}. The most obvious changes are for the radii of the two components, which go from $R_{\rm A} = \er{0.992}{0.058}{0.070}$\Rsun\ and $R_{\rm b} = \er{0.96}{0.06}{0.07}$\Rjup\ \citep{Endl+11apjs} to \wwo{$R_{\rm A} = 1.253 \pm 0.047 \pm 0.020$}{$R_{\rm A} = 1.25 \pm 0.05$}\Rsun\ and \wwo{$R_{\rm b} = 1.289 \pm 0.050 \pm 0.021$}{$R_{\rm b} = 1.29 \pm 0.05$}\Rjup\ (this work). The $\log g_{\rm A}$ I find is in better agreement with the spectroscopic measurement, albeit at low significance. My revisions to the physical properties of the Kepler-15 system decrease the planet's measured density by more than a factor of 2, increase the measured age of the star, and include the first measurement of $g_{\rm b}$, \Teq\ and \safronov.


\subsection{Kepler-17}                                                                                                      \label{sec:teps:kepler17}

\begin{figure} \includegraphics[width=\columnwidth,angle=0]{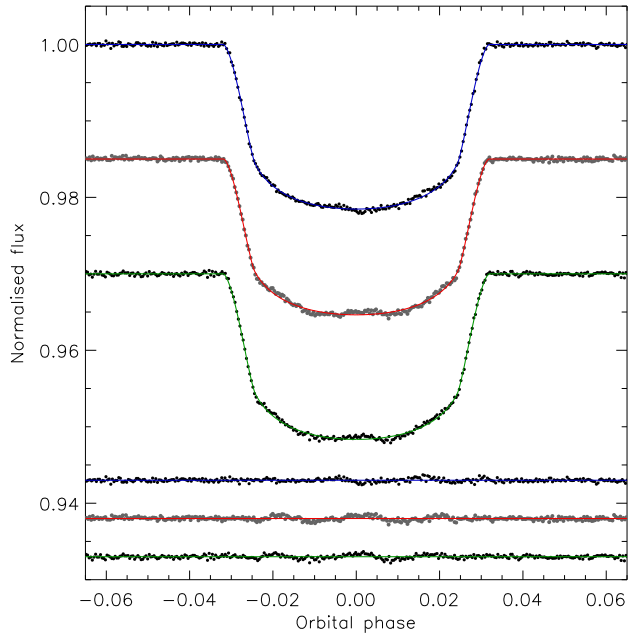}
\caption{\label{fig:kepler17:lc} The \kepler\ Q4-Q6 short-cadence light curves
of Kepler-17 compared to the {\sc jktebop} best fits. Other comments are the
same as in Fig.\,\ref{fig:hatp3:lc}.} \end{figure}

Kepler-17 comprises a hot and massive planet ($\Teq = 1712 \pm 25$\,K, $M_{\rm b} = \er{2.34}{0.09}{0.24}$\Mjup) orbiting an active star. Starspot activity is obvious both in the overall light curve and by spot crossings during the planetary transits, allowing a measurement of the star's rotation period in the discovery paper \citep{Desert+11apjs}. The rotation period is close to eight times \Porb, giving a `stroboscopic' effect whereby some starspots can be identified each eighth transit. This also allows the orbital obliquity to be constrained to be less than 15$^\circ$ \citep{Desert+11apjs}. \spitzer\ observations of an occultation allowed limits to be placed on orbital shape $e < 0.011$ in combination with the \kepler\ photometry and RV observations. An independent discovery of the Kepler-17 system was made by \citet{Bonomo+12aa}, who combined their RVs with those of \citet{Desert+11apjs} to obtain a more precise $K_{\rm A}$. Their other result were in good agreement with those from the discovery paper, but less precise as they only had \kepler\ Q1-Q2 data versus the Q1-Q6 available to \citet{Desert+11apjs}.

As before (see HAT-P-11 in Paper\,IV) I have assumed that the starspot activity under the transit chord is on average the same as that for the rest of the stellar photosphere. This means that the data affected by starspots do not require special treatment, and the effect of the starspot crossings on the solution will simply be to inflate the errorbars as an additional source of correlated noise. This in turn is partially coverted into white noise when the data are phase-binned, and will ultimately cause larger errorbars to be found than if the star were quiet.

The \kepler\ Q4 to Q6 data are available in short-cadence flavour, and these were phase-binned as usual to yield 621, 634 and 621 normal points, respectively. A significant fraction ($\sim$15\%) of transits were normalised using a quadratic function instead of a straight line, in order to cope with the modulation due to starspot activity. 120 datapoints were rejected from the Q5 data by a $4\sigma$ clip, due to an unusually large number of scattered observations, whereas the numbers rejected from Q4 and Q6 data were 3 and 11 respectively.

Third light was tackled by adopting the season-dependent contamination values from the KIC and assigning an uncertainty of 0.5\% to each one. The orbit was assumed to be circular. The full sets of light curve solutions are given in \apptabbb{3} and in each case the RP errors were found to be moderately larger than the MC ones, due to the starspots on the stellar surface. The best fits are plotted in Fig.\,\ref{fig:kepler17:lc}, where the influence of spot activity can be seen in the residuals of the best fit. The light curve solutions are summarised in \apptab, and are in good agreement, except for $k$ where $\chir = 5.8$. This is due both to starspot activity and the need to use quadratic functions to normalise some transits, and has been accounted for in the errorbars. It is not sufficient to cause a disagreements in $r_{\rm A}$ or $r_{\rm b}$.

The light curve solution I find is close to a central transit ($i = \er{89.73}{0.27}{0.94}$\,deg) whereas \citet{Desert+11apjs} prefer a lower-inclination solution ($i = 87.2 \pm 0.15$\,deg) which differs by $2.6\sigma$ from my own. This affects $r_{\rm A}$ (which is $3.9\sigma$ larger according to \citeauthor{Desert+11apjs}) and $k$ ($3.9\sigma$ smaller). The source of this divergence is unclear. \citet{Bonomo+12aa} had access to only a little of the \kepler\ data, so adopted the photometric parameters from \citet{Desert+11apjs} rather than calculate their own. Given the discrepancies, the larger errorbars found in the present study should be preferred to those given in the literature.

When calculating the physical properties of the Kepler-17 system I found slight disagreements within results using the five sets of stellar models, which leads to a significant systematic component in the errorbars of the final system parameters. The values themselves are in reasonable agreement with previous studies (\apptab). As \kepler\ continues to gather photometry of this system, the action with the most benefit would be the procurement of improved spectroscopy for \Teff, \FeH\ and RV measurements.


\subsection{KOI-135}                                                                                                          \label{sec:teps:koi135}

\begin{figure} \includegraphics[width=\columnwidth,angle=0]{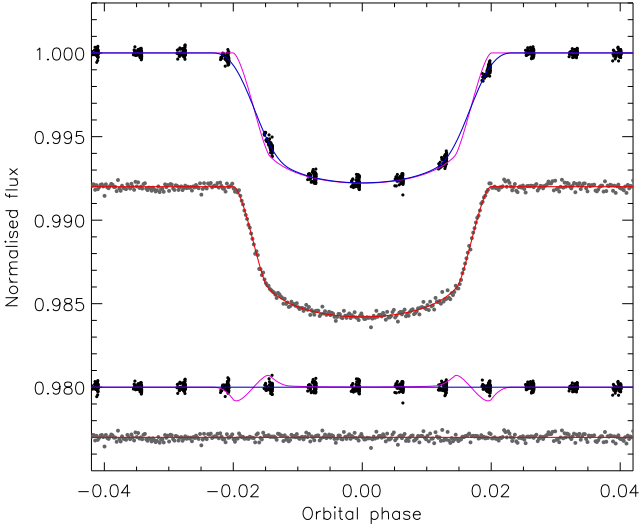}
\caption{\label{fig:koi135:lc} The \kepler\ light curves of KOI-135 compared
to the {\sc jktebop} best fits. The upper dataset is the Q1-Q3 long cadence
data and the lower dataset is the Q3 short cadence data. Other comments are
the same as Fig.\,\ref{fig:corot17:lc}.} \end{figure}

\reff{KOI-135 was discovered by \citet{Bonomo+12aa} to contain} a massive planet ($3.09 \pm 0.21$\Mjup). \citet{Bonomo+12aa} had access to only the Q1 and Q2 long-cadence data. One of the problems they faced is that the \kepler\ long-cadence sampling rate is close to an integer multiple of the \Porb\ of KOI-135, leading to very poor sampling of the shape of the transit (see Fig.\,\ref{fig:koi135:lc}). Q3 short-cadence data have now been released, which should allow the current work to produce a significant improvement in the measured system parameters.

\citet{Bonomo+12aa} found that their spectroscopic measurement of $\log g_{\rm A}$ disagreed with that obtained from analysis of the transits by $2.7\sigma$. They found that using the transit-derived $\log g_{\rm A}$ resulted in an insignificantly different measurement of \Teff\ from their spectra.

My first action was to obtain an improved orbital ephemeris using the Q1 to Q3 data. This was done in the same way as for Kepler-7 in Sect.\,\ref{sec:teps:kepler7}, resulting in:
$$ T_0 = {\rm BJD(TDB)} \,\, 2\,455\,068.235237 (54) \, + \, 3.0240933 (27) \times E $$
The precision of this \Porb\ is a factor of ten better than that based on only Q1 and Q2 data \citep{Bonomo+12aa}.

Constraints on third light were obtained from the KIC as above, and errors of 0.5\% specified. A circular orbit was adopted as \citet{Bonomo+12aa} found $e < 0.025$ (significance level not stated). Only the short-cadence (Q3) data were modelled, for the reasons given above. The original 125\,140 datapoints were slimmed down to 18\,938 close to any of the 27 transits observed by \kepler, 13 were rejected by a $4\sigma$ clip, and the remainder were phase-binned by a factor of 25 to get 759 normal points. \apptab\ shows that the LD-fitted solutions can be used. The RP errors were smaller than the MC errors. The Q3 data yield a $3\sigma$ smaller $r_{\rm A}$ and $r_{\rm b}$ than found by \citet{Bonomo+12aa}, so the availability of improved photometry has resulted in the modification of the measured system properties (\apptab).

Despite the large uncertainties in \Teff\ (143\,K) and \FeH\ (0.11\,dex), I found highly consistent physical properties for the KOI-135 system using different sets of stellar models (\apptab). Both the star and planet are slightly smaller than found by \citet{Bonomo+12aa}, but these changes are within the errorbars. Forthcoming \kepler\ photometry and further spectroscopic measurements of \Teff\ and \FeH\ would be very helpful in refining the system properties.


\subsection{KOI-196}                                                                                                          \label{sec:teps:koi196}

\begin{figure} \includegraphics[width=\columnwidth,angle=0]{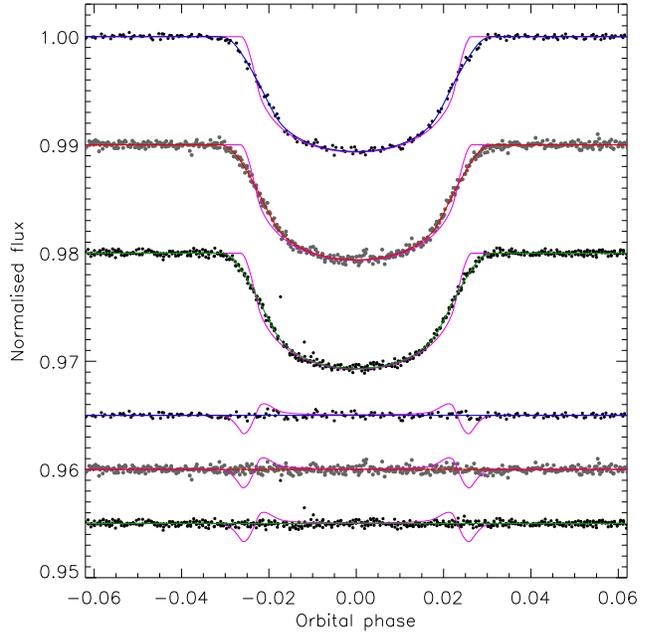}
\caption{\label{fig:koi196:lc} The \kepler\ light curves of KOI-196 compared
to the {\sc jktebop} best fits. From top to bottom the three datasets come
from the Q1, Q2 and Q3 time period. Other comments are the same as
Fig.\,\ref{fig:corot17:lc}.} \end{figure}

\citet{Santerne+11aa2} discovered the planetary nature of this system, based on Q1-Q2 data from the \kepler\ satellite and RVs from the SOPHIE spectrograph. \citeauthor{Santerne+11aa2} also detected occultations in the \kepler\ data, of a depth which suggest the planet has a high albedo and at an orbital phase which is consistent with a circular orbit. Since this study the Q3 data have become available, but like Q1 and Q2 are only long-cadence.

Third light was estimated from the KIC, 0.5\% errors assigned, and a circular orbit assumed. An improved orbital ephemeris was obtained from the Q1-Q3 data using the method in Sect.\,\ref{sec:teps:kepler7}:
$$ T_0 = {\rm BJD(TDB)} \,\, 2\,455\,066.669310 (48) \, + \, 1.8555588 (15) \times E $$

For the Q1 data, 257 of the 1625 datapoints are in the region of one of the 18 transits. For Q2 these numbers are 4802, 648 and 45, and for Q3 they are 4145, 662 and 46. The data were not binned or phased, and $N_{\rm int} = 10$ was used in the modelling process. Two families of solutions were found: one with a nearly central transit ($i \approx 90^\circ$) and one with $i \approx 80$--$82^\circ$. The $i \approx 80$--$82^\circ$ family occurs mainly for the LD-fixed light curve solutions, and results in weird physical properties. In contrast, the LD-fit/fix solutions universally have $i \approx 90^\circ$ and lead to reasonable physical properties. This implies that fixing LD coefficients can have a deleterious effect on a light curve solution, specifically when working with long-cadence data. I adopted the LD-fit/fix solutions (\apptabbb{3}), in agreement with \citet{Santerne+11aa2}, and found that the RP errors are of similar size to the MC errors.

The photometric parameters are given in \apptab\ and the best fits are plotted in Fig.\,\ref{fig:koi196:lc}. The interagreement between the Q1-Q3 data is excellent, but further photometry (preferably short-cadence) is needed to obtain a definitive solution. \citet{Santerne+11aa2} found similar but slightly discrepant photometric parameters, based on fewer data.

As with KOI-135, I found highly consistent physical properties for the KOI-135 system from different sets of stellar models (\apptab) despite the large uncertainties in \Teff\ (100\,K) and \FeH\ (0.11\,dex). Compared to \citet{Santerne+11aa2}, I deduced a slightly smaller and less massive star and planet. Further spectroscopy and forthcoming \kepler\ photometry would allow a useful improvement in measurements of the physical properties of KOI-196.


\subsection{KOI-204}                                                                                                          \label{sec:teps:koi204}

\begin{figure} \includegraphics[width=\columnwidth,angle=0]{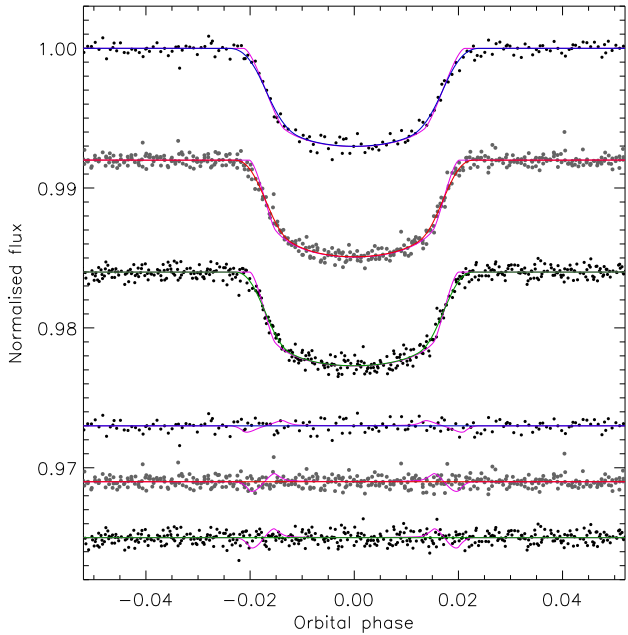}
\caption{\label{fig:koi204:lc} The \kepler\ Q1, Q2 and Q3 long-cadence light
curves of KOI-204 compared to the {\sc jktebop} best fits. Other comments are
the same as Fig.\,\ref{fig:corot17:lc}.} \end{figure}

KOI-204\,\reff{b} was discovered by \citet{Bonomo+12aa} as part of a program to observe the brighter \kepler\ TEP candidate systems with SOPHIE. \citet{Bonomo+12aa} had access to \kepler\ Q1 and Q2 data, and Q3 data are now available. I derived an improved orbital ephemeris using these data and the method in Sect.\,\ref{sec:teps:kepler7}, finding
$$ T_0 = {\rm BJD(TDB)} \,\, 2\,455\,067.02680 (15) \, + \, 3.2467220 (73) \times E $$

The orbit was assumed to be circular (\citeauthor{Bonomo+12aa} found $e < 0.021$) and third light values were obtained from the KIC. These are a little larger than usual, being 0.060, 0.093, 0.154 and 0.044, in seasons 0 to 3 respectively. The individual quarters of data were then fitted using $N_{\rm int} = 10$, and the full solutions are given in \apptabbb{3}. The RP errors are smaller than those from the MC algorithm. The best fits are plotted in Fig.\,\ref{fig:koi204:lc}.

\apptab\ gives the final photometric parameters. The three datasets agree well except for $k$ ($\chir=1.9$), which is likely a symptom of spot crossings duing the transit. These are not visible to the eye, due to the poor time resolution of the long-cadence data. The values from \citet{Bonomo+12aa} are concordant with my own, but their errorbars are smaller despite being based on significantly less data. The same comments apply to the physical properties determined and gathered in \apptab. More observations of all types are needed to characterise this planetary system well.


\subsection{KOI-254}                                                                                                          \label{sec:teps:koi254}

\begin{figure} \includegraphics[width=\columnwidth,angle=0]{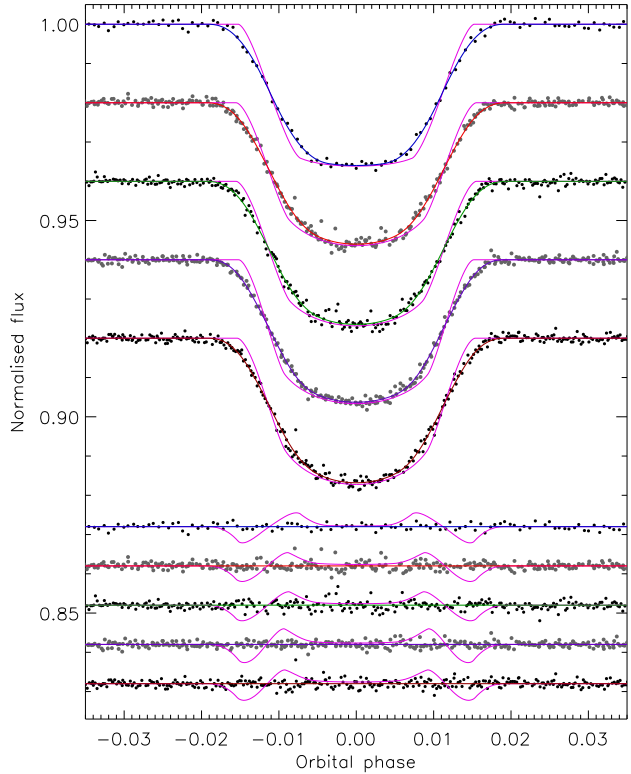}
\caption{\label{fig:koi254:lc} The \kepler\ Q1 to Q5 light curves (from top
to bottom) of KOI-254 compared to the {\sc jktebop} best fits. Other
comments are the same as Fig.\,\ref{fig:corot17:lc}.} \end{figure}

\citet{Johnson+12apj} discovered this planetary system from RV follow-up of \kepler\ KOI objects \citep{Borucki+11apj} classified as M\,dwarfs, in order to probe this host star mass regime. The stellar component is on the border of the M spectral class ($M_{\rm A} = 0.57 \pm 0.06$\Msun\ and $\Teff = 3820 \pm 90$\,K), and the planet is the largest one found around a sub-4000\,K star ($R_{\rm b} = 1.00 \pm 0.07$\Rjup) by some distance (the next is GJ\,436\,b at $0.37 \pm 0.02$\Rjup). KOI-254 is therefore an interesting object in delineating the minimum stellar mass required to host a Jupiter-sized object. \citet{Johnson+12apj} analysed the \kepler\ Q1 and Q2 data, which shows clear spot activity, plus a $Z$-band transit light curve obtained using the 1\,m Nickel telescope at Lick Observatory.

At the time of writing, \kepler\ Q1 to Q5 data were available, all at long cadence. A refined orbital ephemeris was calculated as in Sect.\,\ref{sec:teps:kepler7}:
$$ T_0 = {\rm BJD(TDB)} \,\, 2\,455\,160.956396 (31) \, + \, 2.45524122 (65) \times E $$
which is a factor of ten more precise than the ephemeris based on only the Q1 and Q2 data.

The data from each quarter were modelled assuming an eccentric orbit with $e = \er{0.11}{0.1}{0.09}$ and $\omega = 230^\circ \pm 68^\circ$ \citep{Johnson+12apj}, third light values from the KIC with uncertainties assumed to be 0.5\%, and $N_{\rm int} = 10$ (\apptabbb{5}). \apptab\ shows the agreement to be very good both between the five datasets and with the values from \citet{Johnson+12apj}. The RP errors were all smaller than the MC errors. The best fits are shown in Fig.\,\ref{fig:koi254:lc}.

Precise physical properties become difficult to obtain into the M-dwarf regime, as the interagreement between stellar models deteriorates, but KOI-254 is thankfully not strongly affected. The physical properties I find (\apptab) agree well with those of \citet{Johnson+12apj}, but the availability of additional data allows the errorbars to be beaten down. \kepler\ will deliver further photometry, but more spectroscopy would be useful in nailing down the \Teff, \FeH\ and spectrosocpic orbit of the host star. A significant source of uncertainty in the current results stems from the measurement of $e$ and $\omega$, which is rather imprecise due to the small RV signal exhibited by the star.


\subsection{KOI-423}                                                                                                       \label{sec:teps:koi423}

\begin{figure} \includegraphics[width=\columnwidth,angle=0]{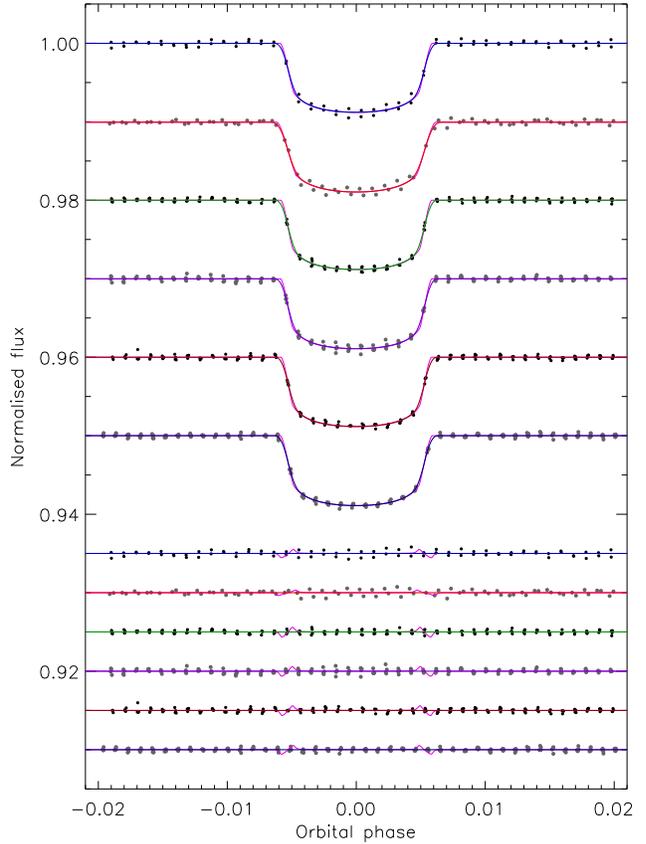}
\caption{\label{fig:koi423:lc} The \kepler\ Q1 (top) to Q6 (bottom) light
curves of KOI-423 compared to the {\sc jktebop} best fits. Other comments
are the same as Fig.\,\ref{fig:corot17:lc}.} \end{figure}

KOI-423 contains an object of brown-dwarf mass ($M_{\rm b} = 17.9 \pm 1.8$) in a long-period (21.1\,d) eccentric orbit ($e = \er{0.121}{0.022}{0.023}$) around a hot and metal-poor ($\Teff = 6260 \pm 140$\,K and $\FeH = -0.29 \pm 0.10$) star \citep{Bouchy+11aa2}. Its mass, however, places it firmly in the planetary rather than the stellar population \reff{according to} \citet{GretherLineweaver06apj}. \citet{Bouchy+11aa2} based their analysis on RVs from the SOPHIE spectrograph, and the \kepler\ Q1 and Q2 data (which encompass only five transits between them).

The Q3 to Q6 data are also now public; all six datasets are not only long-cadence but also sample the transit shape poorly (Fig.\,\ref{fig:koi423:lc}). A revised orbital ephemeris was first procured using the method in Sect.\,\ref{sec:teps:kepler7}:
$$ T_0 = {\rm BJD(TDB)} \,\, 2\,455\,204.55515 (14) \, + \, 21.087168 (21) \times E $$

Each dataset was solved with numerical integration ($N_{\rm int} = 10$), constraints on $L_3$ from the KIC, and adopting $e\cos\omega = \er{0.120}{0.022}{0.024}$ and $e\sin\omega = \er{-0.019}{0.015}{0.012}$ \citep{Bouchy+11aa2}. A quadratic or cubic function was needed to normalise the majority of the transits to unit flux. MC errors were larger than RP errors, and full solutions (except for LD-fitted) can be found in \apptabbb{6}. The total dataset covers 22 transits, of which one had to be rejected due to normalisation problems caused by missing data, which is a significant improvement on the five available to \citet{Bouchy+11aa2}.

\apptab\ shows that the six different light curves agree well. The combined solution has a somewhat larger $i$, $r_{\rm A}$ and $r_{\rm b}$, but these are within $1$--$2\sigma$ of those found by \citet{Bouchy+11aa2}. The physical properties of KOI-423 (\apptab) are tolerably well defined, and are reasonably close to the values found by \citet{Bouchy+11aa2}. \kepler\ short-cadence data would be valuable for this object, as would further spectroscopy to refine \Teff\ and \FeH.


\subsection{KOI-428}                                                                                                       \label{sec:teps:koi428}

\begin{figure} \smfig{\includegraphics[width=\columnwidth,angle=0]{plot-koi428.eps}}{\includegraphics[width=\columnwidth,angle=0]{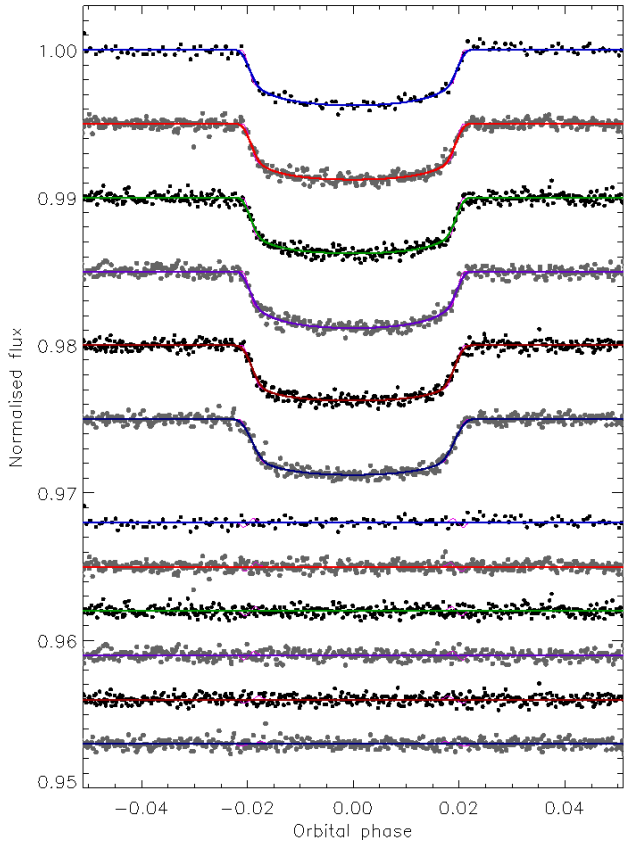}}
\caption{\label{fig:koi428:lc} The \kepler\ Q1 (top) to Q6 (bottom) light
curves of KOI-428 compared to the {\sc jktebop} best fits. Other comments
are the same as Fig.\,\ref{fig:corot17:lc}.} \end{figure}

This planetary system was discovered by \citet{Santerne+11aa}, and studied using the \kepler\ Q1 data. It was subsequently analysed in Paper\,IV, based on the Q1 and Q2 data. I have revisited it because Q3 to Q6 data are now available -- all long-cadence -- and cover 67 transits compared to the four in Q1 and the eight in Q2. The other modification to my previous work concerns the small $L_3$, which in Paper\,IV was ignored but in the current work is taken from the KIC as usual. The new orbital ephemeris based on the Q1 to Q6 data is:
$$ T_0 = {\rm BJD(TDB)} \,\, 2\,455\,204.83996 (20) \, + \, 6.8731697 (96) \times E $$

A circular orbit was assumed, and each quarter of data was tackled separately with $N_{\rm int} = 10$. The full solutions are given in \apptabbb{6} and the best fits displayed in Fig.\,\ref{fig:koi428:lc}. Correlated noise is unimportant, and the six light curves agree within the rather large MC errorbars (\apptab). This underlines the low information content in data with a sampling rate of only one per 30\,min. Short-cadence data is necessary to obtain good results on this object. The revised $r_{\rm A}$ is comfortably larger than found in the two previous studies; the change is within the errors for Paper\,IV but outside the rather small errorbars quoted by \citet{Santerne+11aa}.

The revised physical properties (\apptab) are significantly different, due to their sensitivity to $r_{\rm A}$ via the stellar density. The masses are almost unchanged but the radii both increase by about 10\%. The star is now quite evolved (\wwo{$\log g_{\rm A} = 3.812 \pm 0.048 \pm 0.017$}{$\log g_{\rm A} = 3.81 \pm 0.05$}) and the largest known TEP host star at \wwo{$R_{\rm A} = 2.48 \pm 0.17 \pm 0.20$}{$R_{\rm A} = 2.48 \pm 0.26$}\Rsun\ (the second-largest being Kepler-14\,A at \wwo{$2.09 \pm 0.11 \pm 0.02$}{$2.09 \pm 0.11$}\Rsun; Sect.\,\ref{sec:teps:kepler14}). A significant contribution to the uncertainties in these numbers comes from significant disagreement between the five model sets. The {\it DSEP} and {\it Y$^2$} models prefer a larger system scale and a slightly pre-main-sequence star, whereas the other three models find an age of 1--2\,Gyr. Arbitrarily restricting the star to be on the main sequence would improve the agreement between models, but cannot be scientifically justified. Further \kepler\ photometry and ground-based spectroscopy would be important steps towards confirming and refining the physical properties of this planetary system.


\subsection{OGLE-TR-56}                                                                                                       \label{sec:teps:ogle56}

\begin{figure} \smfig{\includegraphics[width=\columnwidth,angle=0]{plot-ogle56.eps}}{\includegraphics[width=\columnwidth,angle=0]{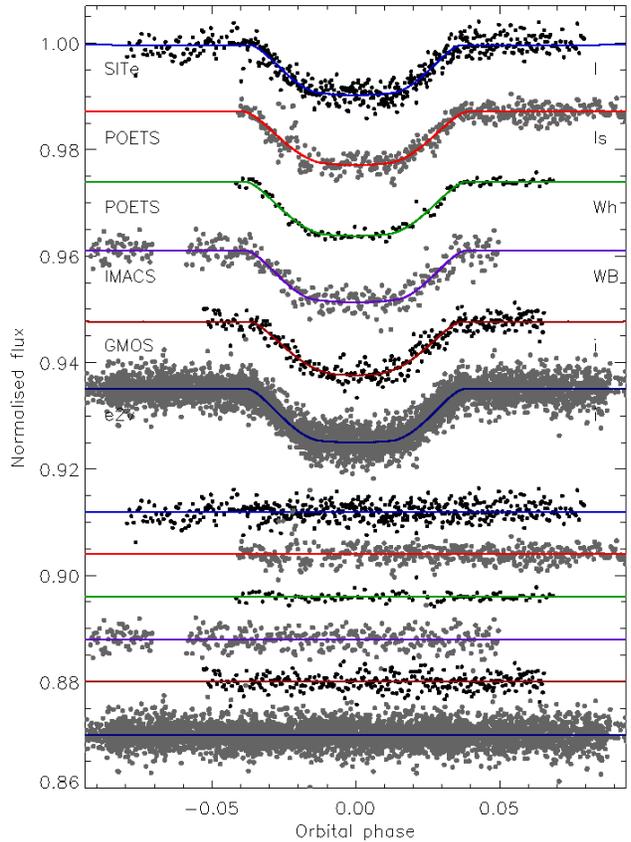}}
\caption{\label{fig:ogle56:lc} The \citet{Adams+11apj2} light curves of
OGLE-TR-56 compared to the {\sc jktebop} best fits. The sources of the data
and the filters used are annotated on the left and right, respectively.
Other comments are the same as Fig.\,\ref{fig:hatp3:lc}.} \end{figure}

OGLE-TR-56 has been studied in previous papers of this series, but qualifies for further attention due to the recent publication of much improved photometric data. Paper\,I summarises the observational history of OGLE-TR-56. \citet{Adams+11apj2} have since obtained 19 new transit light curves, primarily with the high-resolution MagIC camera on the Magellan telescopes. A few transits were also observed using the POETS and IMACS imagers on the same telescopes, and with GMOS on Gemini-South. \citet{Adams+11apj2} found a discrepancy between their transit duration and that obtained by \citet{Pont+07aa} from one transit observed using VLT/FORS2. A re-analysis of the VLT data resulted in a smaller discrepancy but more scattered photometry. One observation by \citet{Adams+11apj2} is of the same transit observed using the VLT, and the transit shape is in good agreement with the others presented by \citet{Adams+11apj2}. I have therefore discounted the VLT data (as analysed in Paper\,I) as possibly suffering from systematic noise, and rederived the properties of OGLE-TR-56 based on the photometry obtained by \citet{Adams+11apj2}. The datasets obtained using the SITe CCD and $B$, $r$ and $z$ filters were ignored as each do not cover a whole transit.

\subsubsection{Light curve analysis}

The data taken by \citet{Adams+11apj2} using the SITe CCD and $I$ filter cover four transits, of which one was only partially observed and suffers from correlated noise so was rejected. 498 datapoints cover the remaining three transits and were fitted with \Porb\ fixed. The RP errors are slightly larger than the MC errors, and the LD-fixed solutions were adopted because the LD-fit/fix alternatives gave unphysical LD coefficients (\apptab).

Two transits were observed using POETS in the $I_s$ band (\reff{a filter} described as `Schuler Astrodon Johnson-Cousins'). There is a short-lived increase in flux just before second contact for the first transit, which could be due to a stellar flare. The observations were taken at a high cadence, resulting in 8961 datapoints. These were binned in time by a factor of 20 to obtain 449 normal points, with four datapoints rejected by a $3\sigma$ clip. Once again, the RP errors were slightly larger than the MC ones, and the LD-fixed solution had to be adopted (\apptab).

One more transit was observed using POETS without a filter, in an attempt to achieve a lower Poisson noise. This transit event was the same one observed by \citet{Pont+07aa} using VLT/FORS2. The observations were taken at high cadence, resulting in 3245 datapoints. Those after time HJD 2453936.68 were rejected as they are well after the completion of the transit and suffer from systematic noise. The remainder were time-binned by a factor of ten to give 95 normal points, then modelled using {\sc jktebop}. LD coefficients appropriate for the Cousins $R$ band were used. The LD-fit/fix solutions are unfortunately poor (\apptab), and the LD-fixed solutions are possibly biased due to the imposition of $R$-band LD coefficients onto an unfiltered dataset. I therefore did not consider these data when calculating the final photometric parameters.

One transit (352 datapoints) was observed using IMACS and a wide-band filter (630--950\,nm), for which I adopted Cousins $I$-band LD coefficients. Correlated noise was unimportant, but once again LD-fixed solutions had to be adopted (\apptab). Another transit (307 datapoints) was obtained using Gemini/GMOS and an $i$ filter. For this one correlated noise was unimportant and it was possible to adopt the LD-fit/fix solution (\apptab).

Finally, nine complete transits were observed in the $i$ band using Magellan/MagIC and its E2V CCD. For these 5186 datapoints I found that correlated noise was moderately important but that it was also possible to adopt the LD-fit/fix solution (\apptab). All light curves are compared in Fig.\,\ref{fig:ogle56:lc} to their respective best-fitting models.

One minor concern is that OGLE-TR-56\,b is sufficiently close to its parent star to be distorted beyond the formal limit of applicability of the {\sc ebop} model. All other work on OGLE-TR-56 has been performed using spherical geometry, such as the analytical models by \citet{MandelAgol02apj} and \citet{Gimenez06aa}. {\sc jktebop} is more sophisiticated than those, in its use of biaxial spheroids to represent the stellar and planetary figures, but it would be worthwhile to check them using a more physically realistic model which incorporates Roche geometry. An excellent discussion of this can be found in \citet{Budaj11aj}.

The agreement between the six different datasets from \citet{Adams+11apj2} is very good, ranging from $\chir = 0.19$ for $i$ to $\chir = 0.98$ for $k$. The combined solution was calculated excluding the unfiltered POETS data, and the errorbars were (as usual) conservatively {\em not} reduced in light of the low \chir\ values. The solutions are summarised in \apptab, which also hosts a comparison with (the many) previous results for OGLE-TR-56. My new photometric parameters are in good agreement with those of \citet{Adams+11apj2} but are in severe disagreement with all previous work. This is expected because my new results are based on the data from \citealt{Adams+11apj2}, whereas all prior analyses had access to only the original survey-quality OGLE data and/or the VLT observations which have turned out to be affected by red noise \citep{Adams+11apj2}.

\subsubsection{Physical properties}

\begin{figure} \includegraphics[width=\columnwidth,angle=0]{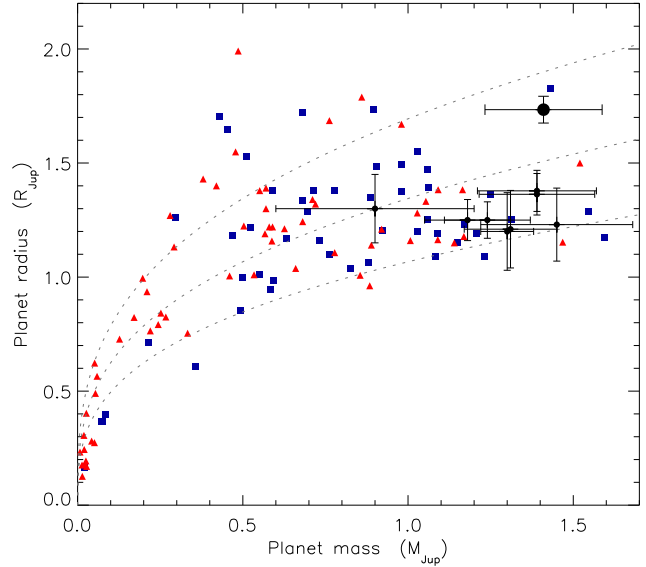}
\caption{\label{fig:ogle56:M2R2} Mass--radius plot showing the known TEPs
(blue \reff{squares} for those in the {\it Homogeneous Studies} series and red
\reff{triangles} for the remainder) with their errorbars suppressed for clarity.
The new result for OGLE-TR-56\,b is shown with a large black filled circle,
whereas previous results are shown with small black filled circles.} \end{figure}

A major change in the photometric parameters \reff{induces} a similar effect in the physical properties of the system. My results (\apptab) show that there is some disagreement between different model sets, in that the {\it Claret} and {\it Teramo} models coverge on solutions with a $1\sigma$ lower \Teff. This causes a significant systematic uncertainty to enter the error budget, but does not obscure the main story. I have found physical properties which are very different from all previous estimates, and point to OGLE-TR-56\,A being somewhat evolved (\wwo{$M_{\rm A} = 1.339 \pm 0.077 \pm 0.066$}{$M_{\rm A} = 1.34 \pm 0.10$}\Msun\ and \wwo{$R_{\rm A} = 1.737 \pm 0.035 \pm 0.029$}{$R_{\rm A} = 1.74 \pm 0.05$}\Rsun). By contrast, all nine previously published measurements of $R_{\rm A}$ were in the range $1.10$--$1.36$\Rsun. The same situation is found for the planet (\wwo{$M_{\rm b} = 1.41 \pm 0.17 \pm 0.05$}{$M_{\rm b} = 1.41 \pm 0.18$}\Mjup), for which the new radius measurement is \wwo{$R_{\rm b} = 1.734 \pm 0.051 \pm 0.029$}{$R_{\rm b} = 1.73 \pm 0.06$}\Rjup\ whereas all previous values are in the range $1.23$--$1.378$\Rjup. This is illustrated in Fig.\,\ref{fig:ogle56:M2R2}. OGLE-TR-56\,b therefore joins the group of the most inflated TEPs. This is unsurprising insofar as its host star is relatively hot and massive. A detailed discussion of the most inflated TEPs is given by  \citet{Me+12c}.

How did such a large change in the measured properties occur? Early analyses of the system \citep{Konacki+03nat,Torres+04apj,Santos+06aa,Bouchy+05aa2} all relied on on the OGLE discovery light curve \citep{Udalski+02aca3} which is of low quality compared to dedicated follow-up observations. Later analyses (\citealt{Pont+07aa,Torres++08apj}; Paper\,II; Paper\,III) rested on the VLT light curve from \citet{Pont+07aa}, which has been shown by \citet{Adams+11apj2} to suffer from substantial systematic noise. It is only with the data gathered by \citet{Adams+11apj2} that it is now possible to derive definitive physical properties of the system.

\citet{Adams+11apj2} themselves studied their light curves in detail, but adopted old values of $M_{\rm A}$ and $R_{\rm A}$ from Paper\,III for calculating the system properties. They therefore did not propagate the stellar density found from their light curves to the physical properties of the system, and so ended up with properties similar those found in previous works. It is only when the newer photometric data are used to specify the stellar density, and therefore its mass and radius in conjunction with theoretical models, that the system parameters experience a significant change. The physical properties of OGLE-TR-56 found in the current work are the first ones to provide a full picture of the system.

One moral from this story is that a single good light curve is not enough to provide a definitive set of physical properties, as undetected sysmatic errors may bias the results. A significant fraction of other TEPs might be subject to this concern, for example HD\,80606 \citep{Hebrard+10aa} and WASP-7 \citep{Me+11aa}. A second moral is that any significant revision to the photometric parameters should be propagated into the physical properties of the system.


\subsection{OGLE-TR-111}                                                                                                     \label{sec:teps:ogle111}

\begin{figure} \smfig{\includegraphics[width=\columnwidth,angle=0]{plot-ogle111.eps}}{\includegraphics[width=\columnwidth,angle=0]{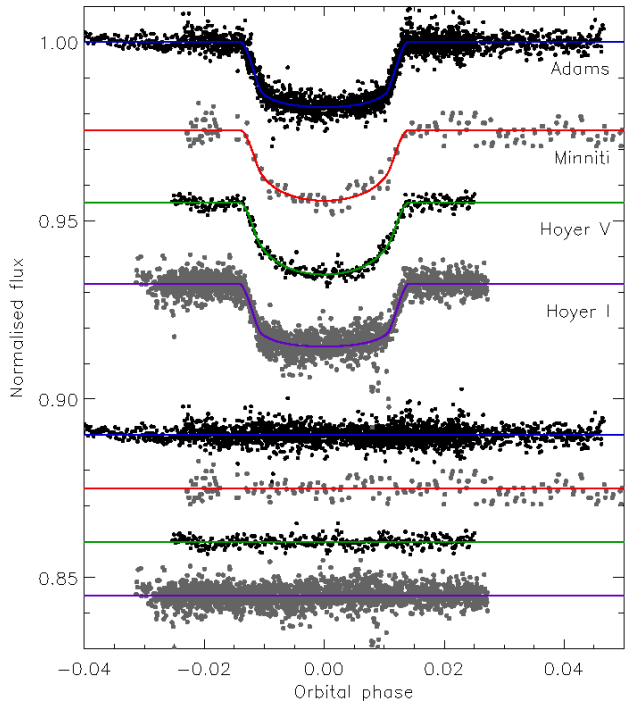}}
\caption{\label{fig:ogle111:lc} The Magellan \citep{Adams+10apj} and VLT
\citep{Minniti+07apj,Hoyer+11apj} light curves of OGLE-TR-111 compared
to the {\sc jktebop} best fits. Other comments are the same as
Fig.\,\ref{fig:hatp3:lc}.} \end{figure}

The transiting nature of OGLE-TR-111 was announced by \citet{Udalski+02aca2} and its planetary nature confirmed by \citet{Pont+04aa}. Its publication history was summarised in Paper\,I (sect.\,3.9), where the data from \citet{Winn++07aj} were analysed. Three new sets of transit light curves have been published since then, prompting \reff{this} rediscussion of the system. \citet{Adams+10apj} observed six transits through an SDSS $i$ filter, using Magellan/MagIC-e2v high-resolution imager, allowing them to improve the measurement of $R_{\rm b}$ and dismiss the existence of the TTVs tentatively found by \citet{Diaz+08apj}. \citet{Hoyer+11apj} studied five transits, one with VLT/FORS1 and a ``$v$-{\it HIGH}'' filter and the remaining four with VLT/FORS2 and a Bessell $I$ filter. I have also included the single $V$-band light curve from \citet{Minniti+07apj} in my analysis.

The six transits observed by \citet{Adams+10apj} are contained in 2205 datapoints, for which I found that correlated noise was moderately important. I fitted for \Porb, and the light curve solutions are given in \apptab.
\citet{Minniti+07apj} obtained 213 datapoints covering one transit in $V$, for which I found no evidence of correlated noise (\apptab).
The data from \citet{Hoyer+11apj} were supplied with timestamps in JD, which I converted to BJD(TDB) before analysis. The $V$-band transit has 323 datapoints and the four $I$-band transits comprise 1967 datapoints. I found the RP errors to be larger than the MC errors for both datasets (\apptabb).

The light curve solutions are summarised in \apptab\ and the best fits are shown in Fig.\,\ref{fig:ogle111:lc}. The agreement between the light curves is good, except for $k$ which has $\chir = 1.9$. This situation has occured frequently within this series of papers. Published photometric properties agree well with my own. A similar story occurs for the physical properties (\apptab), although I find a comparatively low concordance between the results for different stellar models. OGLE-TR-111 would benefit from an improved measurement of $K_{\rm A}$, although its faintness ($V=16.96$) makes this non-trivial.


\subsection{OGLE-TR-113}                                                                                                     \label{sec:teps:ogle113}

\begin{figure} \smfig{\includegraphics[width=\columnwidth,angle=0]{plot-ogle113.eps}}{\includegraphics[width=\columnwidth,angle=0]{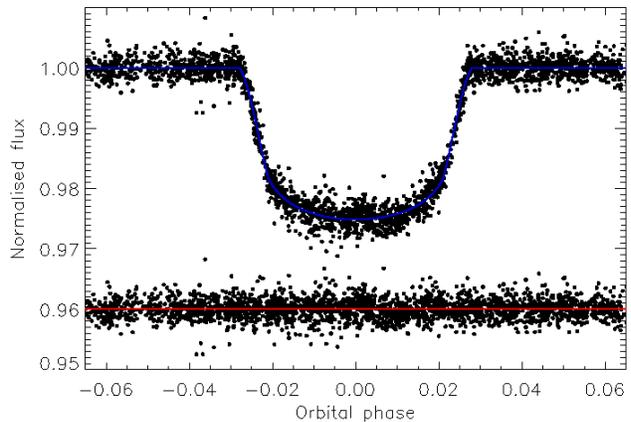}}
\caption{\label{fig:ogle113:lc} The Magellan light curve of OGLE-TR-113
\citep{Adams+10apj2} compared to the {\sc jktebop} best fit.
Other comments are the same as Fig.\,\ref{fig:hatp3:lc}.} \end{figure}

OGLE-TR-113 was also found to be a transiting system by \citet{Udalski+02aca2}, and a planetary system by \citet{Bouchy+04aa}. It was studied in Paper\,III (sect.\,4.3 and fig.\,7), based on the photometry presented by \citet{Gillon+06aa}, \citet{SnellenCovino07mn} and \citet{Diaz+07apj}. Since then \citet{Adams+10apj2} have published six nice transit light curves from Magellan/MagIC, allowing them to revise $R_{\rm b}$ and set limits on TTVs. They found a possible gradual decrease in \Porb, with a significance of $4\sigma$, which should be investigated once a longer temporal baseline is available.

I fitted the data from \citet{Adams+10apj2}, which represents 2697 datapoints, including \Porb\ as a fitted parameter. My initial model was compromised by some poor data which were included in the datafile but not in the analysis by \citet{Adams+10apj2}. 79 datapoints were rejected, allowing a good fit to be found to the remaining data. I found correlated noise to be slightly important. The LD-fit/fix solution was adopted (\apptab\ and Fig.\,\ref{fig:ogle113:lc}).

The new data from \citet{Adams+10apj2} allow a great improvement in the photometric parameters (\apptab) and in the physical properties (\apptab). I found a good agreement between results from different model sets, and against values from the literature. As with OGLE-TR-111, a more precise $K_{\rm A}$ would be useful but difficult to achieve because the system is faint ($V = 16.48$).


\subsection{OGLE-TR-132}                                                                                                     \label{sec:teps:ogle132}

\begin{figure} \smfig{\includegraphics[width=\columnwidth,angle=0]{plot-ogle132.eps}}{\includegraphics[width=\columnwidth,angle=0]{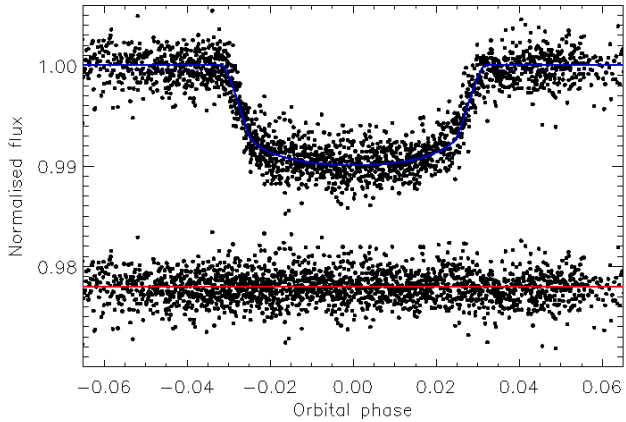}}
\caption{\label{fig:ogle132:lc} The Magellan light curves of OGLE-TR-132 from
\citet{Adams+11apj}, compared to the {\sc jktebop} best fit. Other comments
are the same as Fig.\,\ref{fig:hatp3:lc}.} \end{figure}

\citet{Udalski+03aca} observed transits in this system, and \citet{Bouchy+04aa} confirmed its planetary nature. OGLE-TR-132 was studies in Paper\,I (sect.\,3.10 and fig.\,12) using a nice VLT/FORS2 light curve from \citet{Gillon+07aa3}. I have revisited the system because \citet{Adams+11apj} has published a light curve of 4940 datapoints covering seven transits, using them to measure a revised $R_{\rm b}$ and put upper limits on the presence of TTVs. Two of these transits were observed at a much higher cadence ($10$--$15$\,s) versus the rest ($35$--$130$\,s), so these were time-binned by a factor of five. This is an important step if the data are modelled together and subjected to an RP algorithm.

The resulting 2131 datapoints were analysed, with the findings that modest correlated noise is present and that the LD-fit/fix solution is good (\apptab\ and Fig.\,\ref{fig:ogle132:lc}). Final photometric parameters were calculated from these results and those in Paper\,I. The agreement between these results and also with published values is excellent (\apptab). The ensuing physical properties (\apptab) are an improvement over Paper\,I due to the additional data, but are in good agreement with literature results. OGLE-TR-132 needs more RVs if its measured physical properties are going to be improved significantly.


\subsection{OGLE-TR-L9}                                                                                                       \label{sec:teps:oglel9}

\begin{figure} \smfig{\includegraphics[width=\columnwidth,angle=0]{plot-oglel9.eps}}{\includegraphics[width=\columnwidth,angle=0]{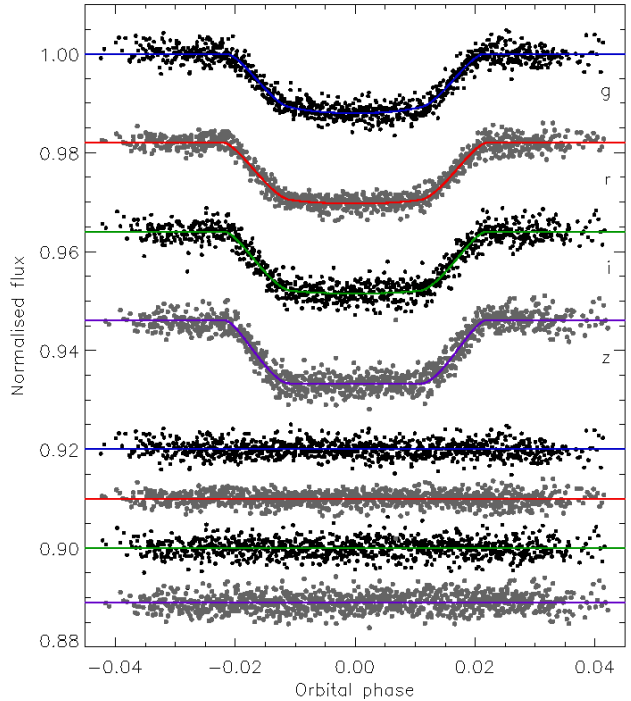}}
\caption{\label{fig:oglel9:lc} The GROND light curves of OGLE-TR-L9 from
\citet{Lendl+10aa} compared to the {\sc jktebop} best fits. Other comments
are the same as Fig.\,\ref{fig:hatp3:lc}.} \end{figure}

This was discovered by \citet{Snellen+09aa} based on archival OGLE data, follow-up four-band photometry from the GROND imager, and UVES spectroscopy. It is of interest as containing a massive planet ($4.4 \pm 1.5$\Mjup) around an unusually hot star ($6933 \pm 58$\,K), and was duly studied in Paper\,III. \citet{Lendl+10aa} have recently obtained five more transits with GROND\footnote{The datafiles as lodged on CDS are incorrect but the data are available on request to M.\ Lendl.} I have modelled the combined dataset of six transits from \citet{Snellen+09aa} and \citet{Lendl+10aa} as they were obtained with the same instrument and observing strategy. The numbers of datapoints is as follows: 922 in $g$, 921 in $r$, 917 in $i$ and 922 in $z$. In all cases the RP errors were smaller than the MC errors. The full sets of solutions can be found in \apptabbb{4}, the summary in \apptab, and the best fits in Fig.\,\ref{fig:oglel9:lc}. The final photometric parameters agree well with those found by \citet{Lendl+10aa}, as expected given the identical datasets used, but not so well with those from \citet{Snellen+09aa} or Paper\,III (both of which used the same smaller dataset).

\citet{Lendl+10aa} did not derive the full physical properties of OGLE-TR-L9, so \apptab\ contains the first physical properties measured using their extensive photometry. The parameters are consistent with previous solutions, but with a useful lowering of the errorbars due to the new light curves. An improved $K_{\rm A}$ and \FeH\ are needed; the light curves are unusually not the observational bottleneck because these spectroscopic constraints are so poor (mainly due to the faintness of the system at $V \approx 15.5$).


\subsection{TrES-4}                                                                                                            \label{sec:teps:tres4}

\begin{figure} \smfig{\includegraphics[width=\columnwidth,angle=0]{plot-tres4.eps}}{\includegraphics[width=\columnwidth,angle=0]{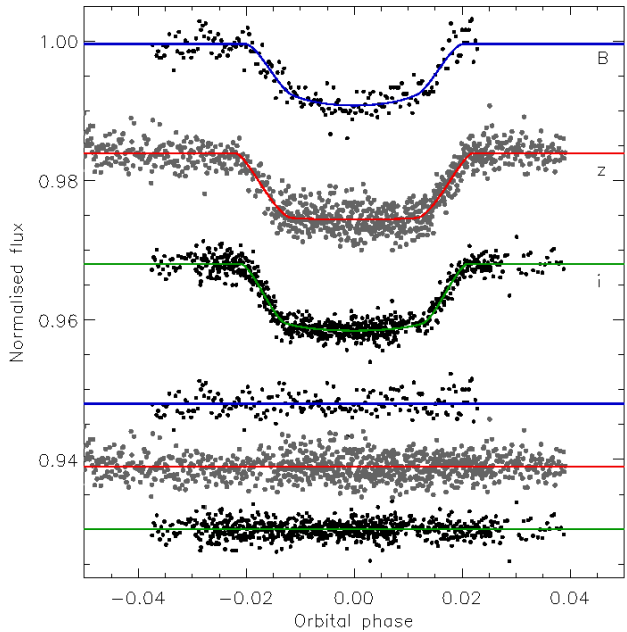}}
\caption{\label{fig:tres4:lc} The $B$ and $z$ \citep{Mandushev+07apj} and $i$
light curves \citep{Chan+11aj} of TrES-4 compared to the {\sc jktebop} best fits.
Other comments are the same as Fig.\,\ref{fig:hatp3:lc}.} \end{figure}

This system, containing a large and low-density planet, was studied in Paper\,III based on the photometry presented in the discovery paper \citep{Mandushev+07apj}. New data are now available from \citet{Chan+11aj}, covering five transits in the $i$ band obtained using KeplerCam. Here I present revised results incorporating these new data. A notable result since Paper\,III is the measurement of the Rossiter-McLaughlin effect by \citet{Narita+10pasj}, showing the system to be aligned to within 1.5$\sigma$.

TrES-4 is straightforward except for the presence of a much fainter star separated by only $1.5\as$ on the sky. This was treated as third light (Paper\,III). The magnitude difference measured in the $i$ band by \citet{Daemgen+09aa}, $\Delta i = 4.560 \pm 0.017$, was converted into a flux ratio to obtain the constraint on $L_3$.

The data from \citet{Chan+11aj} cover five transits, two of them fully, with 765 datapoints. I fitted for \Porb, and found that correlated noise was unimportant and that the LD-fit/fix solutions were good (\apptab). These results were added to the $B$- and $z$-band \citep{Mandushev+07apj} results from Paper\,III, and a comparatively poor agreement was found. The most concordant parameter was, unusually, $k$ ($\chir = 0.4$), whereas the other photometric parameters have $\chir = 1.1$--$1.8$. The new $i$-band data yield results roughly in the middle of those from the $B$ and $z$ data, which are formally discrepant. The final photometric parameters are the weighted average of the individual ones from the $B$, $i$ and $z$ data, with the errorbars inflated to enforce $\chir = 1$ (\apptab). This leads to a significant difference with respect to Paper\,III, where the $B$ data were rejected and the $z$-band results taken as final, but is in reasonable agreement with other studies. The best fits to the three light curves (i.e.\ those studied in Paper\,III as well as the current work) are plotted in Fig.\,\ref{fig:tres4:lc}.

When calculating the physical properties of TrES-4 I found that the different model sets did not agree completely. The {\it Y$^2$} models are a strong outlier, but are also the closest to the spectroscopically-measured stellar \Teff. These results have nevertheless been rejected, and the final values based on those from the other four sets of models (\apptab). They point to the planet having an unexpectedly large radius (\wwo{$R_{\rm b} = 1.735 \pm 0.072 \pm 0.007$}{$R_{\rm b} = 1.74 \pm 0.07$}\Rjup), confirming the conclusions from previous studies. I recommend that new transit photometry and \'echelle spectra are obtained for TrES-4 in order to provide a definitive answer to the question of its physical properties.


\subsection{WASP-12}                                                                                                          \label{sec:teps:wasp12}

\begin{figure} \smfig{\includegraphics[width=\columnwidth,angle=0]{plot-wasp12.eps}}{\includegraphics[width=\columnwidth,angle=0]{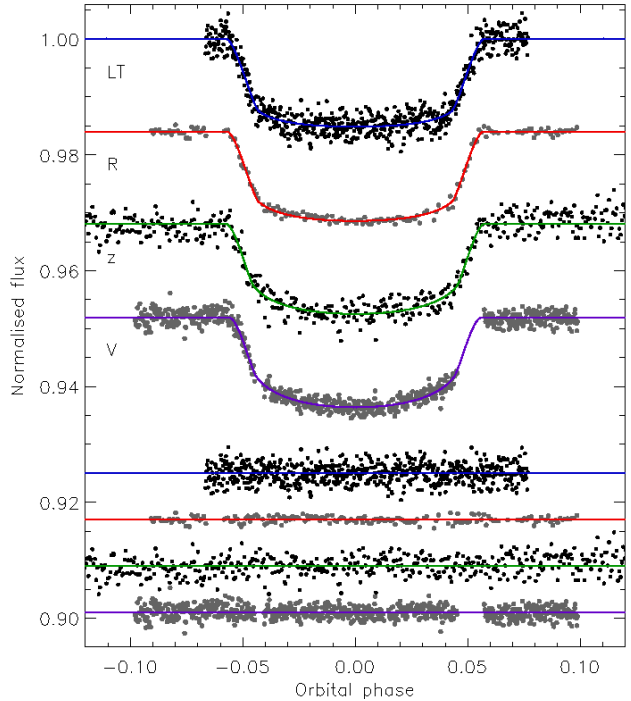}}
\caption{\label{fig:wasp12:lc} Light curve of WASP-12 compared to the {\sc jktebop} best
fits. From top to bottom the datasets are from \citet{Hebb+09apj}, \citet{Maciejewski+11aa},
and two from \citet{Chan+11aj}. Other comments are the same as Fig.\,\ref{fig:hatp3:lc}.}
\end{figure}

At the time of its discovery, WASP-12 contained the hottest, most irradiated and shortest-period planet then known \citep{Hebb+09apj}, making it a key object in our understanding of gas-giant planets. Occultations of the planet by the star have been observed in the $z$ band \citep{Lopez+10apj}, in the $J$, $H$ and $K_s$ bands \citep{Croll+11aj}, and in the 3.6, 4.5, 5.8 and 8.0 $\mu$m {\it Spitzer}/IRAC bands \citep{Campo+11apj}. \citet{Madhusudhan+11nat} interpreted the occultation depths in these bands as evidence of an enhanced ratio of carbon to oxygen in the atmosphere of the planet. Most recently, \citet{Cowan+12apj} used \spitzer\ to obtain photometry at 3.6 and 4.5\,$\mu$m covering full orbits of WASP-1, \reff{and \citet{Swain+12xxx} have measured transit and occultation depths in the infrared using HST/WFC3}. High-precision optical transit photometry has been obtained by \citet{Maciejewski+11aa} and \citet{Chan+11aj} and used to confirm the large planetary radius measurement obtained by \citet{Hebb+09apj}.

The possibility of orbital eccentricity in the WASP-12 system has received much attention. \citet{Hebb+09apj} found a value of $e = 0.049 \pm 0.015$, which was unexpected because tidal effects should have circularised such a short-period orbit \citep{GoldreichSoter66icar}. The probability that this eccentricity is real is 99.5\% according to the test of \citet{LucySweeney71aj}. The occultation timing found by \citet{Lopez+10apj} leads to the constraint $e\cos\omega = 0.0156 \pm 0.0035$, which agrees with the eccentricity detection of \citet{Hebb+09apj}. However, the later occultation observations \citep{Croll+11aj,Campo+11apj} define the eclipse shape much more precisely and constrain $e$ to be very small and probably zero. Further support for a circular orbit comes from continued RV monitoring \citep{Husnoo+11mn}, and from consideration of the effect of tidal deformation of the host star on its measured RV \citep{Arras+12mn}. I have therefore assumed a circular orbit in the analysis below.

\citet{Hebb+09apj} presented photometry of two transits, a $z$-band one from LT/RATCam and a $B$-band one from the 0.8\,m Tenagra telescope in Arizona. The former of the two datasets is of much better quality than the latter, and was included in my analysis. \citet{Maciejewski+11aa} obtained photometry of two transits in the Johnson $R$ band with the CAHA 2.2m/CAFOS. The second dataset suffers from greater scatter, gaps and very little data outside transit, so I considered only the observations of the first transit. \citet{Chan+11aj} suffered from a lot of bad weather, but were able to get good observations of one transit in $i$ using KeplerCam\footnote{The observing log in \citet{Chan+11aj} specifies a $z$-band filter for this dataset but the datafile from ApJ labels it as $i$. J.\ Winn (2011, private communication) \reff{has confirmed} that the latter is correct.} and one transit in $V$ using NOT/ALFOSC. Both datasets were added to my analysis.

One problem is the asphericity of the planet, which is technically outside the range of validity of the {\sc ebop} model (models assuming spherical planets will be more strongly affected). This was discussed in the context of OGLE-TR-56 in Sect.\,\ref{sec:teps:ogle56}; the corollary for WASP-12 is that the volume-equivalent radius is up to 5\% larger than the measured transit radius.

For all datasets I found that the MC errors were bigger than the RP errors and that the LD-fit/fix solutions were good. Results for the $z$-band data (613 points) are given in \apptab; for the $R$-band data (167 points) in \apptab; and for the $i$ and $V$ band data (470 and 670, respectively) in \apptab\ and \apptab. The $R$-band data from \citet{Maciejewski+11aa} are of very high quality, with a scatter of only 0.59\,mmag per point, and for these it was possible to adopt the LD-fitted solution. The best fits are plotted in Fig.\,\ref{fig:wasp12:lc}.

\apptab\ shows that the agreement between the four light curves is poor for $k$ ($\chir = 2.5$) but fine for the other photometric parameters. Literature values are in good agreement with my final results, except for those of \citet{Chan+11aj}. These authors find a lower $k$ than I achieve for any of the four datasets considered.

The physical properties from the five different model sets display a comparatively poor agreement. The {\it VRSS} models are the most discrepant, but not by enough to entertain rejecting them. My final results (\apptab) therefore have a significant systematic error component, which however is still dwarfed by the statistical uncertainties. WASP-12 would really benefit from more precise measurements of the stellar \Teff\ and \FeH; the mass of the star is uncertain by 14\% ($M_{\rm A} = 1.38 \pm 0.19$\Msun) primarily due to the uncertainty in the star's atmospheric parameters. In comparison to literature results, I find \wwo{$R_{\rm b} = 1.825 \pm 0.091 \pm 0.024$}{$R_{\rm b} = 1.83 \pm 0.09$}\Rjup, which is $1\sigma$ smaller than that from \citet{Maciejewski+11aa} and $1\sigma$ larger than that from \citet{Chan+11aj}. WASP-12\,b is therefore confirmed to have an unusually large radius.


\subsection{WASP-13}                                                                                                          \label{sec:teps:wasp13}

\begin{figure} \smfig{\includegraphics[width=\columnwidth,angle=0]{plot-wasp13.eps}}{\includegraphics[width=\columnwidth,angle=0]{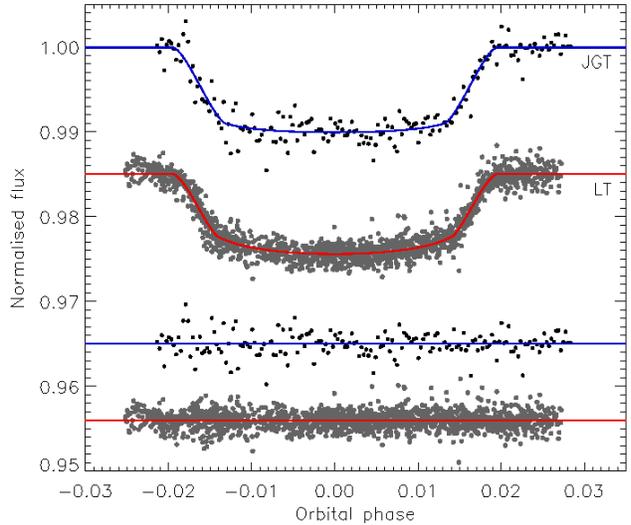}}
\caption{\label{fig:wasp13:lc} The JGT (top) and LT (lower) light curves of
WASP-13 compared to the {\sc jktebop} best fits. Other comments are the same
as Fig.\,\ref{fig:hatp3:lc}.} \end{figure}

\citet{Skillen+09aa} presented the discovery of \reff{the planetary nature of} this system, with an analysis including observation of one full transit from the 0.95\,m James Gregory Telescope at St.\ Andrews, UK. \citet{Barros+12mn} subsequently observed four transits using LT/RISE, although none in their entirety. I have analysed both datasets here, with \Porb\ incuded as a fitted parameter for the LT data (1530 datapoints) but not for the single transit from the JGT (173 datapoints). In both cases I found that correlated noise was significant: the RP errors are greater than the MC errors by a factor of 1.5 for the JGT data and by a factor of 2 for the LT data. The full solutions can be found in \apptab\ and \apptab.

The photometric solutions are summarised in \apptab\ and plotted in Fig.\,\ref{fig:wasp13:lc}. The agreement between the two datasets is good except for $k$, as usual. The parameters found by \citet{Skillen+09aa} disagree with my own, by roughly $2\sigma$ depending on parameter. The reasons for this are not clear, given than I found the JGT and LT data to be consistent with each other, but may be related to inclusion of the SuperWASP data in their analysis. \citet{Barros+12mn} found photometric parameters which agree with my own, but with errorbars a factor of 2 smaller.

The {\it DSEP} models returned results in some disagreement with the other model sets, so were discluded when calculating the final physical properties. The extant \Teff\ and \FeH\ measurements for WASP-13\,A are comparatively poor, and contribute to the uncertainties in the properties of this system. I find a larger and more massive star and planet, the change in each case being approximately the same size as the errorbars (\apptab). My revised system parameters makes WASP-13\,b a rather large (\wwo{$R_{\rm b} = 1.528 \pm 0.084 \pm 0.004$}{$R_{\rm b} = 1.53 \pm 0.08$}\Rjup) and low-density ($\rho_{\rm b} = 0.134 \pm 0.023$\pjup) TEP. The corresponding values found by \citet{Skillen+09aa} and \citet{Barros+12mn} are for a smaller and less massive star and planet. WASP-13 deserves further photometry and spectroscopy.


\subsection{WASP-14}                                                                                                          \label{sec:teps:wasp14}

\begin{figure} \includegraphics[width=\columnwidth,angle=0]{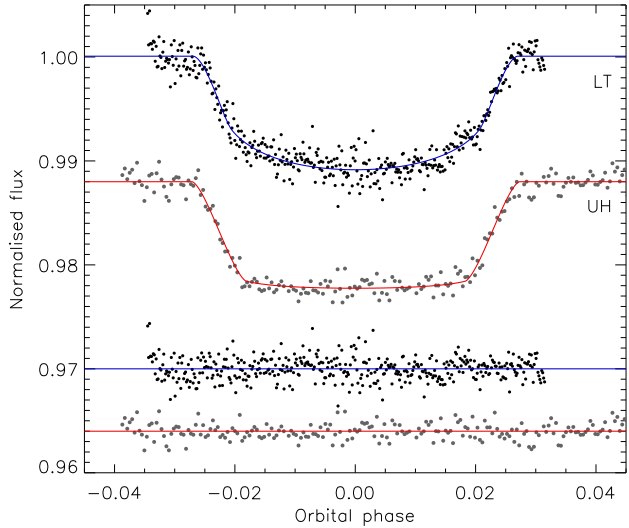}
\caption{\label{fig:wasp14:lc} Light curves of WASP-14 from \citet{Joshi+09mn}
and \citet{Johnson+09pasp} compared to the {\sc jktebop} best fits. Other comments
are the same as in Fig.\,\ref{fig:hatp3:lc}.} \end{figure}

Discovered by \citet{Joshi+09mn}, this planet is massive ($7.9 \pm 0.5$\Mjup) and has a definitely eccentric orbit ($e = 0.087 \pm 0.002$). These two properties are known to preferentially coincide \citep{Me+09apj}. \citet{Johnson+09pasp} has found that the planetary orbit is misaligned with the stellar spin ($\lambda = -33.1 \pm 7.4$ degrees) from Rossiter-McLaughlin observations. \citet{Blecic+11xxx} obtained Spitzer observations of three occutations in order to probe the atmospheric properties of the planet from its infrared flux distribution.

The orbital eccentricity was detected by \citet{Joshi+09mn} and has been confirmed by \citet{Husnoo+11mn} from continued RV monitoring. The most precise value comes from \citet{Blecic+11xxx}, whose three occultations allow a good constraint on $e\cos\omega$. I have adopted the measured $e\cos\omega = -0.02557 \pm 0.00038$ and $e\sin\omega = 0.0831 \pm 0.0021$ as constraints on the {\sc jktebop} models.

\citet{Joshi+09mn} presented an LT/RISE light curve of one transit containing 419 datapoints. The LD-fit/fix solutions are reasonable, and LD-fitted ones were not attempted (\apptab). Correlated noise occurs, causing RP errors to be 1.5 times greater than the MC errors.

\citet{Johnson+09pasp} observed one transit (247 datapoints) using the UH 2.2m and OPTIC CCD, in a non-standard filter centred on 850\,nm and 40\,nm wide. They used this to constrain the orbital ephemeris to aid their Rossiter-McLaughlin measurement, but did not attempt a revision of the other parameters of the system. I adopted LD coefficients for the Cousins $I$ band as this is reasonably close to the actual filter used (given that the predicted LD coefficients are significantly scattered). I found that the MC errors were larger than the RP ones, and that the LD-fit/fix solution is good (\apptab).

The light curve solutions are collected in \apptab\ and the best fits are shown in Fig.\,\ref{fig:wasp14:lc}. The agreement between the two light curves is unimpressive, with $k$ having $\chir = 0.8$ and the other photometric parameters having $\chir = 1.5$--$1.9$. There is no specific reason to prefer one light curve over the other, so two results for the two have been combined and the errorbars inflated to account for the modest discrepancy. WASP-14 would clearly benefit from further photometric observations. \citet{Joshi+09mn} obtained results in agreement with those I find from their data, but in less good agreement with my final photometric parameters; no results are available from the other works cited above.

It is no surprise that I find a significantly different set of physical properties of WASP-14 compared to previous works (\apptab). The star becomes more massive and more evolved ($M_{\rm A} = 1.35 \pm 0.12$\Msun\ and $R_{\rm A} = 1.67 \pm 0.10$\Rsun). The planet's size is similarly revised, to \wwo{$R_{\rm b} = 1.633 \pm 0.092 \pm 0.009$}{$R_{\rm b} = 1.63 \pm 0.09$}\Rjup), an increase of $4\sigma$ over the value of \er{1.281}{0.075}{0.082}\Rjup\ found by \citet{Joshi+09mn}. This new $R_{\rm b}$ is in clear disagreement with standard theoretical models of planets \citep[e.g.][]{Bodenheimer++03apj,Fortney++07apj,Chabrier+09conf}. New light curves are vital in sorting out the moderate discrepancy between the two existing ones, and new stellar atmospheric parameters would allow a great improvement in our understanding of WASP-14.


\subsection{WASP-21}                                                                                                          \label{sec:teps:wasp21}

\begin{figure} \includegraphics[width=\columnwidth,angle=0]{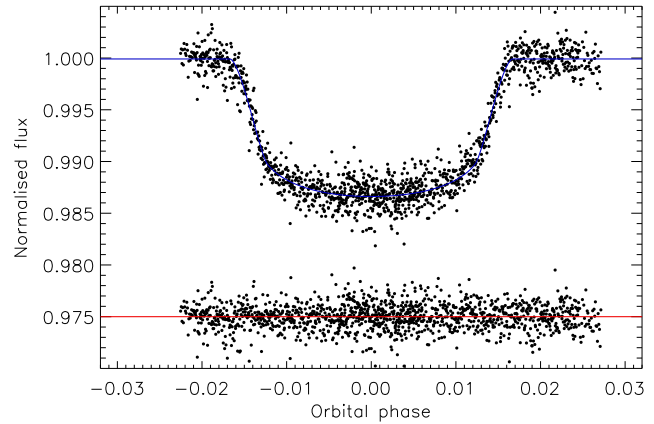}
\caption{\label{fig:wasp21:lc} The LT light curve of WASP-21 compared to the
{\sc jktebop} best fit. Other comments are the same as Fig.\,\ref{fig:hatp3:lc}.}
\end{figure}

\citet{Bouchy+10aa} discovered this planetary system, and obtained its physical properties using their version of the dEB constraint. They obtained adaptive-optics imaging which show no faint stars contaminating the flux from the system. The interest in WASP-21 lies in its Saturn-mass planet ($M_{\rm b} = 0.295 \pm 0.030$\Mjup) and the metallicity which is the lowest known for a TEP host star ($\FeH = -0.46 \pm 0.11$). \citet{Barros+11mn} analysed LT/RISE observations covering three transits, of which the first was originally presented in the discovery paper \citep{Bouchy+10aa}. They used the {\it Y$^2$} models to provide their additional constraint, yielding significant differences compared to those found by \citet{Bouchy+10aa}: $M_{\rm A} = 0.86 \pm 0.04$\Msun\ versus $1.01 \pm 0.03$\Msun, and $M_{\rm b} = 0.27 \pm 0.01$\Msun\ versus $0.300 \pm 0.011$\Msun. WASP-21 is therefore an obvious candidate for inclusion in the {\it Homogeneous Studies} project.

The LT data contain 1302 points over three transits (one each in 2008, 2009 and 2010), of which only one has full coverage. These were modelled with \Porb\ included as a fitted parameter. As with WASP-13, I found the RP errors to be 1.5 times larger than the MC errors, and that the LD-fit/fix solution was the best. The full set of solutions is in \apptab\ and the best fit is plotted in Fig.\,\ref{fig:wasp21:lc}. I find photometric parameters in reasonable agreement with those of \citet{Barros+11mn} but with larger errorbars (\apptab). The parameters from \citet{Bouchy+10aa} are discrepant, as has been discussed by \citet{Barros+11mn}.

The physical properties arising from the five model sets are not in great agreement and this is straightforwardly attributable to the low \FeH, for which the model predictions diverge. I find stellar and planetary masses closer to those from \citet{Bouchy+10aa} than \citet{Barros+11mn}, but stellar and planetary radii which are bigger than found by either study (\apptab). 
The planet is now one of the least dense known (\wwo{$\rho_{\rm b} = 0.137 \pm 0.021 \pm 0.003$}{$\rho_{\rm b} = 0.137 \pm 0.021$}\pjup). A new spectroscopic study of WASP-21 should be given priority, to confirm the low \FeH\ and increase the precision of the \Teff\ and \FeH\ measurements. A new light curve would be useful, as the current one dominates the uncertainty in $R_{\rm A}$ and $R_{\rm b}$.

%
%
%
%
%
%
%
%
%
%

\subsection{TEPs without new photometric studies: \corot-2, WASP-2, WASP-2, WASP-4, WASP-5, WASP-7, WASP-18 and XO-2}          \label{sec:teps:other}

{\bf \corot-2} was studied in Paper\,IV but has since been the subject of a high-resolution spectroscopic study by \citet{Schroter+11aa}. They find improved atmospheric parameters for the star ($\Teff = 5598 \pm 34$\,K and $\FeH = 0.04 \pm 0.02$). For comparison, the values used in Paper\,IV were $5696 \pm 70$\,K and $0.03 \pm 0.06$, respectively. The errorbars on the new measurements are rounded up to $\pm$$50$\,K and $\pm$$0.05$\,dex for the reasons given in Paper\,II. The new \Teff\ yields a slightly smaller system scale (\apptab), which means lower masses and radii for both components. The {\it Claret} models disagree slightly with the others, resulting in comparatively large systematic errorbars for some calculated quantities.

{\bf WASP-1} is in receipt of new results from \citet{Albrecht+11apj2}, who revised $K_{\rm A}$ to $125 \pm 5$\ms\ versus $111 \pm 9$\ms, \Teff\ to $6213 \pm 51$\,K versus $6110 \pm 50$\,K, and \FeH\ to $0.17 \pm 0.05$ versus $0.23 \pm 0.08$. It is slightly unusual that an upward revision in \Teff\ brings about a downward revision in \FeH, as the two are positively correlated in the spectral analysis of F, G and K stars. The change in \Teff\ is also more than the errorbars, so a confirmation of this result would be appropriate. \apptab\ shows that the new measurements modify the physical properties of WASP-1 only within the errorbars compared to Paper\,III. WASP-1\,b's status as a hot and bloated planet is secure.

{\bf WASP-2} was studied by \citet{Me+10mn}. New spectroscopic results (covering both RVs and atmospheric parameters) are now available from \citet{Albrecht+11apj2}. It is not straightforward to deduce the star's \Teff\ as multiple values available in the literature do not agree: \citet{Cameron+07mn} found $5200 \pm 200$\,K; \citet{Ghezzi+10apj} obtained $5227 \pm 46$\,K; \citet{Triaud+10aa} found $5150 \pm 80$\,K; \citet{Albrecht+11apj2} obtained $5206 \pm 50$\,K; and \citet{Maxted++11mn} deduced $5110 \pm 60$\,K. I have therefore adopted $5170 \pm 60$\,K as a value which encompasses all these determinations and leans towards the slightly lower \Teff\ from \citet{Maxted++11mn}. Their value was obtained using the InfraRed Flux Method \citep[IRFM:][]{BlackwellShallis77mn,Blackwell++80aa}, which I consider to be the most reliable method due to its emphasis on the stellar spectral energy distribution (which is conceptually nicer for \Teff\ determinations) and its limited dependence on theoretical calculations. \apptab\ shows that the agreement between models is not great, as the star is substantially less massive than the Sun, but the overall results are a big step forward from previous measurements. This is due to the more precise \Teff\ and \FeH, and means that WASP-2 is now one of the best-understood TEP systems.

{\bf WASP-4} gets a new \Teff\ of $5540 \pm 55$\,K from \citet{Maxted++11mn}, which is a substantial improvement on the $5500 \pm 100$\,K \citep{Gillon+09aa} used in Paper\,III and \citet{Me+09mn2}. The agreement between models is imperfect with respect to the system properties, but the final results achieve a high precision and are in agreement with literature values (\apptab). WASP-4 is one of the best-understood TEP systems. The most effective improvement would be an increase in the precision of \FeH, but this is not a high priority.

{\bf WASP-5} has received a new \Teff\ measurement of $5770 \pm 65$\,K from \citet{Maxted++11mn} which is a useful improvement on the $5700 \pm 100$\,K \citep{Gillon+09aa} adopted in Paper\,III and by \citet{Me+09mn}. The ensuing physical properties (\apptab) show a modest agreement between models but a good agreement overall with literature values. WASP-5 could do with a better light curve and \FeH\ determination. New photometric data of WASP-5 have been obtained by several researchers -- some is already published -- and this will help in the near future.

{\bf WASP-7} has a revised \Teff\ of $6250 \pm 70$\,K from \citet{Maxted++11mn}, compared to the $6400 \pm 100$\,K \citep{Hellier+09apj} used in \citet{Me+11aa}. Note that WASP-7 was recently found to have an almost polar orbit through observations of the Rossiter-McLaughlin effect \citep{Albrecht+12apj}, but this study did not generate a new $K_{\rm A}$. \apptab\ shows that there is little change in the system parameters with the new \Teff. WASP-7 could still do with a better light curve and \FeH\ measurement.

{\bf WASP-18} gets a new \Teff\ of $6455 \pm 70$\,K from \citet{Maxted++11mn}, which is more precise than the $6400 \pm 100$\,K \citep{Hellier+09nat} adopted in Paper\,III. \apptab\ shows that there is not much difference in the physical properties, but reinforces the suggestion that the system is young as a zero age is the preferred solution for three of the five models. The errorbars may be artifically small due to the concomitant edge effects. WASP-18 would benefit from a new spectral-synthesis study as well as additional light curve data.

{\bf XO-2} has been awarded a revised $K_{\rm A}$ by \citet{Narita+12pasj}, who found $92.2 \pm 1.7$\ms\ compared to the $85 \pm 8$\ms\ \citep{Burke+07apj} used in Paper\,III. XO-2\,A remains the TEP host star with the highest \FeH\ at $+0.45 \pm 0.05$\,dex. The models continue to disagree significantly (\apptab), due to the high \FeH, and in such a way that is not accounted for by most analyses of this system. The new results are very similar to the previous ones from Paper\,III.


\section{Physical properties of the transiting extrasolar planetary systems}                                                   \label{sec:properties}

\begin{table*} \caption{\label{tab:absdim:stars} Derived physical properties of the stellar components
of the TEPs studied in this work. For most quantities the first errorbar gives the statistical errors
and the second errorbar gives the systematic errors arising from the dependence on stellar theory.}
\setlength{\tabcolsep}{3pt}
\begin{tabular}{l l@{\,$\pm$\,}l@{\,$\pm$\,}l l@{\,$\pm$\,}l@{\,$\pm$\,}l l@{\,$\pm$\,}l@{\,$\pm$\,}l l@{\,$\pm$\,}l@{\,$\pm$\,}l l@{\,$\pm$\,}l@{\,$\pm$\,}l l@{\,$\pm$\,}l@{\,$\pm$\,}l}
\hline \hline
System & \mcc{Semimajor axis (AU)} & \mcc{Mass (\Msun)} & \mcc{Radius (\Rsun)} & \mcc{$\log g_{\rm A}$ [cm/s]} & \mcc{Density (\psun)} & \mcc{Age (Gyr)} \\
\hline
\corot-2    & 0.02835   & 0.00029   & 0.00037     & 0.997     & 0.030     & 0.039       & 0.901     & 0.015     & 0.012       & 4.527     & 0.016     & 0.006       & \mcc{$1.362 \pm 0.064$}            & \ermcc{1.2}{1.8}{1.7}{0.7}{1.2}    \\
\corot-17   & \ermcc{0.04810}{0.00109}{0.00064}{0.00023}{0.00034}  & \ermcc{1.043}{0.073}{0.041}{0.015}{0.022}            & \ermcc{1.62}{0.47}{0.13}{0.01}{0.01}                 & \ermcc{4.035}{0.069}{0.208}{0.002}{0.003}            & \ercc{0.243}{0.066}{0.126}                           & \ermcc{ 8.0}{ 0.7}{ 3.5}{ 0.3}{ 0.4}                  \\
\corot-18   & 0.02860   & 0.00065   & 0.00019     & 0.861     & 0.059     & 0.017       & 0.924     & 0.057     & 0.006       & 4.442     & 0.043     & 0.003       & \mcc{$1.09 \pm 0.16$}              & \mc{unconstrained}                 \\
\corot-19   & \ermcc{0.0512}{0.0018}{0.0011}{0.0003}{0.0003}       & \ermcc{1.181}{0.125}{0.074}{0.024}{0.022}            & \ermcc{1.576}{0.354}{0.096}{0.011}{0.010}            & \ermcc{4.115}{0.047}{0.152}{0.003}{0.003}            & \ercc{0.302}{0.049}{0.125}                           & \ermcc{ 4.9}{ 0.6}{ 1.7}{ 0.3}{ 0.3}                  \\
\corot-20   & 0.0892    & 0.0028    & 0.0004      & 1.11      & 0.10      & 0.01        & 1.34      & 0.37      & 0.01        & 4.23      & 0.24      & 0.00        & \mcc{$0.46 \pm 0.48$}              & \ermcc{ 5.9}{ 1.6}{11.4}{ 0.8}{ 1.6}    \\
\corot-23   & \ermcc{0.04809}{0.00122}{0.00088}{0.00039}{0.00056}  & \ermcc{1.122}{0.088}{0.061}{0.027}{0.039}            & \ermcc{1.74}{0.26}{0.16}{0.01}{0.02}                 & \ermcc{4.008}{0.076}{0.107}{0.004}{0.005}            & \ercc{0.214}{0.064}{0.066}                           & \ermcc{ 5.4}{ 2.4}{ 1.7}{ 0.5}{ 1.0}                  \\
HAT-P-3     & 0.03842   & 0.00050   & 0.00063     & 0.900     & 0.036     & 0.044       & 0.870     & 0.016     & 0.014       & 4.513     & 0.020     & 0.007       & \mcc{$1.365 \pm 0.078$}            & \ermcc{7.5}{4.2}{3.8}{3.6}{2.7}    \\
HAT-P-6     & 0.05244   & 0.00077   & 0.00051     & 1.295     & 0.057     & 0.038       & 1.518     & 0.070     & 0.015       & 4.188     & 0.035     & 0.004       & \mcc{$0.370 \pm 0.045$}            & \ermcc{2.3}{0.4}{0.8}{0.4}{0.3}    \\
HAT-P-9     & 0.0529    & 0.0014    & 0.0003      & 1.28      & 0.10      & 0.02        & 1.339     & 0.080     & 0.008       & 4.293     & 0.046     & 0.002       & \mcc{$0.534 \pm 0.082$}            & \ermcc{1.6}{1.4}{2.0}{0.4}{0.4}    \\
HAT-P-14    & 0.06108   & 0.00069   & 0.00036     & 1.418     & 0.048     & 0.025       & 1.591     & 0.056     & 0.009       & 4.187     & 0.025     & 0.003       & \mcc{$0.352 \pm 0.031$}            & \ermcc{1.5}{0.4}{0.4}{0.4}{0.2}    \\
Kepler-7    & 0.06324   & 0.00042   & 0.00090     & 1.413     & 0.081     & 0.060       & 2.028     & 0.025     & 0.029       & 3.974     & 0.018     & 0.006       & \mcc{$0.169 \pm 0.007$}            & \ermcc{2.9}{0.0}{0.1}{0.3}{0.3}    \\
Kepler-12   & \ermcc{0.0555}{0.0022}{0.0013}{0.0003}{0.0002}       & \ermcc{1.157}{0.147}{0.078}{0.022}{0.011}            & \ermcc{1.490}{0.061}{0.038}{0.009}{0.005}            & \ermcc{4.155}{0.018}{0.014}{0.003}{0.001}            & \ercc{0.3496}{0.0064}{0.0091}                        & \ermcc{ 5.3}{ 1.5}{ 5.0}{ 0.3}{ 0.3}                  \\
Kepler-14   & 0.0771    & 0.0010    & 0.0006      & 1.318     & 0.052     & 0.029       & 2.09      & 0.11      & 0.02        & 3.918     & 0.040     & 0.003       & \mcc{$0.145 \pm 0.020$}            & \ermcc{2.2}{1.6}{1.9}{0.9}{0.5}    \\
Kepler-15   & 0.0583    & 0.0022    & 0.0009      & 1.08      & 0.12      & 0.05        & 1.253     & 0.047     & 0.020       & 4.275     & 0.017     & 0.007       & \mcc{$0.548 \pm 0.008$}            & \ermcc{6.7}{6.7}{1.6}{0.9}{1.1}    \\
Kepler-17   & \ermcc{0.02605}{0.00018}{0.00124}{0.00037}{0.00050}  & \ermcc{1.066}{0.022}{0.150}{0.045}{0.061}            & \ermcc{0.983}{0.013}{0.047}{0.014}{0.019}            & \ermcc{4.481}{0.005}{0.023}{0.006}{0.008}            & \ercc{1.121}{0.015}{0.034}                           & \ermcc{ 1.5}{ 9.9}{ 1.2}{ 2.6}{ 1.5}                  \\
KOI-135     & 0.0440    & 0.0013    & 0.0002      & 1.24      & 0.11      & 0.02        & 1.332     & 0.046     & 0.007       & 4.281     & 0.017     & 0.002       & \mcc{$0.523 \pm 0.022$}            & \ermcc{2.7}{2.4}{1.4}{1.2}{0.5}    \\
KOI-196     & \ermcc{0.02895}{0.00091}{0.00086}{0.00020}{0.00016}  & \ermcc{0.939}{0.092}{0.082}{0.019}{0.015}            & \ermcc{0.939}{0.062}{0.031}{0.006}{0.005}            & \ermcc{4.466}{0.018}{0.054}{0.003}{0.002}            & \ercc{1.135}{0.046}{0.183}                           & \ermcc{ 6.2}{ 5.2}{ 5.1}{ 0.6}{ 0.5}                  \\
KOI-204     & 0.0457    & 0.0017    & 0.0001      & 1.21      & 0.13      & 0.01        & 1.46      & 0.12      & 0.00        & 4.192     & 0.055     & 0.001       & \mcc{$0.388 \pm 0.071$}            & \ermcc{4.7}{4.4}{6.7}{0.3}{0.4}    \\
KOI-254     & 0.02954   & 0.00068   & 0.00075     & 0.570     & 0.040     & 0.044       & 0.539     & 0.036     & 0.014       & 4.730     & 0.039     & 0.011       & \mcc{$3.63 \pm 0.52$}              & \mc{unconstrained}                 \\
KOI-423     & 0.1539    & 0.0073    & 0.0013      & 1.08      & 0.16      & 0.03        & 1.225     & 0.092     & 0.010       & 4.294     & 0.039     & 0.004       & \mcc{$0.586 \pm 0.072$}            & \ermcc{4.4}{2.2}{4.2}{0.7}{0.5}    \\
KOI-428     & 0.0802    & 0.0032    & 0.0032      & 1.46      & 0.18      & 0.18        & 2.48      & 0.17      & 0.10        & 3.812     & 0.048     & 0.017       & \mcc{$0.096 \pm 0.014$}            & \ermcc{0.8}{3.6}{1.5}{0.7}{0.8}    \\
OGLE-TR-56  & 0.02453   & 0.00046   & 0.00041     & 1.339     & 0.077     & 0.066       & 1.737     & 0.035     & 0.029       & 4.086     & 0.018     & 0.007       & \mcc{$0.256 \pm 0.011$}            & \ermcc{3.0}{2.9}{1.3}{0.3}{0.4}    \\
OGLE-TR-111 & 0.04676   & 0.00092   & 0.00081     & 0.846     & 0.049     & 0.043       & 0.822     & 0.019     & 0.014       & 4.536     & 0.024     & 0.008       & \mcc{$1.52 \pm 0.10$}              & \ermcc{9.7}{9.6}{8.8}{5.3}{2.8}    \\
OGLE-TR-113 & 0.02265   & 0.00099   & 0.00021     & 0.755     & 0.095     & 0.021       & 0.766     & 0.038     & 0.007       & 4.548     & 0.021     & 0.004       & \mcc{$1.679 \pm 0.064$}            & \mc{unconstrained}                 \\
OGLE-TR-132 & 0.03026   & 0.00098   & 0.00029     & 1.29      & 0.12      & 0.04        & 1.338     & 0.075     & 0.013       & 4.297     & 0.036     & 0.004       & \mcc{$0.540 \pm 0.062$}            & \ermcc{1.7}{3.5}{2.2}{2.1}{0.7}    \\
OGLE-TR-L9  & 0.04047   & 0.00093   & 0.00026     & 1.427     & 0.099     & 0.028       & 1.499     & 0.042     & 0.010       & 4.241     & 0.016     & 0.003       & \mcc{$0.424 \pm 0.018$}            & \ermcc{1.0}{0.6}{1.3}{0.3}{0.2}    \\
TrES-4      & 0.0502    & 0.0010    & 0.0002      & 1.339     & 0.084     & 0.016       & 1.834     & 0.087     & 0.007       & 4.038     & 0.033     & 0.002       & \mcc{$0.217 \pm 0.025$}            & \ermcc{2.5}{2.8}{0.7}{0.9}{2.5}    \\
WASP-1      & \ermcc{0.03920}{0.00053}{0.00031}{0.00015}{0.00015}  & \ermcc{1.265}{0.052}{0.030}{0.015}{0.014}            & \ermcc{1.465}{0.051}{0.079}{0.006}{0.005}            & \ermcc{4.209}{0.051}{0.028}{0.002}{0.002}            & \ercc{0.403}{0.069}{0.037}                           & \ermcc{ 2.7}{ 0.4}{ 3.5}{ 0.5}{ 0.4}                  \\
WASP-2      & 0.03092   & 0.00046   & 0.00038     & 0.851     & 0.038     & 0.032       & 0.823     & 0.015     & 0.010       & 4.537     & 0.016     & 0.005       & \mcc{$1.524 \pm 0.067$}            & \ermcc{8.2}{6.3}{3.3}{2.5}{1.7}    \\
WASP-4      & 0.02318   & 0.00037   & 0.00028     & 0.927     & 0.044     & 0.034       & 0.910     & 0.015     & 0.011       & 4.487     & 0.009     & 0.005       & \mcc{$1.230 \pm 0.022$}            & \ermcc{6.4}{2.7}{5.3}{1.9}{1.9}    \\
WASP-5      & 0.02740   & 0.00040   & 0.00019     & 1.033     & 0.045     & 0.021       & 1.088     & 0.040     & 0.008       & 4.379     & 0.030     & 0.003       & \mcc{$0.801 \pm 0.080$}            & \ermcc{5.6}{2.2}{2.2}{1.1}{1.0}    \\
WASP-7      & 0.0624    & 0.0011    & 0.0003      & 1.317     & 0.069     & 0.022       & 1.478     & 0.088     & 0.008       & 4.218     & 0.048     & 0.002       & \mcc{$0.408 \pm 0.068$}            & \ermcc{1.9}{0.7}{0.6}{0.3}{0.3}    \\
WASP-12     & 0.02309   & 0.00096   & 0.00030     & 1.38      & 0.18      & 0.05        & 1.619     & 0.076     & 0.021       & 4.159     & 0.023     & 0.006       & \mcc{$0.325 \pm 0.016$}            & \ermcc{2.0}{0.7}{2.6}{0.3}{0.6}    \\
WASP-13     & 0.0557    & 0.0018    & 0.0001      & 1.22      & 0.12      & 0.01        & 1.657     & 0.079     & 0.004       & 4.086     & 0.033     & 0.001       & \mcc{$0.268 \pm 0.029$}            & \ermcc{5.0}{2.6}{1.7}{1.6}{0.9}    \\
WASP-14     & 0.0372    & 0.0011    & 0.0002      & 1.35      & 0.12      & 0.02        & 1.666     & 0.097     & 0.009       & 4.126     & 0.042     & 0.002       & \mcc{$0.293 \pm 0.042$}            & \ermcc{2.4}{1.5}{1.0}{0.2}{0.2}    \\
WASP-18     & 0.02043   & 0.00029   & 0.00014     & 1.274     & 0.054     & 0.027       & 1.228     & 0.042     & 0.009       & 4.365     & 0.027     & 0.003       & \mcc{$0.687 \pm 0.062$}            & \ermcc{0.4}{1.0}{0.7}{0.4}{0.3}    \\
WASP-21     & 0.0516    & 0.0020    & 0.0012      & 0.98      & 0.12      & 0.07        & 1.186     & 0.081     & 0.028       & 4.281     & 0.031     & 0.010       & \mcc{$0.587 \pm 0.061$}            & \mc{unconstrained}                 \\
XO-2        & \ermcc{0.03616}{0.00065}{0.00182}{0.00100}{0.00167}  & \ermcc{0.924}{0.054}{0.122}{0.076}{0.124}            & \ermcc{0.962}{0.026}{0.054}{0.027}{0.044}            & \ermcc{4.436}{0.037}{0.035}{0.012}{0.020}            & \ercc{1.034}{0.127}{0.058}                           & \ermcc{ 6.5}{23.2}{ 3.5}{ 5.3}{ 3.8}                  \\
\hline \hline \end{tabular} \end{table*}

\begin{table*} \caption{\label{tab:absdim:planets} Derived physical properties of the planets
of the TEPs studied in this work. For many quantities the first errorbar gives the statistical errors
and the second errorbar gives the systematic errors arising from the dependence on stellar theory.}
\setlength{\tabcolsep}{4pt}
\begin{tabular}{l l@{\,$\pm$\,}l@{\,$\pm$\,}l l@{\,$\pm$\,}l@{\,$\pm$\,}l l@{\,$\pm$\,}l@{\,$\pm$\,}l l@{\,$\pm$\,}l@{\,$\pm$\,}l l@{\,$\pm$\,}l@{\,$\pm$\,}l l@{\,$\pm$\,}l@{\,$\pm$\,}l}
\hline \hline
System & \mcc{Mass (\Mjup)} & \mcc{Radius (\Rjup)} & \mcc{$g_{\rm b}$ (\mss)} & \mcc{Density (\pjup)} & \mcc{\Teq\ (K)} & \mcc{\safronov} \\
\hline
\corot-2    & 3.57      & 0.13      & 0.09        & 1.460     & 0.025     & 0.019       & \mcc{$41.5 \pm  1.7$}              & 1.073     & 0.057     & 0.014       & \mcc{$1521 \pm   18$}              & 0.1390    & 0.0048    & 0.0018       \\
\corot-17   & \ermcc{2.46}{0.26}{0.24}{0.02}{0.04}                 & \ermcc{1.007}{0.345}{0.092}{0.005}{0.007}            & \ercc{60}{13}{27}                                    & \ermcc{2.26}{0.77}{1.35}{0.02}{0.01}                 & \ercc{1610}{ 210}{  70}                              & \ermcc{0.225}{0.031}{0.062}{0.002}{0.001}             \\
\corot-18   & 3.27      & 0.17      & 0.04        & 1.251     & 0.083     & 0.008       & \mcc{$51.8 \pm  6.6$}              & 1.56      & 0.30      & 0.01        & \mcc{$1490 \pm   45$}              & 0.173     & 0.012     & 0.001        \\
\corot-19   & \ermcc{1.090}{0.092}{0.077}{0.015}{0.014}            & \ermcc{1.190}{0.260}{0.069}{0.008}{0.007}            & \ercc{19.1}{ 2.4}{ 6.2}                              & \ermcc{0.60}{0.11}{0.27}{0.00}{0.00}                 & \ercc{1630}{ 150}{  40}                              & \ermcc{0.0794}{0.0064}{0.0150}{0.0005}{0.0005}        \\
\corot-20   & 5.06      & 0.36      & 0.04        & 1.16      & 0.26      & 0.00        & \mcc{$93 \pm 46$}                  & 3.0       & 2.5       & 0.0         & \mcc{$1100 \pm  150$}              & 0.70      & 0.17      & 0.00         \\
\corot-23   & \ermcc{3.06}{0.32}{0.30}{0.05}{0.07}                 & \ermcc{1.18}{0.19}{0.11}{0.01}{0.01}                 & \ercc{54}{12}{14}                                    & \ermcc{1.75}{0.63}{0.64}{0.02}{0.01}                 & \ercc{1710}{ 110}{  80}                              & \ermcc{0.222}{0.031}{0.036}{0.003}{0.002}             \\
HAT-P-3     & 0.584     & 0.020     & 0.019       & 0.947     & 0.027     & 0.015       & \mcc{$16.14 \pm  0.90$}            & 0.643     & 0.052     & 0.011       & \mcc{$1189 \pm   16$}              & 0.0526    & 0.0019    & 0.0009       \\
HAT-P-6     & 1.063     & 0.053     & 0.021       & 1.395     & 0.080     & 0.014       & \mcc{$13.5 \pm  1.6$}              & 0.366     & 0.064     & 0.004       & \mcc{$1704 \pm   40$}              & 0.0617    & 0.0043    & 0.0006       \\
HAT-P-9     & 0.778     & 0.083     & 0.009       & 1.38      & 0.10      & 0.01        & \mcc{$10.1 \pm  1.7$}              & 0.275     & 0.066     & 0.002       & \mcc{$1540 \pm   53$}              & 0.0463    & 0.0056    & 0.0003       \\
HAT-P-14    & 2.271     & 0.079     & 0.027       & 1.219     & 0.059     & 0.007       & \mcc{$37.9 \pm  3.7$}              & 1.17      & 0.17      & 0.01        & \mcc{$1624 \pm   32$}              & 0.1603    & 0.0089    & 0.0010       \\
Kepler-7    & 0.453     & 0.067     & 0.013       & 1.649     & 0.030     & 0.024       & \mcc{$4.13 \pm 0.35$}              & 0.0944    & 0.0085    & 0.0014      & \mcc{$1619 \pm   15$}              & 0.0245    & 0.0021    & 0.0004       \\
Kepler-12   & \ermcc{0.430}{0.053}{0.044}{0.005}{0.003}            & \ermcc{1.706}{0.069}{0.039}{0.011}{0.006}            & \ercc{3.66}{0.33}{0.34}                              & \ermcc{0.0810}{0.0075}{0.0081}{0.0003}{0.0005}       & \ercc{1485}{  25}{  25}                              & \ermcc{0.0241}{0.0022}{0.0024}{0.0001}{0.0002}        \\
Kepler-14   & 7.68      & 0.37      & 0.11        & 1.126     & 0.049     & 0.008       & \mcc{$150 \pm  13$}                & 5.04      & 0.67      & 0.04        & \mcc{$1605 \pm   39$}              & 0.798     & 0.047     & 0.006        \\
Kepler-15   & 0.696     & 0.097     & 0.022       & 1.289     & 0.050     & 0.021       & \mcc{$10.4 \pm  1.1$}              & 0.304     & 0.036     & 0.005       & \mcc{$1251 \pm   27$}              & 0.0582    & 0.0068    & 0.0009       \\
Kepler-17   & \ermcc{2.340}{0.060}{0.228}{0.066}{0.089}            & \ermcc{1.310}{0.016}{0.063}{0.018}{0.025}            & \ercc{33.82}{ 0.85}{ 1.01}                           & \ermcc{0.974}{0.054}{0.037}{0.019}{0.014}            & \ercc{1712}{  26}{  25}                              & \ermcc{0.0873}{0.0047}{0.0023}{0.0017}{0.0012}        \\
KOI-135     & 3.09      & 0.21      & 0.03        & 1.115     & 0.041     & 0.006       & \mcc{$61.6 \pm  3.3$}              & 2.09      & 0.16      & 0.01        & \mcc{$1603 \pm   39$}              & 0.1969    & 0.0100    & 0.0011       \\
KOI-196     & \ermcc{0.493}{0.072}{0.070}{0.007}{0.005}            & \ermcc{0.852}{0.071}{0.031}{0.006}{0.005}            & \ercc{16.9}{ 2.3}{ 3.2}                              & \ermcc{0.75}{0.11}{0.18}{0.00}{0.01}                 & \ercc{1554}{  53}{  29}                              & \ermcc{0.0356}{0.0048}{0.0054}{0.0002}{0.0002}        \\
KOI-204     & 1.030     & 0.087     & 0.006       & 1.20      & 0.11      & 0.00        & \mcc{$17.6 \pm  2.9$}              & 0.55      & 0.14      & 0.00        & \mcc{$1568 \pm   59$}              & 0.0647    & 0.0063    & 0.0002       \\
KOI-254     & 0.500     & 0.055     & 0.026       & 0.999     & 0.064     & 0.026       & \mcc{$12.4 \pm  1.8$}              & 0.469     & 0.093     & 0.012       & \mcc{$787 \pm  26$}                & 0.0518    & 0.0059    & 0.0013       \\
KOI-423     & 17.9      &  1.8      &  0.3        & 1.092     & 0.077     & 0.009       & \mcc{$372 \pm  38$}                & 12.9      &  2.1      &  0.1        & \mcc{$851 \pm  25$}                & 4.69      & 0.35      & 0.04         \\
KOI-428     & 2.16      & 0.40      & 0.18        & 1.44      & 0.11      & 0.06        & \mcc{$25.7 \pm  5.0$}              & 0.67      & 0.17      & 0.03        & \mcc{$1744 \pm   51$}              & 0.165     & 0.028     & 0.007        \\
OGLE-TR-56  & 1.41      & 0.17      & 0.05        & 1.734     & 0.051     & 0.029       & \mcc{$11.6 \pm  1.3$}              & 0.253     & 0.031     & 0.004       & \mcc{$2482 \pm   30$}              & 0.0298    & 0.0033    & 0.0005       \\
OGLE-TR-111 & 0.55      & 0.10      & 0.02        & 1.011     & 0.035     & 0.017       & \mcc{$13.2 \pm  2.5$}              & 0.49      & 0.10      & 0.01        & \mcc{$1019 \pm   20$}              & 0.060     & 0.011     & 0.001        \\
OGLE-TR-113 & 1.23      & 0.20      & 0.02        & 1.088     & 0.054     & 0.010       & \mcc{$25.7 \pm  3.4$}              & 0.89      & 0.13      & 0.01        & \mcc{$1342 \pm   22$}              & 0.0678    & 0.0095    & 0.0006       \\
OGLE-TR-132 & 1.17      & 0.15      & 0.02        & 1.229     & 0.075     & 0.012       & \mcc{$19.2 \pm  2.8$}              & 0.59      & 0.11      & 0.01        & \mcc{$1991 \pm   42$}              & 0.0445    & 0.0056    & 0.0004       \\
OGLE-TR-L9  & 4.4       & 1.5       & 0.1         & 1.633     & 0.045     & 0.011       & \mcc{$40 \pm 13$}                  & 0.94      & 0.32      & 0.01        & \mcc{$2034 \pm   22$}              & 0.151     & 0.051     & 0.001        \\
TrES-4      & 0.897     & 0.075     & 0.007       & 1.735     & 0.072     & 0.007       & \mcc{$7.39 \pm 0.76$}              & 0.161     & 0.021     & 0.001       & \mcc{$1805 \pm   40$}              & 0.0387    & 0.0033    & 0.0002       \\
WASP-1      & \ermcc{0.980}{0.047}{0.042}{0.008}{0.007}            & \ermcc{1.493}{0.061}{0.091}{0.006}{0.006}            & \ercc{10.9}{ 1.5}{ 0.9}                              & \ermcc{0.275}{0.058}{0.032}{0.001}{0.001}            & \ercc{1830}{  33}{  49}                              & \ermcc{0.0406}{0.0031}{0.0023}{0.0002}{0.0002}        \\
WASP-2      & 0.880     & 0.031     & 0.022       & 1.063     & 0.025     & 0.013       & \mcc{$19.31 \pm  0.80$}            & 0.685     & 0.042     & 0.008       & \mcc{$1286 \pm   17$}              & 0.0601    & 0.0019    & 0.0007       \\
WASP-4      & 1.249     & 0.043     & 0.030       & 1.364     & 0.024     & 0.016       & \mcc{$16.64 \pm  0.33$}            & 0.460     & 0.014     & 0.006       & \mcc{$1673 \pm   17$}              & 0.0458    & 0.0010    & 0.0005       \\
WASP-5      & 1.595     & 0.048     & 0.022       & 1.175     & 0.055     & 0.008       & \mcc{$28.6 \pm  2.6$}              & 0.92      & 0.12      & 0.01        & \mcc{$1753 \pm   35$}              & 0.0719    & 0.0034    & 0.0005       \\
WASP-7      & 0.98      & 0.13      & 0.01        & 1.374     & 0.094     & 0.008       & \mcc{$12.9 \pm  2.4$}              & 0.353     & 0.087     & 0.002       & \mcc{$1530 \pm   45$}              & 0.067     & 0.010     & 0.000        \\
WASP-12     & 1.43      & 0.13      & 0.04        & 1.825     & 0.091     & 0.024       & \mcc{$10.63 \pm  0.53$}            & 0.220     & 0.019     & 0.003       & \mcc{$2523 \pm   45$}              & 0.0262    & 0.0013    & 0.0003       \\
WASP-13     & 0.512     & 0.060     & 0.002       & 1.528     & 0.084     & 0.004       & \mcc{$5.44 \pm 0.73$}              & 0.134     & 0.023     & 0.000       & \mcc{$1531 \pm   37$}              & 0.0307    & 0.0035    & 0.0001       \\
WASP-14     & 7.90      & 0.46      & 0.09        & 1.633     & 0.092     & 0.009       & \mcc{$73.4 \pm  7.1$}              & 1.69      & 0.25      & 0.01        & \mcc{$2090 \pm   59$}              & 0.265     & 0.015     & 0.001        \\
WASP-18     & 10.38     &  0.30     &  0.15       & 1.163     & 0.054     & 0.008       & \mcc{$190 \pm  16$}                & 6.18      & 0.83      & 0.04        & \mcc{$2413 \pm   44$}              & 0.286     & 0.013     & 0.002        \\
WASP-21     & 0.295     & 0.027     & 0.014       & 1.263     & 0.085     & 0.029       & \mcc{$4.58 \pm 0.45$}              & 0.137     & 0.021     & 0.003       & \mcc{$1340 \pm   32$}              & 0.0245    & 0.0019    & 0.0006       \\
XO-2        & \ermcc{0.593}{0.025}{0.057}{0.033}{0.054}            & \ermcc{0.984}{0.031}{0.071}{0.027}{0.045}            & \ercc{15.2}{ 1.8}{ 0.8}                              & \ermcc{0.582}{0.116}{0.046}{0.028}{0.016}            & \ercc{1328}{  17}{  28}                              & \ermcc{0.0472}{0.0040}{0.0016}{0.0022}{0.0013}        \\
\hline \hline \end{tabular} \end{table*}

The primary result of this work is measurements of the physical properties of 38 transiting planetary systems. These are collected in Table\,\ref{tab:absdim:stars} for the stellar components and in Table\,\ref{tab:absdim:planets} for the planetary components. When combined with previous results from the {\it Homogeneous Studies} project, homogeneous physical properties are available for a total of 82 TEP systems. In the discussion below I have added results for all other known TEP systems as aggregated in the {\it Transiting Extrasolar Planets Catalogue} (TEPCat\footnote{The Transiting Extrasolar Planet Catalogue (TEPCat) is available at {\tt http://www.astro.keele.ac.uk/jkt/tepcat/}}) on 2012/05/02, which comprises 199 TEPs and six \reff{transiting brown dwarf} systems including those objects in the {\it Homogeneous Studies} project.

\begin{figure} \includegraphics[width=\columnwidth,angle=0]{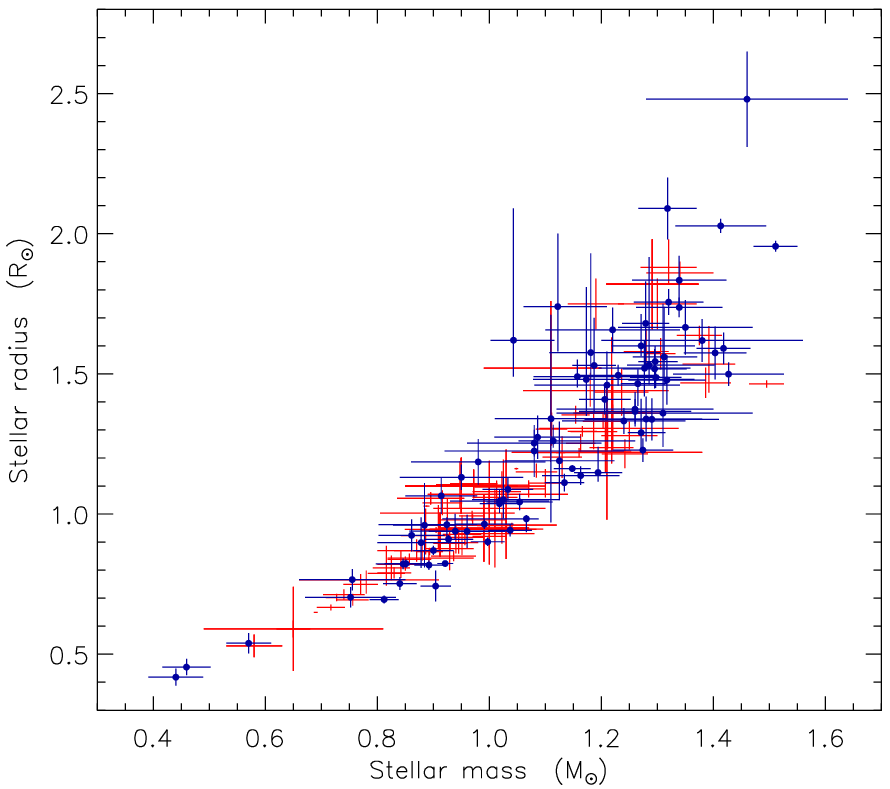}
\caption{\label{fig:absdim:m1r1} Mass--radius plot for the host stars
of the known transiting extrasolar planets. Those objects studied in
this work are shown with (blue) filled circles and numbers taken from
the literature with (red) crosses.} \end{figure}

Fig.\,\ref{fig:absdim:m1r1} is a mass--radius plot for the host stars. This sample is very inhomogeneous due to observational biases and selection effects, particularly because the majority of objects have been discovered by wide-field surveys with small ground-based telescopes. The happy hunting ground is clearly unevolved stars in the mass interval 0.9--1.2\Msun. Lower-mass stars are intrinsically dim and predominate at faint apparent magnitudes, which are inaccessible to wide-field telescopes with coarse pixel scales. Higher-mass stars are brighter and plentiful in such surveys, but exhibit faster rotation and fewer spectral lines, making confirmatory RV measurements difficult. More evolved stars are easier for RV measurements, but their larger radii means transits of planet-sized objects are both longer and shallower, making it difficult to identify the photometric signature of a TEP in the first place. Space-based surveys are less biased towards unevolved stars, so the characteristics of this diagram will change in the near future as new discoveries flood in from \kepler.

\begin{figure} \includegraphics[width=\columnwidth,angle=0]{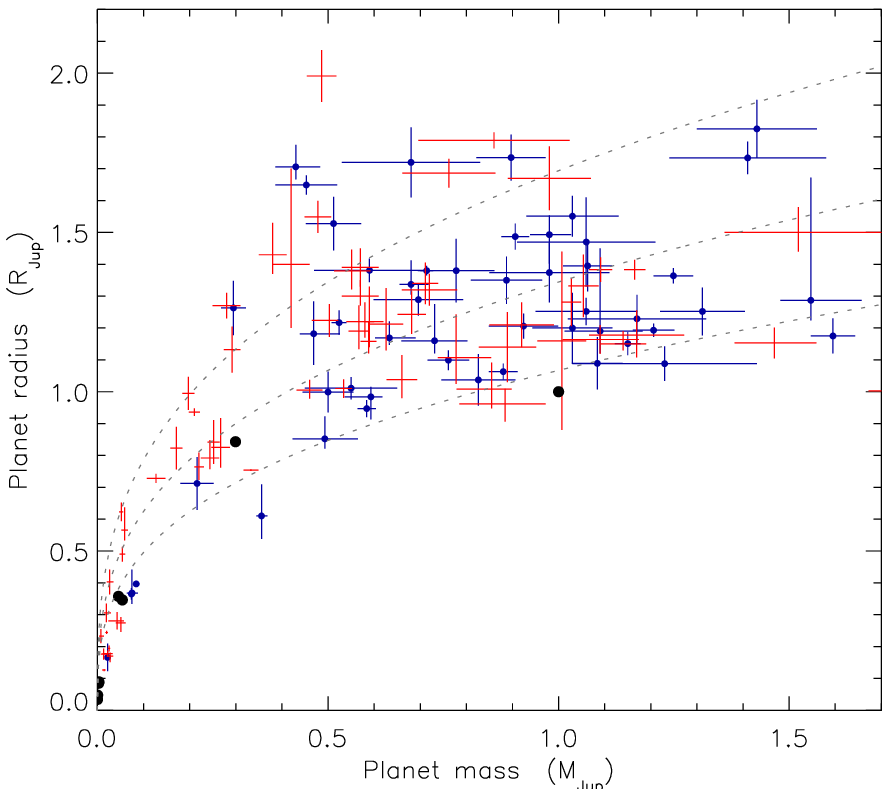}
\caption{\label{fig:absdim:m2r2} Mass--radius plot for the known transiting
extrasolar planets. Symbols are as in Fig.\,\ref{fig:absdim:m1r1} The four
gas giant planets in our Solar System are denoted with (black) filled circles.
Dotted lines show loci where density equals 1.0, 0.5 and 0.25 \pjup.} \end{figure}

Fig.\,\ref{fig:absdim:m2r2} shows the mass--radius plot for the known TEPs in the region of parameter space occupied by Jupiter, with loci of constant density overlaid. Note that Jupiter does not lie on the $\rho_{\rm b} = \pjup$ locus because of the rotationally-induced difference between its equatorial radius (71\,492\,km) and volume-equivalent radius (66\,991\,km\footnote{\tt http://nssdc.gsfc.nasa.gov/planetary/factsheet/ jupiterfact.html}). This Figure shows that the dominant population of TEPs is currently of mass 0.5--1.2\Mjup\ and radius 0.9--1.7\Rjup. The Saturnian objects are currently easily differentiated based on their masses and radii, but future discoveries may fill in the gaps between them, the Jupiter-like planets, and the lower-mass planets.

\begin{figure} \includegraphics[width=\columnwidth,angle=0]{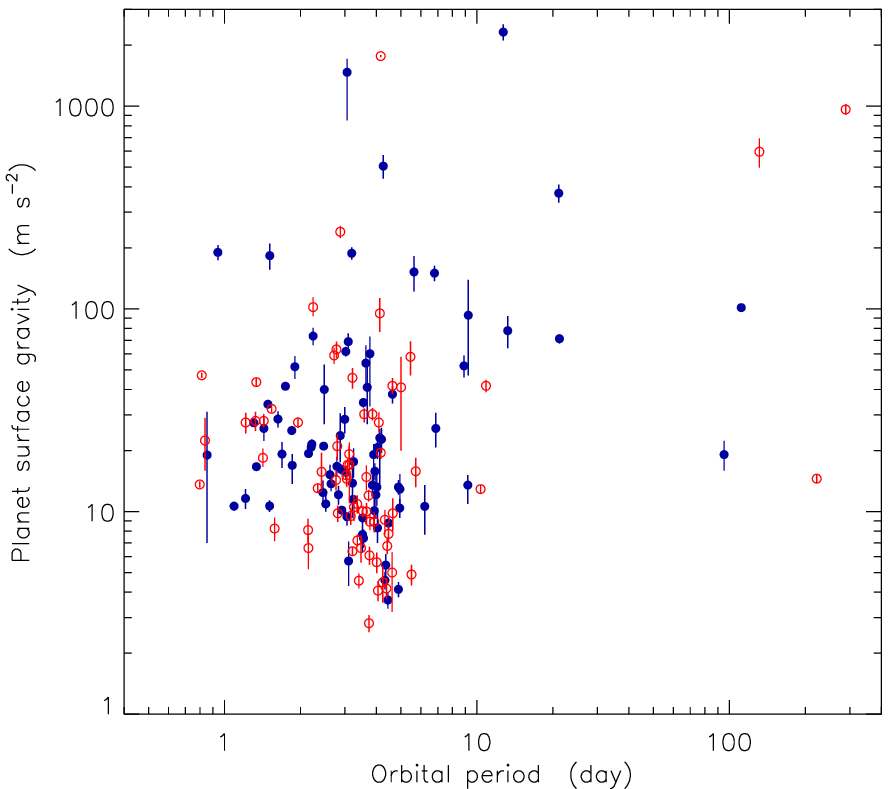}
\caption{\label{fig:absdim:pg2} Plot of the orbital periods versus the
surface gravities of the dominant population of known TEPs. Symbols are
as in Fig.\,\ref{fig:absdim:m1r1}.} \end{figure}

\begin{figure} \includegraphics[width=\columnwidth,angle=0]{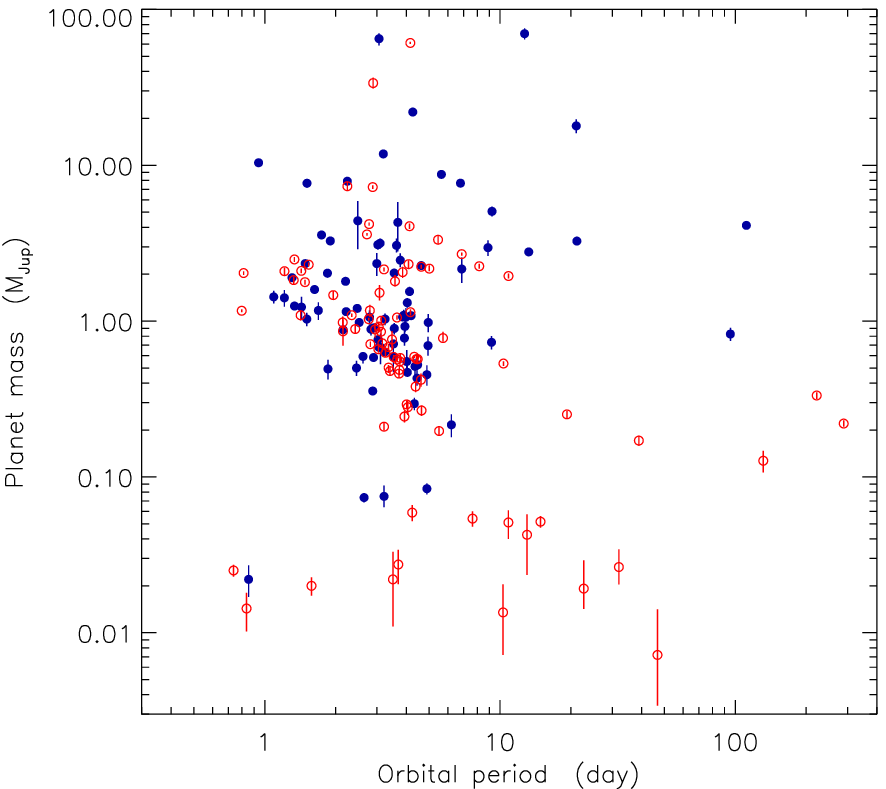}
\caption{\label{fig:absdim:pm2} Plot of the orbital periods versus the
masses of the dominant population of known TEPs. Symbols are as in
Fig.\,\ref{fig:absdim:m1r1}.} \end{figure}

Correlations have previously been noticed between \Porb\ and $g_{\rm b}$ \citep{Me++07mn} and \Porb\ and $M_{\rm b}$ \citep{Mazeh++05mn}. The relevant plots are shown in Figs.\ \ref{fig:absdim:pg2} and \ref{fig:absdim:pm2}. In both cases there are quite a few planets whose properties put them outside the range of \Porb\ shown in these plots. For the following discussion I have neglected those planets which lie outside the relevant plot \reff{due to} large mass or long period. I have also ignored super-Earth planets ($M_{\rm b} < 10$\Mearth) in order to obtain a sample of definitely gaseous objects.

The rank correlation test of \citet{Spearman1904} returns a probability of 99.988\% ($3.9\sigma$) that the $\Porb$--$g_{\rm b}$ correlation is real and 99.987\% ($3.8\sigma$) that the $\Porb$--$M_{\rm b}$ correlation is real. In both cases, inclusion of all TEPs returns an increased significance level of the correlation ($4.8\sigma$ and $5.8\sigma$ respectively). Both correlations are therefore supported by the current data.

\begin{figure} \includegraphics[width=\columnwidth,angle=0]{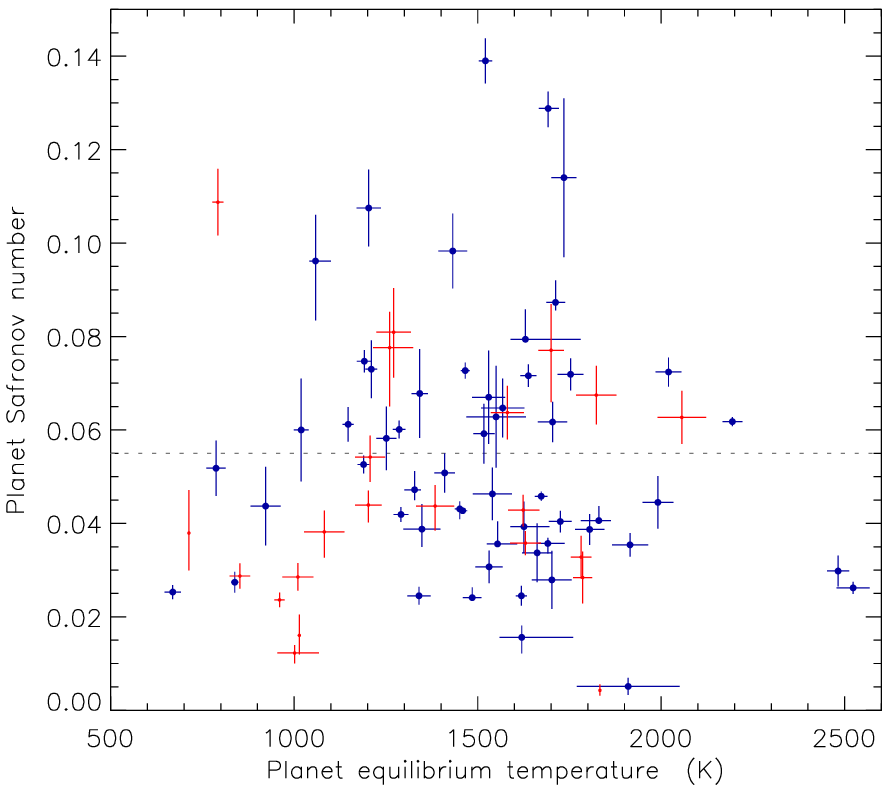}
\caption{\label{fig:absdim:teqsaf} Plot of equilibrium temperature versus
Safronov number for the full sample of planets. Objects shown with (blue)
circles were studied in this work; those which are just (red) errorbars
were not. The dotted line shows the separation between Class A and Class
B proposed by \citet{HansenBarman07apj}.} \end{figure}

\citet{HansenBarman07apj} divided up eighteen of the twenty TEPs then known into two classes based on their position in a diagram of \safronov\ versus $T_{\rm eq}$. An updated version of the diagram can be seen in Fig.\,\ref{fig:absdim:teqsaf}, and agrees with previous conclusions (Paper\,II) that the division between the classes is blurred into insignificance. A dotted line at $\Theta = 0.055$ has been drawn to show the expected boundaries between Class\,I ($\Theta \approx 0.07 \pm 0.01$) and Class\,II ($\Theta \approx 0.04 \pm 0.01$). The previously postulated gap in the distribution is no longer apparent, so the division into two classes can now be safely ignored.

\begin{figure} \includegraphics[width=\columnwidth,angle=0]{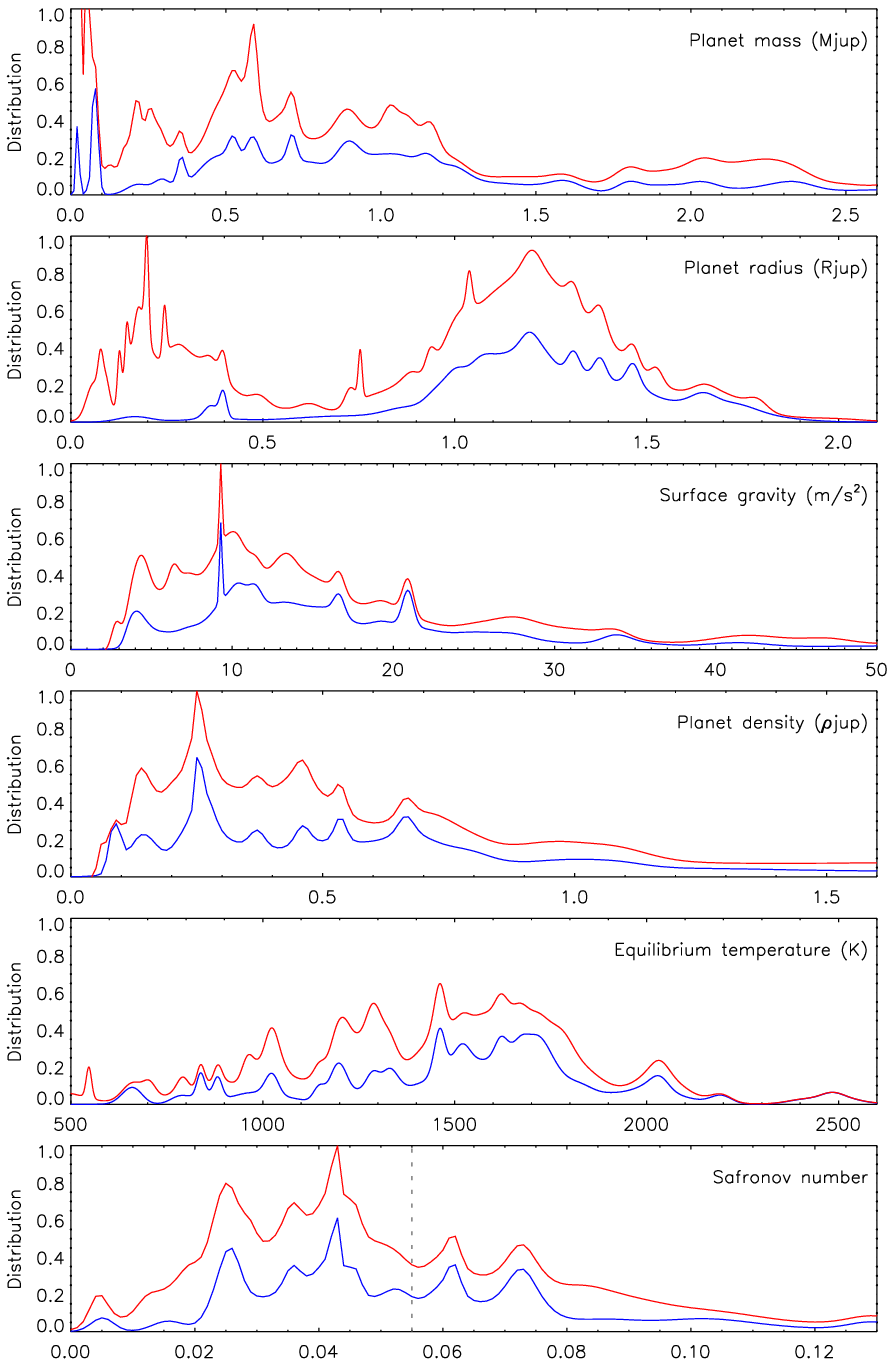}
\caption{\label{fig:absdim:pdist} Plots of the distributions of planet properties.
From top to bottom these are $M_{\rm b}$, $R_{\rm b}$, $g_{\rm b}$, $\rho_{\rm b}$,
\Teq\ and \safronov. In each case the upper (red) line shows the distribution for
all known TEPs and the lower (blue) line shows that for the TEPs considered within
the current series of papers.} \end{figure}

Fig.\,\ref{fig:absdim:pdist} takes a different approach to illustrating the distributions of physical properties of the known TEPs. Distributions have been obtained for each property given in Table\,\ref{tab:absdim:planets}, and for the {\it Homogeneous Studies} objects and for all objects. Each distribution was obtained by calculating a Gaussian function for each planet with a mean and standard deviation equal to the relevant property and its uncertainty, and then taking the sum over all planets in the sample. This produces an alternative to the histogram which avoids the loss of information implicit in the binning process.

Fig.\,\ref{fig:absdim:pdist}a shows that the mass distribution of TEPs has a broad plateau around 0.5--1.2\Mjup, with a peak at the lower edge of this interval. A secondary peak, representing the Saturnian TEPs, occurs around 0.2\Mjup. The distribution then rises towards the lowest masses. Fig.\,\ref{fig:absdim:pdist}b shows the situation for planet radius; the distribution is bimodal with peaks near 1.2\Rjup\ and 0.2\Rjup. The lower-radius peak is almost entirely absent from the {\it Homogeneous Studies} sample, as it is dominated by planets from the \kepler\ satellite which were discovered and characterised by TTVs. Such systems are more complicated to study than normal Hot Jupiters, and there are no plans to include them in the current series of papers.

Fig.\,\ref{fig:absdim:pdist}c shows the distribution of planet surface gravities. The sharp peak around 9\mss\ is due to the extremely precise measurement for HD\,209458. The determination of $g_{\rm b}$ requires only directly measurable photometric and spectroscopic parameters ($r_{\rm b}$, $i$, $K_{\rm A}$ and $e$), so can be measured to extremely high precision in favourable cases. In the case of HD\,209458, $g_{\rm b} = 9.30 \pm 0.08$\mss\ (Paper\,I). The distribution of planetary densities is plotted in Fig.\,\ref{fig:absdim:pdist}d, and has a very similar functional form to that for $g_{\rm b}$ but with greater smoothness. This is because the precision of a $\rho_{\rm b}$ value is very sensitive to the quality of the transit light curves, but is also subject to systematic error from the model-derived additional constraint. In the case of HD\,209458, $\rho_{\rm b} = 0.254 \pm 0.004 \pm 0.002$\pjup.

Fig.\,\ref{fig:absdim:pdist}e depicts the distribution of \Teq, which is noisy but has a broad peak around 1500--1800\,K. Ground-based surveys are observationally biased in favour of short-period planets, so hot planets are over-represented in the known TEP population. Finally, Fig.\,\ref{fig:absdim:pdist}f shows the result for \safronov. The division between Class\,I and Class\,II planets \citep{HansenBarman07apj} is shown with a grey dotted line, and coincides with one of several local minima in the distribution function.


\section{Follow-up observations}                                                                                                 \label{sec:followup}

\begin{table} \centering \caption{\label{tab:absdim:obs} Summary of which
further observations would be useful for the TEPs studied in this work.
$\star$ indicates where additional data would be useful, and $\star\star$
denotes where it would be useful but difficult to either obtain or analyse.
I do not suggest the need for new data when such data are already in the
process of being obtained (e.g.\ \kepler\ photometry).}
\begin{tabular}{lccc}
\hline \hline
System      & Photometric  & Radial       & Spectral     \\
            & observations & velocities   & synthesis    \\
\hline
\corot-2    &              &              &              \\
\corot-17   & $\star\star$ & $\star\star$ & $\star\star$ \\
\corot-18   & $\star\star$ &              & $\star\star$ \\
\corot-19   & $\star$      &              &              \\
\corot-20   & $\star$      &              &              \\
\corot-23   & $\star\star$ & $\star\star$ & $\star\star$ \\
HAT-P-3     &              &              &              \\
HAT-P-6     & $\star$      &              &              \\
HAT-P-9     & $\star$      &              & $\star$      \\
HAT-P-14    & $\star$      &              &              \\
Kepler-7    &              & $\star$      &              \\
Kepler-12   &              & $\star$      & $\star$      \\
Kepler-14   &              &              & $\star\star$ \\
Kepler-15   &              & $\star$      & $\star$      \\
Kepler-17   &              &              &              \\
KOI-135     &              &              & $\star$      \\
KOI-196     &              &              & $\star$      \\
KOI-204     &              & $\star\star$ & $\star\star$ \\
KOI-254     &              & $\star\star$ & $\star\star$ \\
KOI-423     & $\star$      & $\star$      & $\star$      \\
KOI-428     &              & $\star\star$ & $\star\star$ \\
OGLE-TR-56  &              & $\star\star$ &              \\
OGLE-TR-111 &              & $\star\star$ &              \\
OGLE-TR-113 &              & $\star\star$ & $\star\star$ \\
OGLE-TR-132 &              & $\star\star$ &              \\
OGLE-TR-L9  &              & $\star\star$ & $\star\star$ \\
TrES-4      & $\star$      & $\star$      & $\star$      \\
WASP-1      &              &              & $\star$      \\
WASP-2      &              &              &              \\
WASP-4      &              &              &              \\
WASP-5      &              &              & $\star$      \\
WASP-7      & $\star$      & $\star$      & $\star$      \\
WASP-12     &              &              & $\star$      \\
WASP-13     & $\star$      & $\star$      & $\star$      \\
WASP-14     & $\star$      &              & $\star$      \\
WASP-18     &              &              &              \\
WASP-21     & $\star$      &              & $\star$      \\
XO-2        &              &              &              \\
\hline \hline \end{tabular} \end{table}

\begin{table} \centering \caption{\label{tab:absdim:eph} Limits of the current
ephemerides of the known TEPs. The two dates for each TEP indicate the first
transits whose midpoints are uncertain by 1\,hour and by half the transit
duration. RHJD $=$ HJD $-$ 24000000. The \kepler\ planets are not included
because they are the subject of continued observations using the satellite.}
\setlength{\tabcolsep}{4pt}
\begin{tabular}{lcccc}
\hline \hline
TEP          & \mc{1\,hour uncertainty} & \mc{Half-transit uncertainty} \\
             &     RHJD & UT date       &       RHJD & UT date          \\
\hline
\corot-04   & 55181.195810 & 2009\,12\,15 & 56432.674610 & 2013\,05\,20 \\
\corot-14   & 55268.529920 & 2010\,03\,13 & 55186.874360 & 2009\,12\,21 \\
\corot-17   & 55447.075200 & 2010\,09\,07 & 56163.014200 & 2012\,08\,23 \\
\corot-20   & 56550.756250 & 2013\,09\,15 & 56698.641850 & 2014\,02\,10 \\
HAT-P-31    & 56573.327850 & 2013\,10\,07 & 60137.190450 & 2023\,07\,11 \\
\corot-23   & 56823.191100 & 2014\,06\,14 & 58250.292000 & 2018\,05\,11 \\
\corot-10   & 57040.629000 & 2015\,01\,18 & 58377.929600 & 2018\,09\,16 \\
\corot-09   & 57461.558700 & 2016\,03\,14 & 66131.474500 & 2039\,12\,08 \\
OGLE-211 & 58524.988640 & 2019\,02\,10 & 64688.042880 & 2035\,12\,26 \\
\corot-08   & 58780.283621 & 2019\,10\,23 & 60451.414110 & 2024\,05\,20 \\
\corot-15   & 59004.400840 & 2020\,06\,03 & 61639.370800 & 2027\,08\,21 \\
Qatar-1     & 59217.596165 & 2021\,01\,03 & 58499.059467 & 2019\,01\,15 \\
\hline \hline \end{tabular} \end{table}

Most of the TEPs in the current work would benefit from further observations of some sort, and this is summarised in Table\,\ref{tab:absdim:obs}. In many cases the dominant uncertainty stems from the quality of the light curve. This remains true for many of the \corot\ systems, even though they have space-based data. In each case this is due to their relative faintness (remember that the aperture of the \corot\ telescope is only 27\,cm) and/or that few transits were seen due to either a short observing sequence or a long orbital period.

Additional RV measurements are useful too. In many circumstances, particularly for the fainter objects, the RVs are good enough to unambiguously confirm the planetary nature of a system but are the dominant source of uncertainty in the planetary masses. Now over 200 TEPs are known it seems appropriate to concentrate follow-up resources on measuring the physical properties of a golden subset of these to high precision. An additional requirement of RVs is definition of the orbital shape ($e$ and $\omega$), and imprecise measurements of these quantities compromise measurements of the photometric parameters (in particular $r_{\rm A}$).

The physical properties of quite a few of the TEPs are also limited by the precision of the \Teff\ and \FeH\ measurements available. In many cases this can be improved, but in some cases this is not an option because the errorbars are already close to the limit set by our understanding of low-mass stars (taken to be 50\,K in \Teff\ and 0.05\,dex in \FeH). There is no immediate prospect of lowering these thresholds; in fact there is evidence that they are already slightly optimistic \citep{Bruntt+10aa,Bruntt+12mn}.

Quite a few of the TEPs have ephemerides which will become uncertain over the timescale of a few years. To investigate this I compiled a catalogue of ephemerides of all known TEPs and identified the first predicted times of transit which were uncertain by one hour, and by half of one transit duration. A list of objects for which one of these dates is earlier than the year 2022 is given in Table\,\ref{tab:absdim:eph}. The \kepler\ planets are not included in this analysis because they continue to be observed and will benefit from significantly improved ephemerides once these data become available. The list of the other planets is dominated by \corot\ objects, and it is notable that the ephemerides for \corot-4, \corot-14 and \corot-17 are already uncertain by more than one hour. HAT-P-31 is present in the list due to the fact that there is no published follow-up photometry of this object at all, only the discovery data from the HAT survey cameras. Further photometric observations of these objects, \reff{such as those obtained by the TERMS project \citep{Kane+09pasp,Dragomir+11aj}}, are needed before the predictive power of their ephemerides deteriorates much further.


\section{Discovery rate and sky positions of the transiting extrasolar planetary systems}                                         \label{sec:discpos}

\begin{figure*} \includegraphics[width=\textwidth,angle=0]{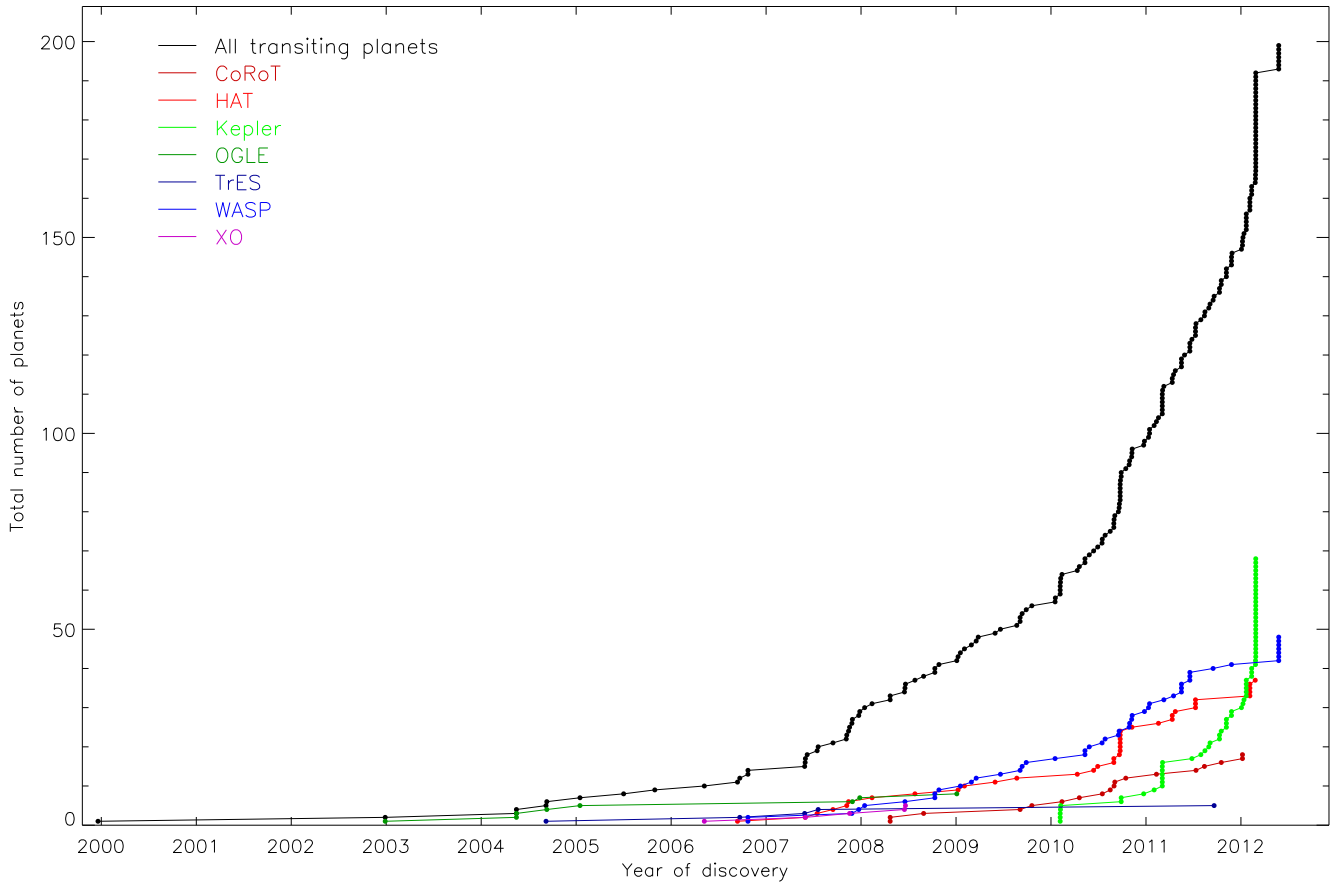}
\caption{\label{fig:discrate} Plot of the discovery rate of the known TEPs.
The total number is shown by the upper black line, whereas the discovery rate
of selected individual surveys are indicated below this using coloured lines
as indicated in the key on the top left of the plot (please see the electronic
version of this paper for the colour figure).} \end{figure*}

Whilst scientifically less relevant than the analyses described above, the discovery rate of the transiting planets is of sufficient interest to be worthy of discussion. Fig.\,\ref{fig:discrate} shows the discovery rate of the known TEPs. For discovery time I have taken the date at which important details\footnote{I define the `important details' to comprise confirmation of the planetary nature of the system (which in turn requires an assessment of the physical properties of the putative planet) and sufficient information to allow the execution of follow-up observations (i.e.\ the full sky position and orbital ephemerides).} of that system became available in a refereed journal article. If the article was lodged on the arXiv\footnote{\tt http://arxiv.org/} preprint server prior to this, I have instead taken the date at which this preprint became public.

The population of known TEPs continues to increase at an exponential rate; a plot of Fig.\,\ref{fig:discrate} but with a logarithmic ordinate axis (not shown) exhibits an approximately straight line. The early front-runners were the OGLE survey (including the follow-up observations which were performed by a range of associated and independent consortia), but the SuperWASP consortium \citep{Pollacco+06pasp} held the lead during the years 2009-11. The greatest number of known TEPs is now attributable to data obtained by the \kepler\ satellite, and once again this number includes both associated and independent groups of researchers. \kepler\ has now been used to find over 2000 candidate TEPs \citep{Batalha+12xxx} and possible as many as 5000 \citep{Tenenbaum+12apjs}; the rate of false positives is probably small but remains under discussion \citep{MortonJohnson11apj,Santerne+12xxx,Colon+12mn}. The candidate list provides the ammunition to continue the exponential rise in the number of known TEPs.

\begin{figure*} \includegraphics[width=\textwidth,angle=0]{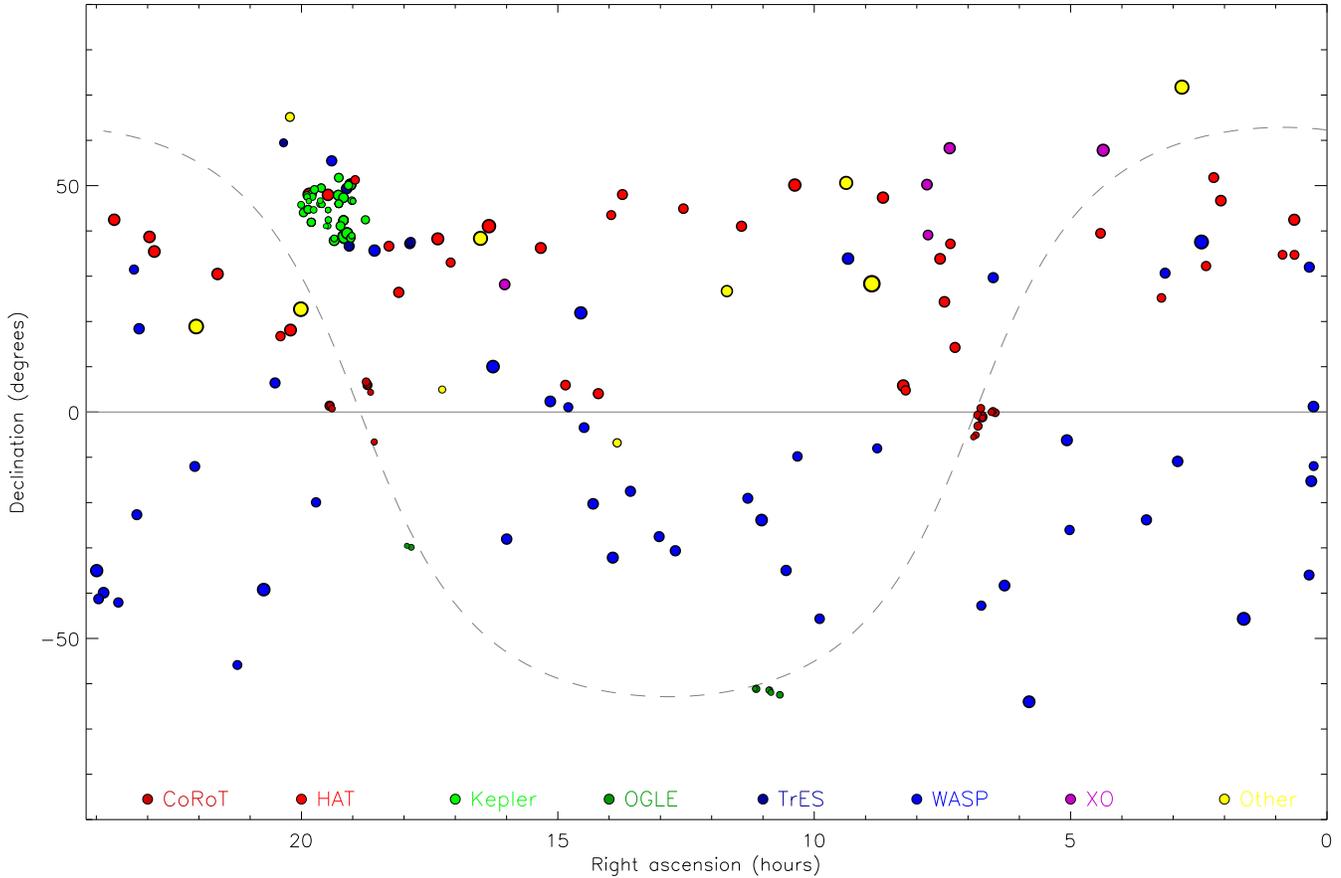}
\caption{\label{fig:skypos} Plot of the sky positions of the known TEP
systems. The celestial equator is shown using a solid grey line and the
plane of the Milky Way galaxy with a dashed grey line. The TEPs are
colour-coded to indicate those found by selected discovery surveys (please
see the electronic version of this paper for the colour figure).} \end{figure*}

Fig.\,\ref{fig:skypos} shows the sky positions of the known TEPs, overlaid with the celestial equator and galactic plane, and is a good tool to illustrate the strength and diversity of the selection effects which afflict this population. The most obvious feature in Fig.\,\ref{fig:skypos} is a strong clustering around 19\,hr right ascension and $+42^\circ$ declination, which is the \kepler\ field. The \kepler\ satellite \citep{Borucki+10sci} is currently the most successful TEP discovery machine, and has a modest field of view (105\,deg$^2$). This concentration of points is the result.

Additional clusters of TEPs occur around the intersections of the Galactic plane with the celestial equator, the two areas targeted by the \corot\ satellite \citep{Baglin+06conf}. \corot\ aims for these positions in order to obtain a high surface density of stars (to give plenty of targets in its small 8.2\,deg$^2$ field of view) which are easily followed up using ground-based telescopes in both hemispheres. Two other small groups of TEPs come via the OGLE survey \citep{Udalski+02aca} observations of the Galactic centre (18\,hr R.A.\ and $-30^\circ$ Dec) and Carina regions (11\,hr R.A.\ and $-60^\circ$ Dec).

The wider spread of objects in Fig.\,\ref{fig:skypos} is dominated by HAT \citep{Bakos+02pasp} in the north and SuperWASP \citep{Pollacco+06pasp} in the south\footnote{Declaration of bias: the author is a member of the SuperWASP consortium.}. These, plus other similar projects, operate multiple small wide-field telescopes with very coarse pixel scales. They are thus able to cover large amounts of sky but must avoid the crowded fields towards the Galactic plane, due to problems with blending. Blending makes it more difficult to measure good photometry and also elevates the rate of false alarms in a search for TEPs.

The surveys operating small ground-based telescopes tend to point away from the Galactic plane, which introduces a bias towards intrinsically faint stars. This is advantageous because it reduces the fraction of giants in the studied sample; these surveys cannot find planets around giant stars but are able to detect larger transiting object which are of lesser scientific interest. Conversely, the \kepler\ satellite studies stars with fainter apparent magnitudes which are located close to the Galactic plane. These stars have slightly different population characteristics to those studied by other surveys with different designs; \citet{SchlaufmanLaughlin11apj} show that the host stars of the \kepler\ planet candidates are preferentially metal-rich. Such biases must be accounted for when studying the population characteristics of the known TEPs and their host stars.

\section{Summary}                                                                                                            \label{sec:teps:summary}

The {\it Homogeneous Studies} project aims to provide measurements of the physical properties of a large sample of transiting extrasolar planetary systems, using consistent methods and with careful attention paid to the estimation of robust statistical and systematic errors. The transit light curves are modelled using the {\sc jktebop} code, and uncertainties are gauged with Monte Carlo and residual-permutation algorithms. Attention is paid to the treatment of limb darkening, orbital eccentricity, and contaminating `third' light. To the resulting photometric parameters are added measured spectroscopic quantities: the \Teff, \FeH\ and velocity amplitude of the host star. One additional constraint is needed, and is supplied either by one of five sets of theoretical stellar evolutionary models or by a semi-empirical calibration of low-mass star properties based on detached eclipsing binaries. The statistical errors are propagated using a perturbation algorithm, leading to complete error budgets for each output quantity. The inclusion of multiple stellar models allows systematic errors to be deduced too.

The current paper has presented complete analyses of thirty TEP systems, and updated the physical properties of eight more systems based on newly published spectroscopic results. Combined with previous work, this yields a total of 82 transiting planets and host stars with homogeneously-measured physical properties. In many cases these results are based on more numerous datasets than previous measurements, so should be preferred over literature values even if homogeneity is not of specific importance to the matter in hand. Headline results from the current paper are summarised below.

Analyses in the current work were the first to consider all available good light curves for HAT-P-3 (five datasets), HAT-P-6 (three), HAT-P-9 (four), HAT-P-14 (four), WASP-12 (four) and WASP-14 (four). For WASP-14 this has resulted in a change in planetary radius ($R_{\rm b}$) from \er{1.281}{0.075}{0.082}\Rjup\ to $1.633 \pm 0.092 \pm 0.009$\Rjup, moving this object away from the parameter space occupied by theoretical predictions.

The current study is also the first to present results based on short-cadence data for Kepler-14, Kepler-15 and KOI-135. The measured mass of Kepler-14\,A has decreased by $2.6\sigma$, and the measured radii of both components in the Kepler-15 system are both significantly larger. The density deduced for Kepler-15\,b is lower by a factor of two compared to previous studies.

Additional data was modelled for KOI-196, KOI-204, KOI-254, KOI-423 and KOI-428. In two cases (KOI-423 and KOI-428) the available data is more than a factor three more extensive than used in published studies. For KOI-428 the radii of the components are larger than previously found; the star is now comfortably the largest known known to host a TEP at \wwo{$R_{\rm A} = 2.48 \pm 0.17 \pm 0.20$}{$R_{\rm A} = 2.48 \pm 0.26$}\Rsun.

The physical properties of OGLE-TR-56 have been heavily revised based on the first complete analysis of newly-published high-quality photometry \citep{Adams+11apj2}. The radius of the planet is $R_{\rm b} = 1.73 \pm 0.06$\Rjup, which is much larger than all previous determinations (1.23--1.38\Rjup). This radius makes OGLE-TR-56\,b one of the largest known planets, as expected for its high equilibrium temperature and hot and massive host star. Its measured density is smaller by a factor of three. Previous analyses were based either on survey-quality photometry, a single transit light curve which is now known to be strongly affected by correlated noise, or on unnecessary assumptions. Two morals can be deduced from this. Firstly, one should always use a newly-derived stellar density in calculating revised physical properties of a TEP system. Secondly, results based on only one transit light curve cannot be definitive as systematic errors may lurk undetected in the data.

New results have also been calculated for OGLE-TR-111, OGLE-TR-113, OGLE-TR-132 and OGLE-TR-L9. These comprise the first complete analyses of recently published high-quality photometry for each system.

For WASP-13 and WASP-21 my analysis has resulted in the upward revision of the planetary radii, making them two of the least dense planets known. For WASP-13 $R_{\rm b}$ changes from \er{1.389}{0.045}{0.056}\Rjup\ to $1.528 \pm 0.084 \pm 0.004$\Rjup, and for WASP-21 the change is from \er{1.143}{0.045}{0.030}\Rjup\ to $1.263 \pm 0.085 \pm 0.029$\Rjup. Finally, the inclusion of an improved \Teff\ measurement makes WASP-2 one of the best-understood TEP systems.

Previously found correlations between \Porb\ and planet gravity, and \Porb\ and planet mass, have been revisited and found to be of improved statistical significance ($3.9\sigma$ and $3.8\sigma$ respectively). The division of TEPs into two classes based on their Safronov number is shown to be spurious.

I recommend further observations of specific types for the majority of the objects studied here. These are summarised in Sect.\,\ref{sec:followup}. In particular, some orbital ephemerides are of low precision and will soon be of limited use. The affected systems would benefit from new transit photometry to solve this problem. Finally, the discovery rate and distribution of sky positions of the known TEPs are plotted and discussed.

The main results from this work will be made available in a convenient format in the {\it Transiting Extrasolar Planets Catalogue} (TEPCat\footnote{The Transiting Extrasolar Planet Catalogue (TEPCat) is available at {\tt http://www.astro.keele.ac.uk/jkt/tepcat/}}).


\section*{Acknowledgments}

I acknowledge financial support from STFC in the form of an Advanced Fellowship.
I am grateful to Barry Smalley for many conversations and to Conrad Vilela for discussions about the \kepler\ archive.
I am grateful for data provided by Joel Hartmann (HAT-P-6 and HAT-P-9), Neale Gibson (HAT-P-3), Guillaume H\'ebrard (\corot-18), Jason Dittmann (HAT-P-9), Pedro Sada (HAT-P-9), Kamen Todorov (HAT-P-9) and Monika Lendl (OGLE-TR-L9).
I thank the CDS, MAST, IAS and NSTeD websites for archiving the many datasets now available for transiting planets. The following internet-based resources were used in research for this paper: the ESO Digitized Sky Survey; the NASA Astrophysics Data System; the SIMBAD database operated at CDS, Strasbourg, France; and the ar$\chi$iv scientific paper preprint service operated by Cornell University.


\bibliographystyle{mn_new}


\end{document}